\documentclass{iopart}
\usepackage{iopams}

\pdfoutput=1

\usepackage{iopams}
\usepackage{amssymb} 
\usepackage{amsfonts}
\newcommand{\text}[1]{\ensuremath{\mbox{#1}}}

\usepackage{amsthm} 
\usepackage{latexsym}
\usepackage{epsfig}
\usepackage{verbatim}
\usepackage[postscript=dvips]{diagrams}
\usepackage{color}

\newtheorem{theorem}{Theorem}[section]
\newtheorem{proposition}[theorem]{Proposition}
\newtheorem{lemma}[theorem]{Lemma}
\newtheorem{corollary}[theorem]{Corollary}
\newtheorem{definition}[theorem]{Definition}
\theoremstyle{definition}\newtheorem{example}[theorem]{Example}
 \theoremstyle{definition}\newtheorem{remark}[theorem]{Remark}

\def\bR{\begin{color}{red}}
\def\bB{\begin{color}{blue}}
\def\bM{\begin{color}{magenta}} 
\def\bC{\begin{color}{cyan}}
\def\bW{\begin{color}{white}}
\def\bBl{\begin{color}{black}}
\def\bG{\begin{color}{green}}
\def\bY{\begin{color}{yellow}}
\def\e{\end{color}}

\def\bc{\begin{center}}
\def\ec{\end{center}}

\usepackage{stmaryrd}
\usepackage{xspace}
\usepackage{graphicx}
\usepackage{xy}
\xyoption{all}

\usepackage{url}

\newcommand{\ie}{i.e.\ }

\newcommand{\sizeof}[1]{
  \left|#1\right|}

\newcommand{\catC}{\ensuremath{{\bf C}}\xspace}
\newcommand{\catD}{\ensuremath{{\bf D}}\xspace}

\newcommand{\catZX}{\ensuremath{{\bf ZX}}\xspace}

\newcommand{\diag}[1]{\ensuremath{{\mathfrak{#1}}}}
\newcommand{\diagA}{\diag{A}}
\newcommand{\diagB}{\diag{B}}

\newcommand{\diagD}{\diag{D}}
\newcommand{\diagU}{\diag{U}}

\newcommand{\calQ}{\ensuremath{{\cal Q}}\xspace}

\newcommand{\fdhilb}{
\ensuremath{\textbf{FdHilb}}\xspace}
\newcommand{\fdHilb}{
\ensuremath{\textbf{FdHilb}}\xspace}

\newcommand{\bra}[1]{
    \ensuremath{\left\langle #1 \right|}\xspace}

\newcommand{\ket}[1]{
    \ensuremath{\left|  #1 \right\rangle}\xspace}

\newcommand{\innp}[2]{
    \ensuremath{\langle #1 \mid #2 \rangle}}

\newcommand{\outp}[2]{
    \ensuremath{\left|\left. #1 \rangle\right.\langle\left. #2 \right|\right. }}

\newcommand{\state}[1]{%
  \ensuremath{\left\llbracket #1 \right\rrbracket}\xspace}

\newcommand{\CZ}{\ensuremath{\wedge {\sf Z}}\xspace}
\newcommand{\CX}{\ensuremath{\wedge {\sf X}}\xspace}

\newcommand{\bit}{\begin{itemize}}
\newcommand{\eit}{\end{itemize}\par\noindent}
\newcommand{\ben}{\begin{enumerate}}
\newcommand{\een}{\end{enumerate}\par\noindent}
\newcommand{\beq}{\begin{equation}}
\newcommand{\beqq}{\begin{equation}\hspace{-1.2cm}}
\newcommand{\eeq}{\end{equation}\par\noindent}
\newcommand{\beqx}{\begin{equation*}}
\newcommand{\beqqx}{\begin{equation*}\hspace{-1.2cm}}
\newcommand{\eeqx}{\end{equation*}\par\noindent}
\newcommand{\beqa}{\begin{eqnarray*}}
\newcommand{\eeqa}{\end{eqnarray*}\par\noindent}
\newcommand{\beqn}{\begin{eqnarray}}
\newcommand{\eeqn}{\end{eqnarray}\par\noindent}

\newarrow{Is}=====

\def\PICT{\begin{center}{\Large Picture:} }

\def\EPICT{\end{center}}

\def\HH{{\cal H}}

\def\II{{\rm I}}

\newcommand{\id}[1]{\ensuremath{\mathrm{1}_{#1}}}

\newcommand{\suck}{\hspace{-1.2cm}}

\newcommand{\inlinegraphic}[2]{
  \dimendef\grafheight=255\dimendef\grafvshift=254
  \grafheight=#1
  \grafvshift=-0.5\grafheight
  \advance\grafvshift by 0.5ex
  \raisebox{\grafvshift}{\includegraphics[height=\grafheight]{images/#2}\xspace}
}

\newcommand{\ninlinegraphic}[2][1.0]{
  \dimendef\grafheight=255\dimendef\grafvshift=254
  \setbox0 = \hbox{\scalebox{#1}{\includegraphics{images/#2}}}
  \grafheight=\the\ht0
  \grafvshift=-0.5\grafheight
  \advance\grafvshift by 0.5ex
  \raisebox{\grafvshift}{\includegraphics[height=\grafheight]{images/#2}\xspace}
}

\newcommand{\ruleT}{\ensuremath{\mathbf{T}}}
\newcommand{\ruleS}{\ensuremath{\mathbf{S}}}
\newcommand{\ruleSi}{\ensuremath{\mathbf{S1}}}
\newcommand{\ruleSii}{\ensuremath{\mathbf{S2}}}
\newcommand{\ruleB}{\ensuremath{\mathbf{B}}}
\newcommand{\ruleBp}{\ensuremath{\mathbf{B'}}}
\newcommand{\ruleBi}{\ensuremath{\mathbf{B1}}}
\newcommand{\ruleBii}{\ensuremath{\mathbf{B2}}}
\newcommand{\ruleK}{\ensuremath{\mathbf{K}}}
\newcommand{\ruleKi}{\ensuremath{\mathbf{K1}}}
\newcommand{\ruleKii}{\ensuremath{\mathbf{K2}}}
\newcommand{\ruleC}{\ensuremath{\mathbf{C}}}
\newcommand{\ruleDi}{\ensuremath{\mathbf{D1}}}
\newcommand{\ruleDii}{\ensuremath{\mathbf{D2}}}

\newcommand{\eruleT}{\ensuremath{(\ruleT)}\xspace}
\newcommand{\eruleS}{\ensuremath{(\ruleS)}\xspace}
\newcommand{\eruleSi}{\ensuremath{(\ruleSi)}\xspace}
\newcommand{\eruleSii}{\ensuremath{(\ruleSii)}\xspace}

\newcommand{\eruleBp}{\ensuremath{(\ruleBp)}\xspace}
\newcommand{\eruleBi}{\ensuremath{(\ruleBi)}\xspace}
\newcommand{\eruleBii}{\ensuremath{(\ruleBii)}\xspace}
\newcommand{\eruleK}{\ensuremath{(\ruleK)}\xspace}
\newcommand{\eruleKi}{\ensuremath{(\ruleKi)}\xspace}
\newcommand{\eruleKii}{\ensuremath{(\ruleKii)}\xspace}
\newcommand{\eruleC}{\ensuremath{(\ruleC)}\xspace}
\newcommand{\eruleDi}{\ensuremath{(\ruleDi)}\xspace}
\newcommand{\eruleDii}{\ensuremath{(\ruleDii)}\xspace}

\newcommand{\dsmc}{$\dag$-\textsc{smc}\xspace}
\newcommand{\smc}{\textsc{smc}\xspace}
\newcommand{\dsmcs}{$\dag$-\textsc{smc}s\xspace}
\newcommand{\smcs}{\textsc{smc}s\xspace}
\newcommand{\coss}{\textsc{cos}\xspace}

\newcommand{\zxcalculus}{\textsc{zx}-calculus\xspace}

\begin{document}

\title[Interacting Quantum Observables]{Interacting Quantum Observables: \\Categorical Algebra and Diagrammatics}

\author{Bob Coecke$^1$ and Ross Duncan$^2$}
\address{
  $^1$Oxford University Computing Laboratory\\
  Wolfson Building, Parks Road, Oxford OX1 3QD, UK
}
\address{
  $^2$Laboratoire d'Information Quantique,  
Universit\'{e} Libre de Bruxelles\\ 
  Boulevard du Triomphe, B-1050, Bruxelles, Belgium
}
\ead{
 $^1$\mailto{coecke@comlab.ox.ac.uk}  \ 
 $^2$\mailto{rduncan@ulb.ac.be}
}

\begin{abstract}
  This paper has two tightly intertwined aims: (i) To introduce an 
  intuitive and universal graphical calculus for multi-qubit
  systems, the \zxcalculus, which greatly simplifies derivations in the
  area of quantum computation and information.  (ii) To axiomatise
  complementarity of quantum observables within a general framework
  for physical theories in terms of dagger symmetric monoidal
  categories.  We also axiomatize phase shifts within this framework.


  Using the well-studied canonical correspondence between graphical
  calculi and dagger symmetric monoidal categories, our results provide a
  purely graphical formalisation of complementarity for quantum
  observables.  Each individual observable, represented by a
  commutative special dagger Frobenius algebra, gives rise to an abelian
  group of phase shifts, which we call the phase group.  
 We also identify a strong form of complementarity, satisfied by the $Z$ and
  $X$ spin observables, which yields a scaled variant of a bialgebra.

\end{abstract}

\tableofcontents
\maketitle

\section{Introduction}

Quantum theory is arguably the single most successful scientific
theory.  While it is now almost a century old, many new results have been
discovered by approaching quantum theory from an computational and/or
information theoretic perspective, signalling the potential for a
quantum information technology revolution.  This approach has also led
to important progress in more traditional areas of physics, for
example, in condensed matter physics and statistical physics,
e.g.~\cite{AmicoVedral_etal}, and it has provided a breath of fresh
air for quantum foundations research
\cite{Hardyaxioms,Fuchs,Spekkens,Barrett}.  Most importantly, this
recent wave of progress has clearly shown that much still remains to
be discovered  concerning the quantum world, and how we reason about it.

Since von Neumann's seminal book in 1932, the \em language \em in
which quantum theory is explained and is understood has been (and
still is) that of Hilbert spaces.  It is in this language that we
understand key quantum mechanical concepts such as observables and
complementarity thereof.  While quantum information and computation (QIC) has proposed new
concepts and paradigms to approach the quantum world, it has not
augmented the language of quantum theory accordingly.  This is in
sharp contrast with the typical practice in computer science, where
new perspectives and concepts are tightly intertwined with
corresponding high-level language features.
To make a blunt analogy, we can think of the Hilbert space
formalism, where states mainly boil down to arrays of complex numbers,
on the same footing as the arrays of 0's and 1's used during the
stone age years of computer science.  So one may wonder:
\[
\hspace{0.8cm}{\mbox{high-level languages}\over b_1 b_2  \ldots b_n\in \mathbb{B}^n}\simeq{\mbox{``our aim''}\over
(c_1\  c_2\    \cdots\ c_n )^T\in \mathbb{C}^n}
\]
where $\mathbb{B}^n$ stands for strings of Booleans $\{0,1\}$ and
$\mathbb{C}^n$ for vectors of complex numbers.

A related issue is that of \em axiomatizing \em quantum theory.
Despite its obvious correctness, as a language to describe quantum
theory, the Hilbert space formalism seems somewhat ad hoc from a
conceptual perspective.  The first to acknowledge this was von
Neumann himself, who for this reason denounced his own
Hilbert space formalism in 1935 (see \cite{Birk2}), 
only three years after he published
it.  There have been many attempts to approach quantum theory in terms
of mathematical structures other then Hilbert spaces \cite{CMW}, in
the hope that this would enhance conceptual insight, but it is fair to
say that none of these has provided a sufficient payoff, if any at
all.

The recent advent of QIC has shed significant new light on this
issue. None of the axiomatic approaches of the previous century
provided an adequate mathematical vehicle for the description of
compound systems, even when given the description of individual
systems.  On the other hand, focussing on compoundness has produced
immense progress within QIC.  This includes important foundational
insights such as the \em no-cloning theorem \em \cite{Dieks,WZ},
physical phenomena such as \em quantum teleportation \em \cite{BBC},
quantum algorithms such as \em polynomial time factoring \em
\cite{Shor:PolyTimeFact:1997}, and computational schemes such as \em
measurement-based quantum computing \em \cite{RB01}.  Historically
speaking, it was Schr\"odinger who emphasised compoundness as early as
1935 \cite{Schrodinger}.

In this paper we aim to catch two flies at once. We introduce a
simple, intuitive, graphical high-level language, in which the atomic
primitives correspond to a pair of complementary observables, and we
perform an axiomatic analysis of complementarity within the very
general framework of symmetric monoidal categories (\smcs).  These two are
related by the fact that there is a tight correspondence between
graphical languages and \smcs \cite{JS,LNPS},
tracing back to Penrose's work on tensor networks \cite{Penrose}.

The diagrammatic notation is intuitive in use, but also formally
rigorous (see Section~\ref{DagSMS_scalars}), and can lead to great
simplications in proofs.  From a pragmatic point of view, the
graphical language provides a compact syntax for manipulating the
linear operations which are the basic elements of quantum mechanics,
and it can replace more special purpose notations such as \em quantum
circuits \em \cite{NieChu} or the \em measurement calculus \em for
measurement-based quantum computing \cite{DKP}, and unify these in one
setting.

From an axiomatic point of view, monoidal categories are the most
general mathematical framework where composing \emph{systems} (cf.~the
tensor product `$\otimes$' in the Hilbert space framework) is a
fundamental action -- see \cite{ContPhys} for a more detailed
discussion.  Since its inception in \cite{AC1}, formulating quantum
mechanics within monoidal categories and developing corresponding
diagrammatic languages has become an active area of research.

The bottom line is: crafting a simple intuitive graphical high-level
language on the one hand, and performing an axiomatic study which
places composition of systems at the forefront on the other hand, are
in fact one and the same thing!
 
 
Our particular focus here is \em complementarity \em of quantum
observables.  In classical physics all observables are \em
compatible\em: they admit sharp values at the same time.  In contrast,
quantum observables are typically \em incompatible\em, and cannot be
assigned sharp values simultaneously. In most axiomatic approaches
incompatibility is a negative property, captured in mathematical terms
by the fact that some equality fails to hold:  operators which do not
commute \cite{Haag},  probabilities which fail to obey Kolmogorov's
axioms \cite{Pitowski}, convex sets which fail to provide a simplex
structure \cite{Ludwig1,Ludwig2}, and lattices which do not enjoy
distributivity
\cite{BvN,Jauch}. 

In this paper we will take a more constructive stance and study the
positive \emph{capabilities} of a pair of maximally incompatible
observables, called \emph{complementary} or \emph{unbiased}, and show
how these capabilities are exploited in QIC.  Doing so will lead to an unexpected connection between
quantum computation and the area of \em Hopf algebras\em and
\em quantum groups \em \cite{Cartier,Kassel}, where graphical methods
have also proved to be very fruitful \cite{StreetBook}.   

All together, we obtain a rich theory from rather
minimal hypotheses.  Many computations with elementary quantum
logic gates can be carried out within this theory of \emph{interacting
  observables}, as can many algorithms and protocols.  To give one
very basic example, the fact that the composite of two \CX-gates is
the identity boils down to the graphical derivation: \bc
\includegraphics[width=162pt]{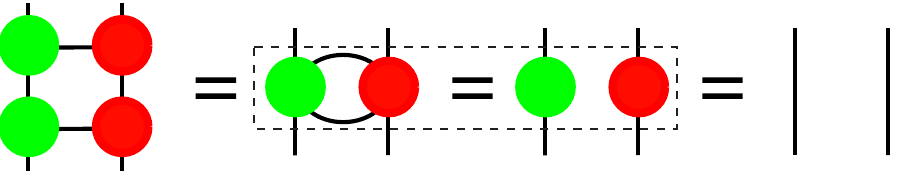}
\ec
where the dotted area is a purely graphical characterisation of
complementarity. 
%
%
%

In the example above, we reasoned by \emph{rewriting}: that is, by locally
replacing some part of a diagram with a diagram equal to it.  This is
one of the distinctive methods of equational reasoning in graphical
lanaguages.  The notion of rewriting as formal mathematical tool has a
long history in computer science (the text books
\cite{Franz-Baader:1999aa} and \cite{Ehrig:2006aa} provide detailed
references), and the \zxcalculus introduced in this paper has indeed
been implemented in a software tool
\cite{Lucas-Dixon:2009yq,Dixon:2010aa,quantomatichome}.  

Specific physical concepts  give rise to specific kinds of
equations over diagrams.  As the example above shows, complementary
observables introduce changes in topology, characterised by
disconnecting components between the red and green dots.  On the other
hand, in the case of compatible observables, connected components can
be contracted \cite{Lack,CPaq}.  The following table illustrates this:
the green components are defined in terms of one observable, and the
red ones in terms of a complementary one.
 \begin{center} 
\begin{tabular}{|r|c|c|} 
\hline
\em compatible (self-)interaction: \em
     & \inlinegraphic{5.0em}{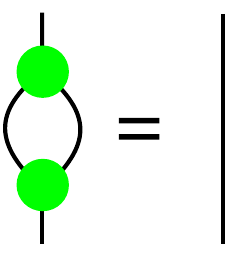}
     & \inlinegraphic{3.0em}{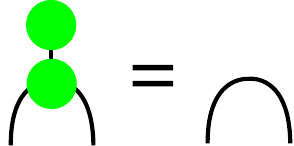}
     \\ \hline
\em complementary interaction: \em
     &\inlinegraphic{5.0em}{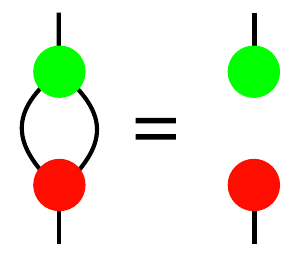} 
     & \inlinegraphic{3.0em}{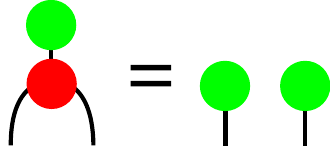}
     \\ \hline
\end{tabular}
\end{center} 
For both of the depicted interactions, complementarity yields two
disconnected components, while for compatible observables
connectedness is preserved.  This topological distinction has very
important implications for the capabilities of complementary
observables in quantum informatics.  The disconnectedness of the
graphical form shows the absence of information flow from one
component to the other, a \emph{dynamic} counterpart to the fact that
knowledge of one observable in a pair of complementary observables
yields no knowledge of the other observable.  
 
 We also provide an axiomatic account of \em phase shifts \em relative
to an observable.  This leads to the mathematical concept of a \em
phase group\em. Together, our account on complementarity and phase
groups provides a universal language for reasoning about multiple
two-level systems, or in modern language, qubits.
For example,
  \begin{equation*}\fl\qquad
    \ninlinegraphic[0.63]{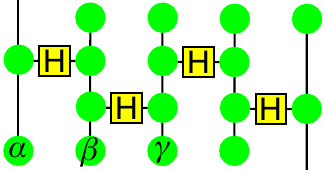} =
    \ninlinegraphic[0.63]{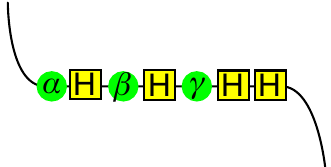} =
    \ninlinegraphic[0.63]{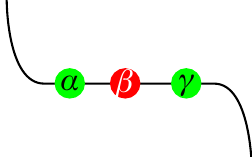} 
  \end{equation*}
is an important computation in the context of measurement based quantum
computing \cite{RBB}, which in Hilbert space terms would involve
computations with $32\times 32$ matrices.  This example provides a
straightforward  translation between quantum computational models,
transforming a measurement-based configuration into a circuit.

From a mathematical perspective, we formalise observables in terms of
algebras: Frobenius algebras, 
bialgebras, etc.  These structures do not depend on having an
underlying Hilbert space, or indeed any linear structure whatsoever,
therefore we can study complementary observables at a much greater
level of generality than the usual Hilbert space formulation of
quantum mechanics.  The results will apply in any `quantum-like'
theories which bear the necessary algebraic structures.  The minimal
mathematical environment to support these structures is generally a
\emph{dagger \smc} or \dsmc \cite{AC2, Selinger}.  By working in an
\smc, we can study the central features of quantum mechanics and
quantum computation, without reference to Hilbert space at all. This
research program was initiated by Abramsky and one of the authors in
\cite{AC1}.

In previous work it was already established that the observables
themselves correspond to certain \em commutative Frobenius algebras
\em \cite{CPav,CPV}. We now explain how conceptual analysis leads
to this algebraic structure,
via a contrapositive of the no-cloning theorem \cite{Dieks,WZ}.
%
%

While the no-cloning theorem suggest a fundamental limitation of QIC
compared to its classical counterpart, a positive reading of it
reveals 
that quantum states may be copied \emph{if they are known to lie in a
  given basis}.  In other words, a quantum state may be treated as
classical data, and therefore copied freely, if it is an eigenstate of
a known, non-degenerate observable.
(Throughout this paper
  we will treat ``orthonormal basis'' and ``non-degenerate
  observable'' as synonyms, and commit abuses like ``measuring against a
  basis'' and so on.)
More concretely, given a finite dimensional Hilbert space $\HH$ with a basis
${\cal A} = \{ \ket{a_i}\}_i$, the \em copying operation \em
 \[
  \delta : \ket{a_i} \mapsto \ket{a_i} \otimes \ket{a_i}
 \]
 encodes the basis ${\cal A}$ as those states that it effectively copies; 
the no-cloning theorem guarantees that the basis vectors are the only states with this property.
 Note here that $\delta$
 may be realised as a unitary map on $\HH \otimes \HH$ with one input
 fixed, for example, by $U:\ket{a_i}\otimes \ket{a_j} \mapsto \ket{a_i} \otimes \ket{a_{i+j}}$ where the sum is taken in $\mathbb{Z}_n$.  

 Now, let $\epsilon$ be the linear functional on $\HH$ defined
 by $\ket{a_i} \mapsto 1$ for each $i$. In more conceptual terms,  $\epsilon$ uniformly
\em  erases \em the elements of the basis ${\cal A}$.  Further, when
 $\epsilon$ is applied to an output of $\delta$ we get the identity
 map:
 \[
  (\id{\HH} \otimes \epsilon) \circ \delta = \id{\HH} = (\epsilon
  \otimes \id{\HH} ) \circ \delta\,.
 \]
 In algebraic terms, $\epsilon$ is the \emph{co-unit} for the
 \emph{co-multiplication} $\delta$.  
 
 Together the pair $(\delta,
 \epsilon)$ form a \emph{special commutative $\dag$-Frobenius algebra}
 on $\HH$.  Previous work established the remarkable fact that
 \emph{every} algebra of this 
 kind on a finite dimensional Hilbert space arises as pair of copying
 and erasing operations for some orthonormal basis \cite{CPav,CPV}.  Since
 these algebras correspond precisely to non-degenerate quantum
 observables, we refer to them as \emph{observable structures}.
 Observable structures $(\delta, \epsilon)$ and $(\delta', \epsilon')$
 which correspond to complementary observables enjoy a special relationship:
 the main body of this paper is dedicated explicating just that
 relationship, and a great deal of additional algebraic structure that
 follows.







\medskip

{{\bf Structure of this paper.}}  This paper contains two self-contained
parts, each of which could be read independently of the other:

{\bf\em Part I.}
Comprising Sections~\ref{Sec:ZXCalc} and
\ref{sec:zx-calculus-use}, the first part is an informal
presentation of a graphical calculus based on the interaction of
complementary observables.  Effectively we begin at the end, by
presenting a calculus that demonstrates many of the
key ideas of the theory, but without presenting the theory
itself until Part II.  It also serves to familiarise the
reader with graphical reasoning, a tool that we will use throughout
this paper.  We rely here on some familiarity with
quantum computing terminology for the examples, but no other
background.

Section 2 introduces the \emph{\zxcalculus}, a graphical language and a set of
equational rules which are based on the Pauli $Z$ and $X$ spin
observables, and  specially tuned for use in quantum
computation.  Quantum systems are represented as diagrams,  and
these can be rewritten according to the equations in order to
prove statements about the corresponding quantum systems.
This language is universal in the sense that any operation
on $n$ qubits can be expressed in it, as shown in Section
\ref{sec:universality}.   

In Section~\ref{sec:zx-calculus-use} we demonstrate a variety
applications:  simulating quantum circuits, and transforming
measurement-based computations into equivalent circuits for example.
These examples are small, but the \zxcalculus is appropriate for real
use, and has been used to prove non-trivial results in this area
\cite{DunPer2010}.

{\bf\em Part II.}
The main body of the paper, Sections~\ref{DagSMS_scalars} to
\ref{SEC:GroupActionEtc} provide an axiomatic analysis of
complementary observables within the general framework 
of \em \dsmc\em.  Throughout, we will use graphical
notation as much as possible.

From this point onwards, the Pauli $Z$ and $X$ spin
observables will only be one example among all possible pairs of 
complementary observables.  This will reveal  additional properties
enjoyed by the $Z$ and $X$ observables, as compared to other pairs of
complementary observables. Also from this point onwards, Hilbert spaces are
simply one particular model of the axiomatic abstract algebra, and
since interpretations in other models may be useful, concepts will
be introduced in full generality.  For example, the observables in 
Spekkens' toy theory \cite{Spekkens},  are also captured by our analysis.

Section~\ref{DagSMS_scalars} reviews the necessary category theoretic
background, in particular \dsmcs, and
their graphical notation.  We rely on the work of Joyal and Street
\cite{JS} and Selinger \cite{LNPS} to establish the validity of the
graphical calculus as a rigorous mathematical syntax, and not simply a
sketch.

Returning briefly to the concrete Hilbert space setting,
Section~\ref{sec:quant-mech-concr} defines the notions of \emph{state
  basis} and \emph{coherent unbiased basis} for Hilbert spaces, and
studies their relation to quantum observables. These concepts play a
key role in this paper, in abstract form, and to our knowledge have
not appeared in the literature
yet.  

The technical core of the paper begins with
Section~\ref{sec:algebras-observables}, which provides the definition
of observable structure---a.k.a. special commutative $\dag$-Frobenius
algebra---and establishes its basic properties, including the `spider
theorem', giving the normal form for expressions in the language of
observable structures.  Before arriving at the definition of
complementarity, in Section~\ref{sec:gener-spid-theor} we provide a
category-theoretic account of an important related concept, namely the
\emph{phase} relative to an observable.  Every observable structure
gives rise to an abelian group of phases, which behave
particularly well with respect to the normal form theorem for diagrams
involving observables.  We refer to this result as the `decorated
spider' theorem.

In Section~\ref{SEC:Complementarity} we characterize complementarity
for observable structures. 
In Section~\ref{sec:bialgebra} we identify a special kind of
complementary observables, which we refer to as \emph{closed}. These
include the complementary observables that are relevant to quantum
computing.  We moreover provide further, equivalent, characterisations
of these closed complementary observables.  All of these equivalent
characterisations take the form of some sort of commutativity, be it
either commutativity of multiplication and a comultiplication,
commutativity of a multiplication and an operation, or commutativity
of operations.  These commutation properties present a remarkable
contrast with the usual characterisation of incompatibility as
non-commutativity.  The technical development concludes in
Section~\ref{SEC:GroupActionEtc}, by examining how the phase groups of
complementary observables act on each other to produce `interference'
phenomena.

{\bf\em  Part III: Coda.}
\label{sec:coda}
Section~\ref{SEC:Customising} returns to the beginning by
demonstrating how the general theory expounded in the Part II produces the
\zxcalculus of Part I.  We note which rules hold on other pairs of
complementary observables, and show where the particular features of
the  $Z$ and $X$ observables appear in the calculus.

Finally, Section~\ref{sec:non-determ-mixed} addresses the most obvious
omission thus far;
it deals with non-determinism and classical data flow.

\medskip

{{\bf About this paper.}} The genesis of the current paper was an
attempt to apply observable structures \cite{CPaq,CPav}---then called
\emph{classical structures}---to a diagrammatic notation for
measurement-based quantum computation \cite{Duncan:thesis:2006}.  An
initial report on these results was first presented at the
\textsc{icalp} conference in 2008 \cite{CoeDun2008}, albeit under
severe space restrictions.  During the intervening period the theory
was under active development in Oxford, and several papers have
appeared making use of the key ideas and applying them in various
settings: in measurement-based quantum computation
\cite{DunPer,DunPer2010}, in the study of Spekkens' toy theory and non-locality
\cite{Spek,CES}, quantum protocols \cite{CPer}, complementarity in the
category of relations \cite{Spek,Dusko,Evans:2009aa}, among others.  This
paper is the first complete presentation of our categorical treatment
of complementary observables, and it corrects several errors in the
earlier paper.

\section{The ZX (or green-red) graphical calculus}\label{Sec:ZXCalc}

The state space of the elementary quantum  computational unit,  the  \emph{qubit}, is denoted by ${\cal
  Q} :=\mathbb{C}^2$. The vectors of the \em computational  basis \em or \em $Z$-basis\em, are written $\ket{0}, \ket{1}$,   
  while those of the \em $X$-basis \em are written 
\[
\ket{+}=\frac{1}{\sqrt{2}}(\ket{0}  + \ket{1})\,,
\qquad
\ket{-}=\frac{1}{\sqrt{2}}(\ket{0}- \ket{1})\,.
\] 
On the Bloch sphere these bases can be represented as follows:
\bc
\inlinegraphic{8em}{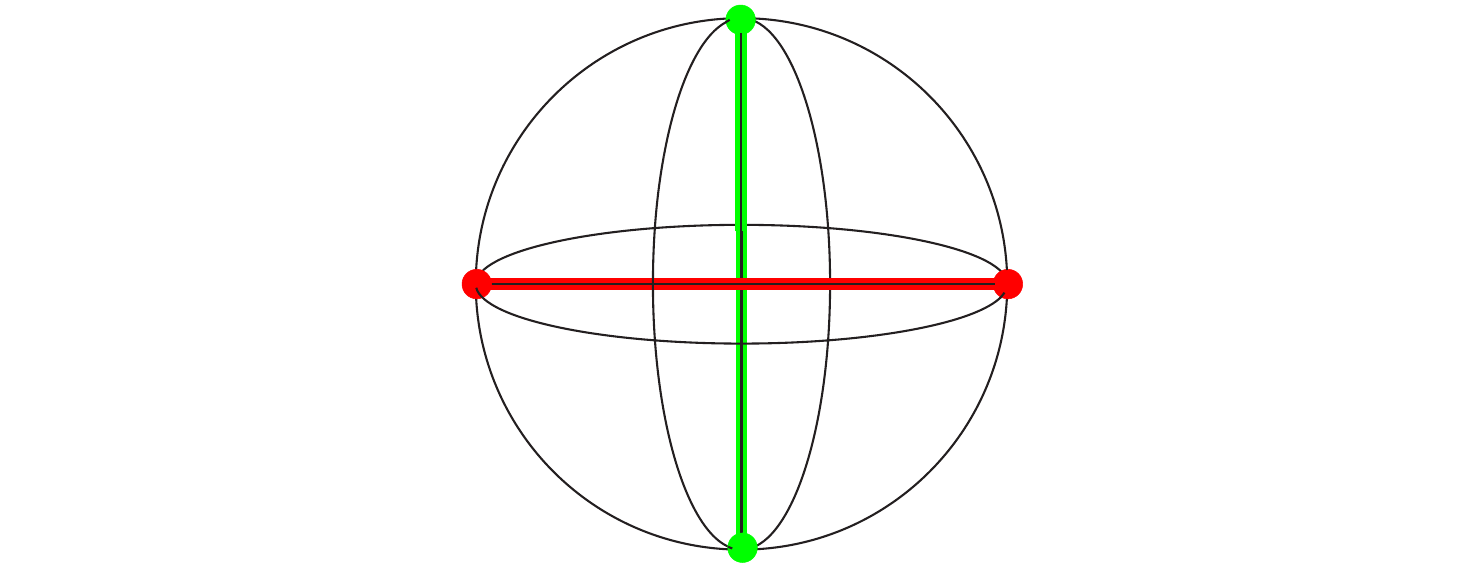}
\ec
where the green dots represent the elements of the $X$-basis and the red dots represent those of the $Z$-basis.

These bases consist of the eigenvectors of the Pauli spin matrices,
\beqq
{\sf Z} = \left(
   \begin{array}{rr}
     1&0\\ 0&-1
   \end{array}
\right)
\qquad
{\sf X} = \left(
   \begin{array}{rr}
     0&1\\ \ 1&\ 0
   \end{array}
\right)\;,
\eeq
and correspond to the possible outcomes upon measuring the spin of
the electron along the $Z$ and $X$ axes respectively.  
Our interest in these particular spins stems from the fact that they
are the simplest example of \emph{complementary observables}. 

In this section we will present a graphical calculus, specific to the
$Z$- and $X$-spin observables, which is a special case of the
general theory which we develop later in this paper.  As well as
demonstrating the main features of the full theory, this simplified
calculus is sufficiently powerful to carry out many calculations
useful in the context of quantum computation, as the examples in
Section~\ref{sec:zx-calculus-use} will demonstrate.

This framework refers exclusively to the mathematics
underlying quantum computation and not to any details of how the
operations are implemented, which makes it ideal for unifying
various approaches to quantum computation.  
For example, we can demonstrate equivalence between
different quantum computational models.

\subsection{The ZX language: networks of wires and dots}

The \zxcalculus consists of components joined by wires,
similar to electronic ciruit diagrams or flow charts.
The simplest non-trivial diagram in the language is simply a wire
running from top to bottom:
\bc
$\id{\cal Q} = \inlinegraphic{22mm}{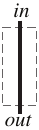}$
\ec
We think of diagrams as being enclosed in a box with a
certain number of points through which wires enter and leave; that
is, each diagram has a fixed \emph{interface}.  Exactly one wire
must be present at each point of the interface, and we must distinguish
which wire is connected to which point.  Indeed, this
is the only difference between the two diagrams below.
\bc
$\id{{\cal Q}\otimes{\cal Q}} = \inlinegraphic{22.5mm}{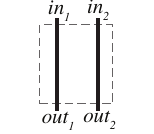}
\qquad
\sigma_{\cal Q} = \inlinegraphic{22mm}{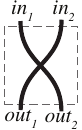}$
\ec
It is not important whether crossing wires pass over or
under (\ie we are in a symmetric setting, not a \em braided \em one \cite{JS2}).
Wires may bend, linking two outputs to form a cap, or two inputs to
form a cup.
\bc
$ \eta_{\cal Q} = \inlinegraphic{18.7mm}{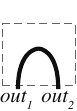}
\qquad 
\epsilon_{\cal Q} = \inlinegraphic{18.7mm}{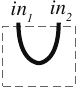}  
$
\ec
From here on, the inputs and outputs will not be named, and
are distinguished simply by their ordering from left to right.  We
write $\diagD : m \to n$ to indicate that the  diagram $\diagD$ has
$m$ inputs and $n$ outputs.

Aside from wires, the \zxcalculus contains four kinds of component:
\begin{itemize}
\item $Z$ vertices (green dots), labelled by an angle $\alpha \in
  [0,2\pi)$, called the \emph{phase}.
These can have any number of inputs or outputs (including none).
\item $X$ vertices (red dots), labelled by a phase.
These too can have any number of inputs or outputs (including none).
\item $H$ vertices (yellow squares).  These must have exactly one
  input and one output.
\item $\sqrt{D}$ vertices (black diamonds).  These may not have any inputs
  nor any outputs.
\end{itemize}
\bc $
Z^n_m(\alpha) = \underbrace{\overbrace{\ninlinegraphic{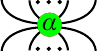}}^n}_m
\quad
X^n_m(\alpha) = \underbrace{\overbrace{\ninlinegraphic{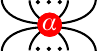}}^n}_m
\qquad
H = \inlinegraphic{3.0em}{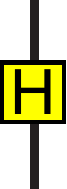}
\qquad
\sqrt{D} = \inlinegraphic{4.5mm}{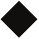}
$ \ec
We refer to the $Z$ and $X$ vertices as ``spiders'', and make the
convention that if $\alpha = 0$ the angle is omitted.  

Diagrams are built from these generators---straight, crossing, and
bending wires, and $Z$, $X$, $H$, and $\sqrt{D}$ vertices---in two manners.
\begin{itemize}
\item Placing them side-by-side:
  \begin{center}
    \ninlinegraphic{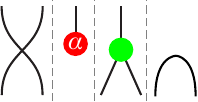}
  \end{center}
  \emph{Notation:} Given $\diagD: m \to n$ and $\diagD':m' \to n'$
  their \emph{tensor product} is denoted $\diagD\otimes\diagD':m+m'
  \to n+n'$.
\item Connecting outputs to inputs:
  \begin{center}
    \ninlinegraphic{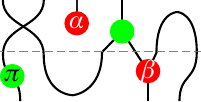}
  \end{center}
  \emph{Notation:} Given $\diagD_1: m \to n$ and $\diagD_2:n \to k$
  their \emph{composition} is denoted $\diagD_2\circ\diagD_1:m\to k$.
\end{itemize}
Therefore the terms of graphical ZX lanaguage are networks of vertices
of each type, straight, crossing, and bent wires:
\bc
$\inlinegraphic{6.6em}{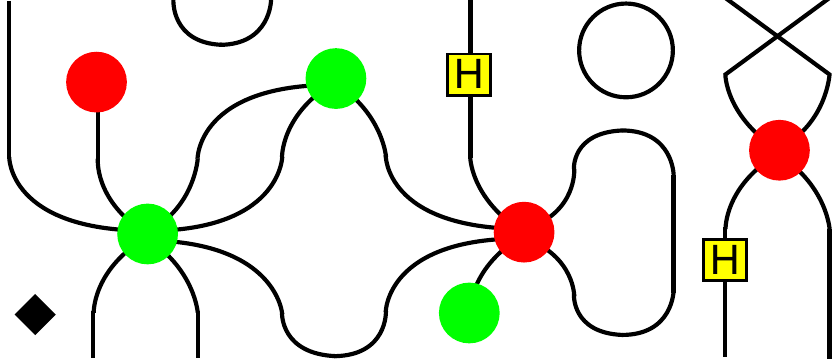}\ $.
\ec
In such a network, there can be no ``loose wires'':  every wire must
terminate at a vertex, or else be an input or output.

Important examples are those spiders with 2 inputs and 1 output
(cf.~a binary operation), with no input and 1 output (cf.~initiation of a
value) which we will call a \em point\em, with 1 input and 2 outputs
(cf.~copying) and with 1 input and no output (cf.~erasing):
\bc
\inlinegraphic{2.00em}{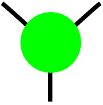}\qquad\quad\inlinegraphic{1.73em}{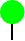}\qquad\quad
\inlinegraphic{2.00em}{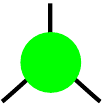}\qquad\quad\inlinegraphic{1.73em}{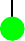}
\ec
As we will see shortly, these unlabelled spiders play a special role in the
calculus, as do those labelled by $\pi$.


\subsection{The ZX equational rules}\label{sec:simplegreenredrules}

\begin{figure}[htb]
  \centering
  \framebox[\linewidth]{
  \begin{minipage}{0.98\linewidth}
    \begin{tabular}{cc}
      \emph{``Only the topology matters''}
      & \eruleT \\[1em]
      \inlinegraphic{5.13em}{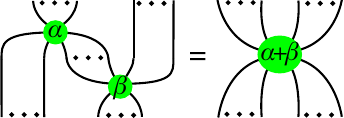} 
      \quad
       \inlinegraphic{5.13em}{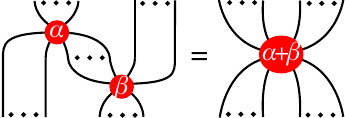}$\!\!\!\!\!\!$
      & \eruleSi \\[3em]
      \inlinegraphic{2.7em}{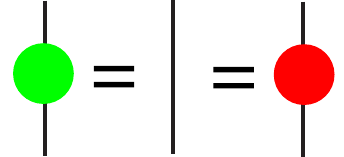} 
      \qquad
      \inlinegraphic{3.50em}{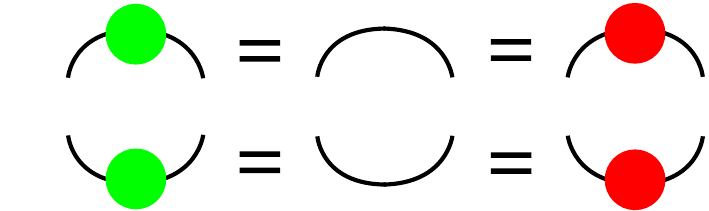}
      & \eruleSii      \\[2em]
      \inlinegraphic{8.5mm}{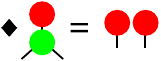}
      \quad
      \inlinegraphic{8.5mm}{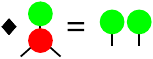}
      \quad
       \eruleBi 
      \quad\ \
      \inlinegraphic{4.0em}{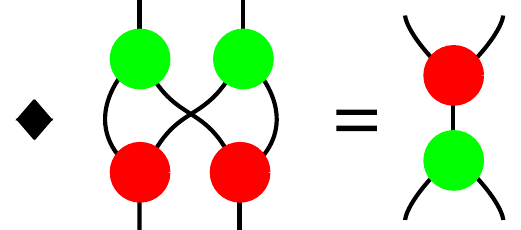}\quad\  
      & \eruleBii \\[2em]
     $\!$ \inlinegraphic{11.5mm}{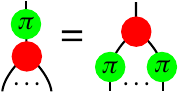}
      \  \, $\!\!$
      \inlinegraphic{11.5mm}{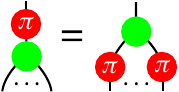}
      \  $\!$
       \eruleKi
       \quad\ \ \
      \inlinegraphic{10.2mm}{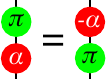}     
      \ \ 
      \inlinegraphic{10.2mm}{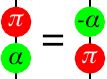}$\!\!$
      & \eruleKii \\[2em]
        $\inlinegraphic{8mm}{zx-gens-redspider-alpha} = \ 
        \inlinegraphic{12.6mm}{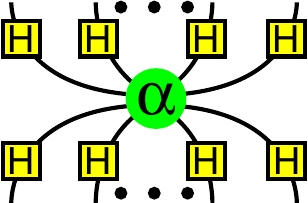}$
      & \eruleC \\[2em]
  \  \qquad  \qquad  $\inlinegraphic{8mm}{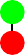} =\ 
  \inlinegraphic{3mm}{black-diamond}$
      \qquad \qquad\quad \eruleDi \qquad \quad
      \inlinegraphic{2.0em}{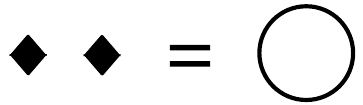}
      & \eruleDii \\[1em]
    \end{tabular}
  \end{minipage}}
\caption{Rules for the \zxcalculus}  \label{fig:ZX-rules}
\end{figure}

In addition to the rules for constructing diagrams, the calculus
consists of a set of equations which specify how one diagram may be
transformed into another.  These rules are presented in
Figure~\ref{fig:ZX-rules}.  We now expand on these rules and give some
examples of their use.


\subsubsection{The \textbf{T}-rule.}
\label{sec:t-rule}

The informally stated {\bf T}-rule  will be made more precise
in Section~\ref{subsec:Graph_calc}--\ref{sec:corrgraphreasonZX}.
For practical purposes, the intuitive reading of ``only the topology
matters'' suffices:  the wires of the diagram may 
be arbitrarily stretched, bent, twisted, tied in knots, etc.,
without altering the meaning of the diagram, provided the connections
are maintained. More precisely, after identifying (e.g.~by
enumerating) the inputs and the outputs, any topological deformation
of the internal structure of the network yields a  network that is
equal to the given one.

Two important examples of such `homotopic rewrites'
are: 
\bc
$\inlinegraphic{3.8em}{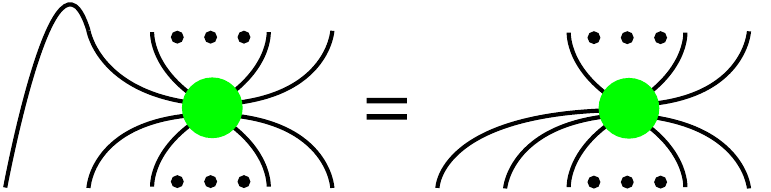}\ \ ({\bf T_1})\qquad\qquad\inlinegraphic{3.4em}{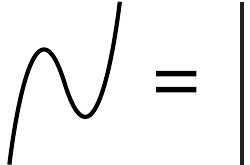}\ \ ({\bf T_2})\ $. 
\ec
In fact, these two rules can also be seen as consequences of the {\bf
  S}-rules, when introducing a green dot on the caps and cups as in
(\textbf{S2}); see Example \ref{ex:Trulesderivation} below.  The reason for considering them within the
{\bf T}-rule will become clear in Section
\ref{sec:corrgraphreasonZX}.  

\begin{remark}
  Since wires can be stretched without consequence, adding a straight
  length of wire to the input or output of diagram has strictly no
  effect.  Hence bundles of straight wires act as \emph{identity
    elements} in the algebra of diagrams.
\end{remark}

\begin{remark}
  While the slogan says ``only the topology matters'', this does not
  imply that the topology is always preserved.  The other rules may
  change the topology of the diagram in various ways, for example to
  remove loops, or to disconnect previously connected vertices.
\end{remark}

\subsubsection{The \textbf{S}-rules.}
\label{sec:s-rules}

The ``spider'' rules govern how dots of the same colour
interact. Rule  \eruleSi states that connected dots of the same
color can be merged, summing the phases; conversely, a dot can be
`decomposed' along one or more connecting wires.  Notice that the
number of connecting wires is irrelevant.

The equations \eruleSii specify when spiders are trivial: dots of
degree 2 with phase $\alpha = 0$ can be removed, or conversely,
introduced.

\begin{example}\label{ex:zx-commutative-monoid}
  If we view the dot $Z^2_1 : 2 \to 1$ as a binary operation, \eruleSi
  tells us that it is associative:
  \bc $
  \ninlinegraphic{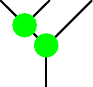} \stackrel{\eruleSi}{=} 
  \ninlinegraphic{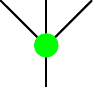} \stackrel{\eruleSi}{=} 
  \ninlinegraphic{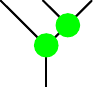}
   $ \,.
  \ec
  Less obviously, \eruleSi implies that this operation is commutative:
  \bc $
  \ninlinegraphic{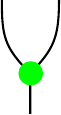} \stackrel{\eruleSi}{=} 
  \ninlinegraphic{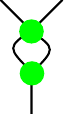} \stackrel{\eruleT}{=} 
  \ninlinegraphic{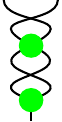} \stackrel{\eruleSi}{=} 
  \ninlinegraphic{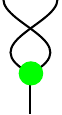}
   $\,.
   \ec
   We leave the reader the (easy!) exercise of showing that $Z^0_1$ is a
   unit for this operation, and hence that we have a \em commutative 
   monoid\em.\footnote{I.e.~a set with a commutative associative unital operation.} 
\end{example}
  
\begin{example}\label{ex:Trulesderivation}
  The $({\bf T_2})$ rule can be derived using the \ruleS-rules:
  \bc 
  $
  \ninlinegraphic{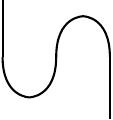} \stackrel{\eruleSii}{=}
  \ninlinegraphic{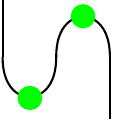} \stackrel{\eruleSi}{=}
  \ninlinegraphic{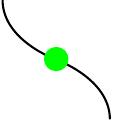} \stackrel{\eruleSii}{=}
  \ninlinegraphic{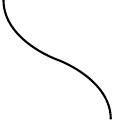} $
  \ec
  The $({\bf T_1})$ rule is derived similarly.
\end{example}

Mathematically, the two \ruleS-rules state that each family of
coloured dots forms a \emph{special commutative dagger-Frobenius algebra},
equipped with a \emph{phase group}.  This will be elaborated upon in
Sections~\ref{sec:algebras-observables} and \ref{sec:gener-spid-theor}.

\subsubsection{The \textbf{B}-rules.}
\label{sec:b-rules}

The  \ruleBi-rule can be read loosely as
``green copies red points'' and ``red copies green points'', in both
cases ``up to a diamond''.

The \ruleBii-rule is a powerful commutation principle, and
generates a whole family of equations, allowing alternating cycles of
red and green dots to be replaced with simpler graphs;  see \cite{DunPer}.

\begin{example}\label{ex:ZX-derive-hopf}
  An important equation derivable from the \ruleB-rules 
  is the following:
  \bc\qquad\ \  $
  \ninlinegraphic{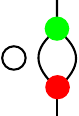} = \
  \ninlinegraphic{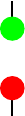}    \quad \ \ \ ({\bf
    B}')$   
  \ec
This equation is obtained as follows: 
\begin{eqnarray*}
  \fl  
  \ninlinegraphic[0.75]{hopf-proof-informal-1}  \stackrel{\eruleT}{=} 
  \ninlinegraphic[0.75]{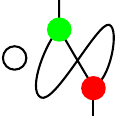}  \stackrel{\eruleS}{=} 
  \ninlinegraphic[0.75]{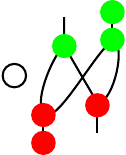}  \stackrel{\eruleDii}{=} 
  \ninlinegraphic[0.75]{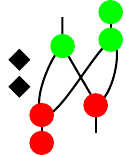}   \stackrel{\eruleBii}{=} 
  \ninlinegraphic[0.75]{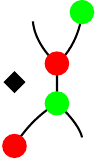}  \stackrel{\eruleBi}{=} 
  \ninlinegraphic[0.75]{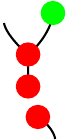}  \stackrel{\eruleS}{=} 
  \ninlinegraphic[0.75]{hopf-proof-informal-7}
\end{eqnarray*}
  Note that the step labelled \eruleBi in fact applies a version of
  that rule deformed by \eruleT, without altering the topology.  We
  could do this more explicitly using the ${\bf T}_1$ and ${\bf
    T}_2$ examples as follows: 
  \bc$
  \ninlinegraphic{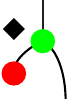} \stackrel{\eruleT}{=}
  \ninlinegraphic{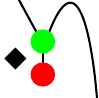} \stackrel{\eruleBi}{=}
  \ninlinegraphic{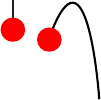} \stackrel{\eruleT}{=}
  \ninlinegraphic{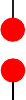}
  $
  \ec
\end{example}

Rules \eruleBp and \eruleBii are known informally as the
\emph{Hopf law} and the \emph{bialgebra law}.  Together, the
\ruleB-rules state that the interaction of different coloured
spiders produce 
a structure we call a \emph{scaled
  bialgebra}, which differs from a bialgebra only by a normalising
factor.  The fact that these structures naturally arise whenever we
have complementary observables is one of the main insights of this
paper, and will be developed further in Section~\ref{SEC:Complementarity}.

\subsubsection{The \textbf{K}-rules.}
\label{sec:k-rules}

These rules are concerned with special properties of spiders with phase
$\alpha = \pi$.  Rule \eruleKi states that dots labelled by $\pi$
commute with spiders of the other colour, i.e., $X^1_1(\pi)$ is a
homomorphism of the comultiplication $Z^1_2(0)$, and vice versa.
\begin{example}
  Thanks to rule \eruleKi, points with phase $\pi$ can also be
  copied just like points with phase zero:
  \bc 
  $
  \ninlinegraphic{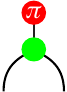} \stackrel{\eruleSi}{=}
  \ninlinegraphic{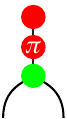} \stackrel{\eruleKi}{=}
  \ninlinegraphic{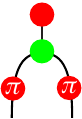} \stackrel{\eruleBi}{=} \;
  \ninlinegraphic{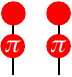} \; \stackrel{\eruleSi}{=} \;
  \ninlinegraphic{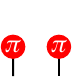} $
  \ec
\end{example}

Since the points labelled with $\pi$ or $0$ can be copied we call
these \emph{classical points};  then $Z^1_1(\pi)$ and $X^1_1(\pi)$ are
called \emph{classical maps}.\footnote{That is, classical relative to a particular observable structure; these classical maps then act as a permutation on the classical points of the observable structure \cite{CPaqPav}.}  (Of course, $\mathbf{K}$ stands for ``\emph{k}lassical''.) In the next section, we will see that
$Z^1_1(\pi)$ and $X^1_1(\pi)$ are interpreted by the familiar {\sf Z} and {\sf X} gates respectively.

Rule \eruleKii states that dots labelled by $\pi$ invert the
phase of dots of the other colour.  

\begin{example}
By  rules \eruleSi and \eruleSii, the degree 2 spiders
$Z^1_1(\alpha)$ form an abelian group, and by \eruleKii, conjugation by $X^1_1(\pi)$--- note here that $X^1_1(\pi)$ is self-inverse since $\pi +\pi=0$ ---sends each element to its inverse.
\end{example}

%

\subsubsection{The \textbf{C}-rule.}
\label{sec:c-rule}

This rule allows the $H$ vertex to function as an explicit colour
changing operation which transforms ``green structures'' into ``red
structures'' and vice versa.  In the next section, we will see that
the $H$ vertex is interpreted by the familiar Hadamard gate,
exchanging the $X$ and $Z$ bases.

\begin{example}
  Some special cases of this rule are: 
  \bc $ 
  \inlinegraphic{2.0em}{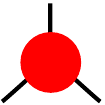} = \inlinegraphic{4.8em}{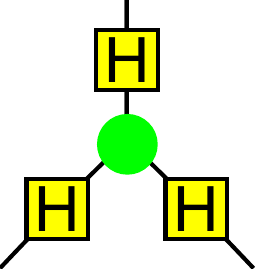} 
  \ \, ({\bf C}_1) \qquad\quad
  \ninlinegraphic{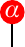} =\
  \ninlinegraphic{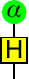} 
  \ \ \, ({\bf C}_2) \qquad\quad
  \inlinegraphic{4.5em}{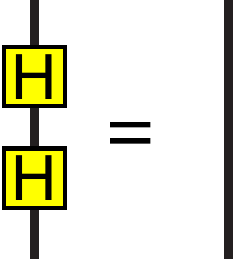} 
  \ \ \, ({\bf C}_3) $.  
  \ec
\end{example}

Notice that $(\mathbf{C}_3)$ asserts that $H$ is self-inverse.
The $\mathbf{C}$-rule effectively allows $H$ vertices to commute with coloured
dots, changing their colour in the process.

\subsubsection{The \textbf{D}-rules.}
\label{sec:d-rule}
The \eruleDii rule states that two black diamonds are equal to a loop of wire,
itself the result of composing a cup and a cap.  We will see in the
next section that the loop represents the dimension of the underlying
Hilbert space, and spacial juxtaposition is a form of multiplication,
justifying the name $\sqrt{D}$ for the diamond.

The \eruleDi rule `almost follows' from the other rules:
  \[
  \ninlinegraphic{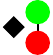} \, \stackrel{\eruleSi}{=} \,
  \ninlinegraphic{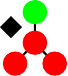} \, \stackrel{\eruleBi}{=} \,
  \ninlinegraphic{zx-example-innp-mub-5} \,
  \ninlinegraphic{zx-example-innp-mub-5} 
  \]
  which would yield the desired result if \inlinegraphic{5mm}{zx-example-innp-mub-5}$\!\!\!$
  could be cancelled.  
  

\subsection{Interpreting the \zxcalculus in Hilbert space}
\label{sec:interpr-zx-calc}


Given a diagram $\diagD$ with $n$ inputs and $m$ outputs, we
construct a corresponding linear map $D : {\cal Q}^n\to{\cal Q}^m$ as
follows.  

\begin{definition}[Interpretation of generators]\em
  If $\diagD$ consists of just a single generator---that is, one of
  $\id{\cal Q}$, $\sigma_{\cal Q}$, $\eta_{\cal Q}$, $\epsilon_{\cal
    Q}$, $Z^n_m(\alpha)$, $X^n_m(\alpha)$, $H$, or $\sqrt{D}$---then
  its corresponding linear map is as shown below:
  \begin{center}
  \qquad     $\ninlinegraphic{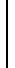} =  
      \left(\begin{array}{cc}
          1&0\\0&1
        \end{array}\right)$
\qquad\quad
      $ \ninlinegraphic{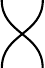} = 
      \left(\begin{array}{cccc}
          1&0&0&0\\0&0&1&0\\0&1&0&0\\0&0&0&1
        \end{array}\right)$
       \\[2em]
      $\ninlinegraphic{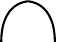}=\ket{00}+\ket{11}$
\qquad\quad
      $\ninlinegraphic{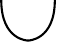}=\bra{00}+\bra{11}$
      \\[2em]
      $\underbrace{\overbrace{\ninlinegraphic{zx-gens-greenspider-alpha}}^n}_m::
      \left\{
      \begin{array}{ccr}
        \overbrace{\ket{0\,\ldots \,0}}^n 
        &\!\! \mapsto\!\! 
        &\ \ \ \ \, \overbrace{\ket{0\,\ldots\, 0}}^m\\
        \ket{1\,\ldots \,1} &\!\! \mapsto\!\! & \rme^{\rmi\alpha}\ket{1\,\ldots\, 1}\\
        \mbox{others}&\!\! \mapsto\!\! & 0\hspace{1.0cm}
      \end{array}
    \right.$
      \\[1.6em]
    \ $\underbrace{\overbrace{\ninlinegraphic{zx-gens-redspider-alpha}}^n}_m::
    \left\{
    \begin{array}{ccr}
      \overbrace{\ket{+\ldots +}}^n 
      &\!\! \mapsto\!\! 
      &\ \ \ \ \, \overbrace{\ket{+\ldots +}}^m\\
      \ket{-\ldots -} &\!\! \mapsto\!\! & \rme^{\rmi\alpha}\ket{-\ldots -}\\
      \mbox{others}&\!\! \mapsto\!\! & 0\hspace{1.1cm}
    \end{array}
  \right.$
        \\[1.6em]
        $\inlinegraphic{3.2em}{hadamard} =
      \frac{1}{\sqrt{2}}\left(
        \begin{array}{cc}
          1 &1 \\ 1 & -1
        \end{array}
      \right)$ 
\qquad\qquad
      $\inlinegraphic{4.5mm}{black-diamond} = \sqrt{2}$
\end{center}
\end{definition}

\begin{example}
The generators $Z^1_1(\pi)$ and $X^1_1(\pi)$ are the Pauli ${\sf Z}$
and ${\sf X}$ matrices:
\begin{center}
$\inlinegraphic{2.9em}{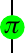} = \left(
   \begin{array}{rr}
     1&0\\ 0&-1
   \end{array}
\right)
\qquad
\inlinegraphic{2.9em}{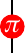} = \left(
   \begin{array}{rr}
     0&1\\ \ 1&\ 0
   \end{array}
\right)\;,$
\end{center}
\end{example}

\begin{example}
The generators $Z^1_2$ and $X^1_2$ are interpreted as follows:
\begin{equation*}
\inlinegraphic{2.00em}{delta}::
\left\{
\begin{array}{ccc}
|0\rangle\mapsto|00\rangle\\
|1\rangle\mapsto|11\rangle
\end{array}
\right.\quad\mbox{and}\quad
\inlinegraphic{2.00em}{red-delta}::
\left\{
\begin{array}{ccc}
|+\rangle\mapsto|++\rangle\\
|-\rangle\mapsto|--\rangle
\end{array}
\right.
\end{equation*}
giving the maps which copy the $Z$-basis vectors and the $X$-basis vectors
respectively.  

Consider $Z^0_1$.  Notice that its corresponding
linear map sends $1\mapsto \ket{0}$ and also $1\mapsto \ket{1}$, hence
by linearity we obtain $1 \mapsto \ket{0}+\ket{1} = \sqrt{2}\ket{+}$.
The complete set of Z- and X-basis vectors is show below.
\begin{equation*}
\inlinegraphic{1.73em}{epsilondag}=\sqrt{2}\ket{+} \,, \ \
\inlinegraphic{1.73em}{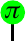}=\sqrt{2}\ket{-} \,,\ \ 
\inlinegraphic{1.73em}{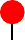}=\sqrt{2}\ket{0} \,, \ \
\inlinegraphic{1.73em}{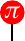}=\sqrt{2}\ket{1} \,. 
\end{equation*}
\end{example}

\begin{definition}[Interpretation of compound diagrams]\em
  If $\diagD$ consists of several generators there are two cases:
  \begin{itemize}
  \item if $\diagD = \diagD_1 \otimes \diagD_2$ then $D = D_1\otimes
    D_2$;
  \item if $\diagD = \diagD_1 \circ \diagD_2$, then $D = D_1 \circ
    D_2$.
  \end{itemize}
\end{definition}

  The order in which we divide the diagram into pieces does not matter
  to the final result, so long as the ``cuts'' do not pass through any
  vertices, nor any points where wires cross, nor any points of
  inflection   of a wire
  ---more accurately: just those inflection points
    were the gradient of the wire changes sign.
    (These
  last two may be thought of as the ``vertices'' defining
  $\sigma_{\cal Q}$, and $\eta_{\cal Q}$ and $\epsilon_{\cal Q}$
  respectively.)

\begin{example}\label{ex:tensor-comp-commute}
  The following diagram can be divided up as follows:
  \begin{center}$
    \begin{array}{rcl}
      \ninlinegraphic{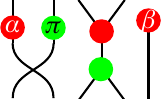} & = &
      \ninlinegraphic{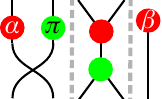} \; =  \;
      \left( \ninlinegraphic{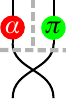} \right) \otimes 
      \left( \ninlinegraphic{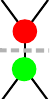} \right) \otimes 
      \left( \ninlinegraphic{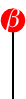} \right)
    \\
    &\;=\;&
    \left( \ninlinegraphic{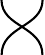} \circ
      \left( \ninlinegraphic{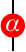} \otimes 
        \ninlinegraphic{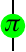} \right)\right)
    \otimes \left( \ninlinegraphic{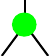} \circ
       \ninlinegraphic{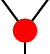} \right)\otimes 
       \ninlinegraphic{zx-factoring-5} 
    \end{array}
   $ \end{center}
 giving the linear map
 \begin{eqnarray*}
   \fl      D  = 
   \left(  
     \left(\begin{array}{cccc}
         1&0&0&0\\0&0&1&0\\0&1&0&0\\0&0&0&1
       \end{array}
     \right)
     \left( 
       \rme^{-\rmi\frac{\alpha}{2}}
       \left(\begin{array}{cc}
           \cos\frac{\alpha}{2} & \rmi\sin\frac{\alpha}{2}\\
           \rmi\sin\frac{\alpha}{2} & \cos\frac{\alpha}{2}
         \end{array}
       \right)
       \otimes
       \left(\begin{array}{cc}
           1&0\\0&-1
         \end{array} 
       \right)
     \right)
   \right)
   \\
   \otimes
   \frac{1}{\sqrt{2}}
   \left(
     \left(
       \begin{array}{cc}
         1&0\\0&0\\0&0\\0&1
       \end{array}
     \right)
     \left(\begin{array}{cccc}
         1&0&0&1\\0&1&1&0
       \end{array} 
     \right)
   \right)
   \otimes
   \rme^{-\rmi\frac{\beta}{2}}
   \left(
     \begin{array}{c}
       \rmi \sin \frac{\beta}{2} \\ \cos \frac{\beta}{2}
     \end{array}
   \right)  \;.
\end{eqnarray*}  
Unlike the diagram, the resulting matrix is rather large  ($16
\times 32$) so it is not shown here.  Any other factorisation of the
diagram, for example, 
  \bc $
  \ninlinegraphic{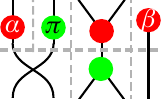} = 
  \left(
    \ninlinegraphic{zx-factoring-9}
    \otimes
    \ninlinegraphic{zx-factoring-10}
    \otimes
    \ninlinegraphic{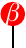}
  \right)
  \circ
  \left(
    \ninlinegraphic{zx-factoring-11}
    \otimes
    \ninlinegraphic{zx-factoring-12}
    \otimes
    \ninlinegraphic{zx-factoring-7}
    \otimes
    \ninlinegraphic{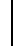}
  \right) 
  $
  \ec
  produces the same interpretation.
\end{example}

\begin{example}
  According to the $\mathbf{T}$-rule, the diagram of
  Example~\ref{ex:tensor-comp-commute} above is
  equivalent to the one shown below:
  \bc
  \ninlinegraphic{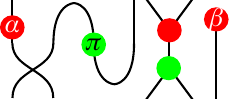}
  \ec
  As one might hope, this gives the same interpretation.
\end{example}

\begin{remark}\label{remark:zx-scalars}
  A linear map $f :\mathbb{C} \to \mathbb{C}$ is completely determined by
  the  value $f(1)$.  For this reason, and since ${\cal Q}^0 =
  \mathbb{C}$, the Hilbert space interpretation of a diagram with no
  inputs or outputs---a map from $\mathbb{C}$ to itself--- is simply a
  complex number. 
\end{remark}

\begin{proposition}[Soundness]\label{prop:soundness-of-zx}
  If diagrams $\diagD_1$ and  $\diagD_2$  are equal according to the equational
  rules of the \zxcalculus  then ${D_1} = {D_2}$.
\end{proposition}

This proposition can largely be verified by computing the maps
corresponding to left and right sides of each of the equational rules given in
Figure~\ref{fig:ZX-rules}, and observing that they are equal.
However, to show that the $\mathbf{T}$-rule is correct, different
techniques are required.  We will return to this point, 
and the (non)-issue of the factorisation order,  in
Section~\ref{subsec:Graph_calc}.

The converse of Proposition~\ref{prop:soundness-of-zx} is false:
there exist diagrams $\diagD_1$ and $\diagD_2$ which
represent the  same linear map but which are not equal by the rules of
the \zxcalculus.   For  example, the following diagrams are not
equivalent in the calculus:
\bc
$\inlinegraphic{30pt}{hadamard} \neq \ninlinegraphic{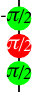}$,
\ec
but their interpretation as linear maps is the Euler-angle decomposition,
\begin{equation*}
H =  Z^1_1(\frac{\pi}{2})\circ X^1_1(\frac{\pi}{2})\circ Z^1_1(-\frac{\pi}{2})\;.
\end{equation*}
This equation is equivalent to Van
den Nest's theorem on local complementation of graph states
\cite{VandenNest}, as shown elsewhere by Perdrix and one of the
authors \cite{DunPer}.

We remark upon this fact for two reasons.  Firstly, as warning that not
every true fact about Hilbert space quantum mechanics can be derived
using the \zxcalculus, although a great many equations used in quantum
information processing can be.  Secondly, since
the equational theory of the \zxcalculus is strictly weaker than that
of Hilbert spaces, it is more general.  Therefore there are models of
the calculus which are distinct
from the usual Hilbert space interpretation of quantum mechanics.
All such models contain a large fragment of quantum mechanics---viewed
as an equational theory---but facts like Van den Nest's theorem
need not hold.

\begin{remark}
The points in the calculus are not normalized. This is required for reasons of simplicity; if we were to normalize $\sigma_{\cal Q}$ and $\eta_{\cal Q}$, then the (\textbf{T1}) rule would require additional scalar multipliers, and hence so would the (\textbf{S1}) rule, and so on.
\end{remark}





\subsection{Universality of the \zxcalculus}\label{sec:universality}

We claim that we now have enough expressive power to write down any
arbitrary linear map from $n$ qubits to $m$ qubits.  The green and red
phases, respectively:
\beqn
\suck
\ninlinegraphic{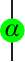}= Z^1_1(\alpha) =\left(
  \begin{array}{cc}
    1 & 0 \\ 0 & \rme^{\rmi \alpha}
  \end{array}\right)
  \\
  \suck
  \ninlinegraphic{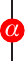}= X^1_1(\alpha) =
  \rme^{-\rmi{\alpha/2}}
  \left(
    \begin{array}{cc}
      \cos\frac{\alpha}{2} & \rmi\sin\frac{\alpha}{2}\\
      \rmi\sin\frac{\alpha}{2} & \cos\frac{\alpha}{2}
    \end{array}
  \right)
  \eeqn
  correspond with rotations of angle ${\alpha}$ respectively around the Z- and X-axis  on the Bloch sphere:
 \[
  \inlinegraphic{6em}{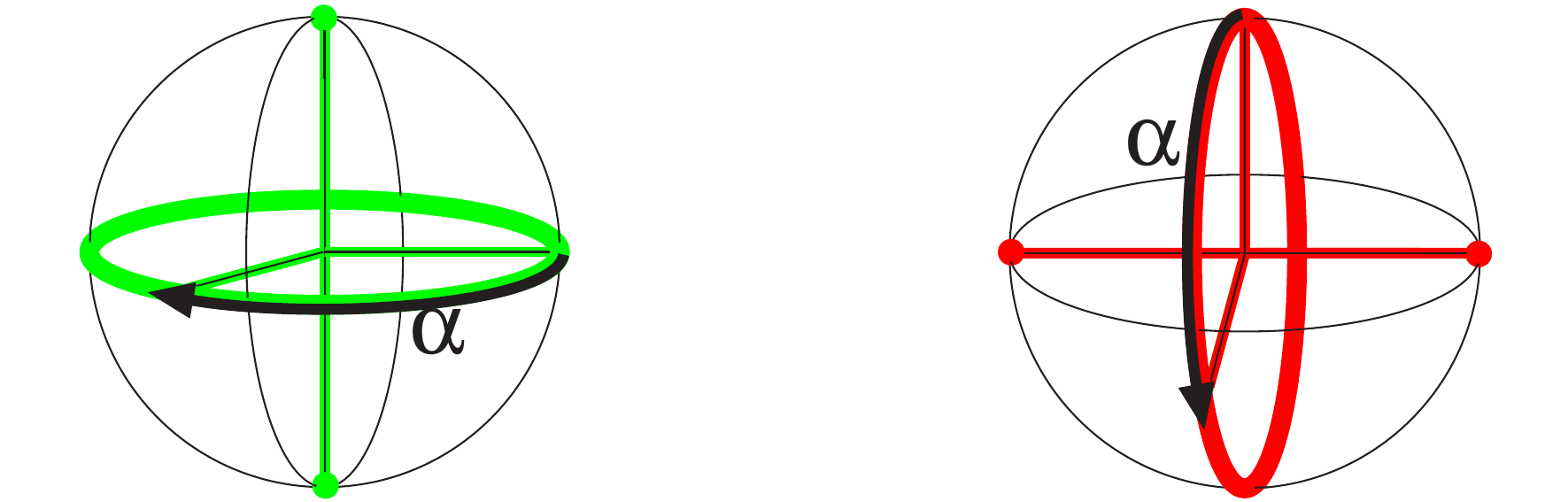}.
  \]
Combining both the `green' and the `red' phases allows us to write down any arbitrary one-qubit unitary in terms of its Euler-angle decompositions on the Bloch sphere:
\beqq\label{eq:euler-phase}
\ninlinegraphic{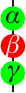}
=
Z^1_1(\gamma) \circ X^1_1(\beta) \circ Z^1_1(\alpha) 
\eeq
The controlled-NOT gate is defined by
\beqq\label{eq:infCNOT}
\raisebox{0.05cm}{\inlinegraphic{3.2em}{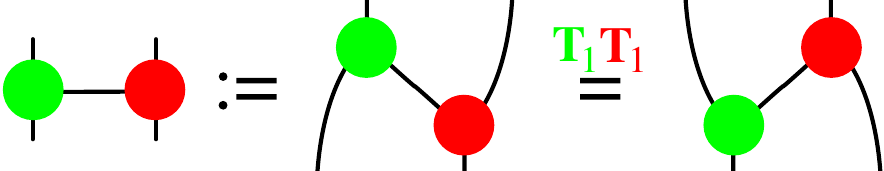}}\ = \ 
\left(\begin{array}{cccc}
1 & 0 & 0 & 0\\
0 & 1 & 0 & 0\\
0 & 0 & 0 & 1\\
0 & 0 & 1 & 0
\end{array}\right)= \CX 
\,.
\eeq
Standard results in quantum computing \cite{NieChu} state that $\CX$ gates
and arbitrary one-qubit unitaries suffice to construct any $n$-qubit
unitary map.  As equations~\ref{eq:euler-phase} and  \ref{eq:infCNOT}
show, the \zxcalculus contains this universal gate set, and hence can
represent any $n$-qubit unitary map.  Arbitrary $n$-qubit states can
therefore be represented as the image of any $n$-qubit state---for example 
$\inlinegraphic{1.73em}{epsilondag}\ldots
\inlinegraphic{1.73em}{epsilondag}$---under a well-chosen unitary.  Finally,
$\inlinegraphic{1.00em}{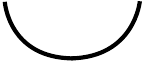}=\langle00|+\langle11|$ allows us to obtain any arbitrary linear map $f$ from
$n$ qubits to $m$ qubits from some $n+m$ qubit state $|\Psi\rangle$,  by relying on the diagrammatic incarnation of map-state duality \cite{AC1}:
\beqq
\begin{array}{l}
\ \overbrace{\ \ \ \ \ \ \ \  }^n\\
\inlinegraphic{3.8em}{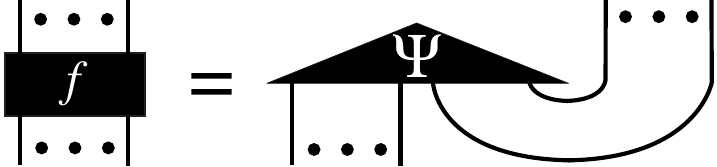}\,.\vspace{-1.8mm}\\ 
\ \underbrace{\ \ \ \ \ \ \ \  }_m
\end{array}
\eeq
Summarising all this:

\begin{proposition}\label{prop:universality-of-zx}
  Let $A:{\cal Q}^n \to {\cal Q}^m$ be a linear map;  then there
  exists a diagram $\diagA$ in the \zxcalculus whose Hilbert space interpretation
  is $A$.
\end{proposition}

\begin{remark}
  Since the converse to Proposition~\ref{prop:soundness-of-zx} does
  not hold, there is no reason why the diagram $\diagA$ should
  be unique.  There could be many inequivalent diagrams all
  of which denote the same linear map.
\end{remark}

\section{The \zxcalculus in use}
\label{sec:zx-calculus-use}

\subsection{Adjoints and inner products}
\label{sec:adjo-inner-prod}

\begin{definition}\label{def:zx-dagger}\em 
  Let $\diagD:m\to n$ be a diagram;  then its \emph{adjoint},
  $\diagD^\dag:n \to m$, is a diagram constructed by reflecting
  $\diagD$ in the horizontal axis, and negating all the angles
  which occur in \diagD.
\end{definition}

A diagram $\diagD$ is called \emph{self-adjoint} if $\diagD =
\diagD^\dag$, and \emph{unitary} if $\diagD \circ \diagD^\dag =
\id{\calQ}^{\otimes n}$ and $\diagD^\dag \circ \diagD =
\id{\calQ}^{\otimes m}$.

\begin{example}\label{ex:zx-defining-adjoint}
  Given the diagram $\diagD$, we form its adjoint $\diagD^\dag$ as
  shown:
  \bc
  $\diagD = \ninlinegraphic{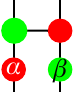}
  \qquad  \qquad  
  \diagD^\dag = \ninlinegraphic{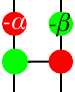}
  $
  \ec
We claim that   \diagD{} is unitary.  Half of the required proof is shown below.
  \begin{equation*}
    \fl \qquad \ 
    \ninlinegraphic[0.9]{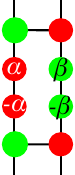} \stackrel{\eruleSi}{=}
    \ninlinegraphic[0.9]{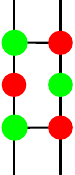} \stackrel{\eruleSii}{=}
    \ninlinegraphic[0.9]{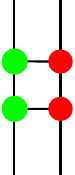} \stackrel{\eruleSi}{=}
    \ninlinegraphic[0.9]{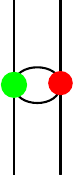} \stackrel{\eruleBp}{=}
    \ninlinegraphic[0.9]{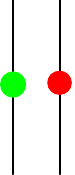} \stackrel{\eruleSii}{=}
    \ninlinegraphic[0.9]{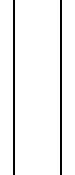} 
  \end{equation*}
The `horizontal application' of the ${\bf B}'$-rule can be decomposed
as follows:
\begin{equation*} \fl \qquad
\ninlinegraphic[0.8]{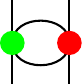} \stackrel{\eruleT}{=}
\ninlinegraphic[0.8]{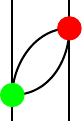} \stackrel{\eruleSi}{=}
\ninlinegraphic[0.8]{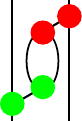} \stackrel{\eruleBp}{=}
\ninlinegraphic[0.8]{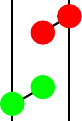} \stackrel{\eruleSi}{=}
\ninlinegraphic[0.8]{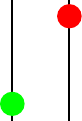} \stackrel{\eruleT}{=}
\ninlinegraphic[0.8]{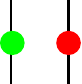} 
\end{equation*}
from which it follows that pairs of wires between green and red dots
can be eliminated.  It remains to
  show that $\diagD \circ \diagD^\dag = \id{{\cal Q}^2}$.

\end{example}

The following is self-evident:
\begin{proposition}
  Let \diagD{} be some diagram.   Then (i)   $\diagD^{\dag\dag} =
  \diagD$; (ii) if $\diagD = \diagA \circ \diagB$, then $\diagD^\dag = \diagB^\dag
  \circ \diagA^\dag$;  and (iii) if $\diagD = \diagA \otimes \diagB$, then
  $\diagD^\dag = \diagA^\dag \otimes  \diagB^\dag$.
\end{proposition}

\begin{proposition}
  If $\diagD:m\to n$ denotes the linear map $D:\calQ^m \to \calQ^n$,
  then the adjoint diagram $\diagD^\dag$ denotes $D^\dag$, the usual
  linear algebraic adjoint of $D$.
\end{proposition}

\begin{corollary}
  If a diagram is self-adjoint or unitary so is its corresponding
  linear map.
\end{corollary}

Recall that any diagram $\diagD : 0 \to n$ has a (possibly
unnormalised) $n$-qubit state as its Hilbert space
interpretation; such diagrams therefore correspond to kets
$\ket{\diagD}$ in Dirac notation.  Since Dirac's bra is the adjoint of
a ket, we now see how to define the inner product of two diagrams.  Given
$\diagA,\diagB : 0 \to n$ we have
\[
\innp{\diagA}{\diagB} = \diagA^\dag \circ \diagB\,.
\]
The resulting diagram $(\diagA^\dag \circ \diagB)$ has no inputs or
outputs, hence by Remark \ref{remark:zx-scalars}, it denotes a complex
a number, as required.

\begin{example}
  We can compute the  inner product of $Z^0_1(\alpha )$ with itself.
  \[
  \ninlinegraphic{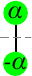}  \stackrel{\eruleSi}{=} 
  \ninlinegraphic{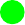}  \stackrel{\eruleSi}{=} 
  \ninlinegraphic{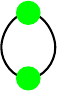}  \stackrel{\eruleSii}{=} 
  \ninlinegraphic{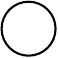}  \stackrel{\eruleDii}{=} 
  \ninlinegraphic{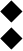}
  \]
  The result is 2 because the ``states'' are not normalised.
\end{example}

\begin{example}
  Let $j,k \in \{0, \pi\}$.  We compute the inner product of
  $Z_1^0(k)$ and $X_1^0(j)$.
  \[
  \ninlinegraphic{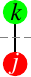} \, \stackrel{\eruleSi}{=} \,
  \ninlinegraphic{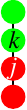} \, \stackrel{\eruleKii}{=} \,
  \ninlinegraphic{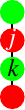} \, \stackrel{\eruleKi}{=} \,
  \ninlinegraphic{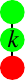} \, \stackrel{\eruleKi}{=} \,
  \ninlinegraphic{zx-example-innp-mub-5} \, \stackrel{\eruleDi}{=} \,
  \ninlinegraphic[0.6]{black-diamond}
  \]
  Since the result is independent of $j$ and $k$,  
  this calculation shows that the $X$ and $Z$ bases are mutually unbiased.  
%
\end{example}

\begin{example}
  Suppose that $\diagU$ is a diagram encoding some 
  complicated unitary operation $U$, acting on $n+1$ qubits.  Suppose
  its input is $\ket{00\cdots 0}$:  what is the amplitude for
  observing the output $\ket{1}$ at its last output?  We need to
  compute:
  \[
  \bra{00\cdots0} U^\dag (\id{{\cal Q}^n} \otimes X) U \ket{00\cdots}
  =
  \ninlinegraphic{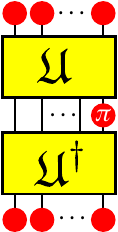}
  \]
  When \diagU{} is presented using the generators of the \zxcalculus, 
  great simplification is (usually) possible, making this expression
  (usually) easy to compute.  
\end{example}

\subsection{Quantum Circuits}
\label{sec:quantum-circuits}

As we have already seen in Section~\ref{sec:universality}, the
\zxcalculus can represent the  basic gates used in quantum circuits.  The
rules of the calculus can give short graphical proofs of many circuit
identities.  

\subsubsection{The \CX gate.}
\label{sec:cx-gate}

We have already seen the controlled-NOT gate:
\[
\CX = \ninlinegraphic{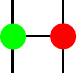}\,.
\]
It is manifestly self-adjoint.  We can prove that it is also unitary:
\begin{equation*} \fl \qquad
\CX\circ\CX = 
\ninlinegraphic{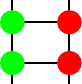} \, \stackrel{\eruleSi}{=} \,
\ninlinegraphic{CX-squared-proof-2} \, \stackrel{\eruleBp}{=} \,
\ninlinegraphic{CX-squared-proof-3} \, \stackrel{\eruleSii}{=} \,
\ninlinegraphic{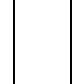} =
\id{\calQ}^{\otimes 2}\ ,
\end{equation*}
  
An elementary exercise is to show that a sequence of three \CX gates
can be used to swap to qubits.  A graphical proof of this fact is given below.
\begin{equation*} \fl \qquad
\ninlinegraphic[0.8]{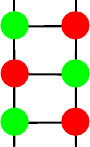} \stackrel{\eruleT}{=}
\ninlinegraphic[0.8]{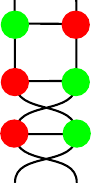} \stackrel{\eruleBii}{=}
\ninlinegraphic[0.8]{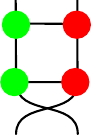} \stackrel{\eruleSi}{=}
\ninlinegraphic[0.8]{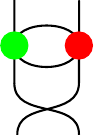} \stackrel{\eruleBp}{=}
\ninlinegraphic[0.8]{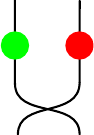} \stackrel{\eruleSii}{=}
\ninlinegraphic[0.8]{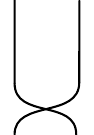}
\end{equation*}
While this is a well-known property for \CX, our proof holds in much
greater generality than qubits, since  as we will see in the remainder
of this paper, the graphical  calculus applies in much greater
generality.  This example relies on the bialgebra law ({\bf B2}), which is a
stronger principle than the Hopf law ({\bf B'}) used in the previous example.  The
relationship between these two laws will be spelled out in
Section~\ref{sec:bialgebra}.

In Section~\ref{sec:universality} the \CX gate was introduced by
checking that its diagram denoted the correct linear map.  However we
can describe \CX by the following ``behavioural specification'':
\emph{when the control input is $\ket{0}$, the target qubit is left
  unchanged; when the  control qubit is $\ket{1}$, the target qubit is
  flipped}. Letting $\inlinegraphic{1.73em}{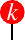}\!$  represent one of
the two red classical points, that is, either
$\inlinegraphic{1.73em}{red-epsilondag}=\ket{0}$  or
$\inlinegraphic{1.73em}{red-pi-short-point}=\ket{1}$, we can supply a
qubit to the control input (the left input, connected to the green
dot), and obtain the following proof:
\begin{equation*} \fl \qquad
\ninlinegraphic{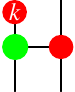} \stackrel{\eruleKi + \eruleBi}{=}
\ninlinegraphic{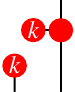} \ \stackrel{\eruleSi}{=}
\ninlinegraphic{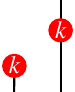}  =
\left\{\begin{array}{ccl}
  \raisebox{-1mm}{ \ninlinegraphic{red-epsilondag}} \  \ninlinegraphic{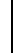}  
  & \mbox{iff}  
  & \ninlinegraphic{red-k-short-point}\!\! =
  \ninlinegraphic{red-epsilondag}
  \\
  \\
 \raisebox{-1mm}{ \ninlinegraphic{red-pi-short-point}} \  \ninlinegraphic{red-pi-short}  
  & \mbox{iff}  
  & \ninlinegraphic{red-k-short-point}\!\! =
  \ninlinegraphic{red-pi-short-point}
\end{array}\right.
\end{equation*}
Notice that in each case the control qubit passes
through the gate unchanged, while the target input is either the
identity, or the Pauli X, depending on the value of the control qubit,
thus meeting the specification.  Further, the colour symmetry of this
proof demonstrates that, if we operate in the Z-basis (i.e., $\ket{+}
= \inlinegraphic{1.73em}{epsilondag}$ and $\ket{-} =
\inlinegraphic{1.73em}{green-pi-short-point}$) the role of left and
right are exchanged.

\subsubsection{The \CZ gate.}
\label{sec:cz-gate}

Since ${\sf Z} = H{\sf X}H$ we can obtain the \CZ gate from the \CX
gate by conjugating the target qubit with $H$ gates, as shown below:

\begin{equation*}\fl\qquad
\CZ = \left(
  \begin{array}{cccc}
    1&0&0&0\\
    0&1&0&0\\
    0&0&1&0\\
    0&0&0&-1
  \end{array}
\right) = \ninlinegraphic{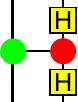}  = \ninlinegraphic{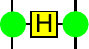}
\;.
\end{equation*}

We can immediately read off two  properties of this gate from
its graphical representation:  it is self-adjoint, and it is symmetric
in its inputs.  It is also unitary:
\begin{equation*}\fl \qquad
  \ninlinegraphic[0.8]{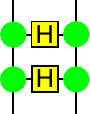} \stackrel{\eruleSi}{=}
  \ninlinegraphic[0.8]{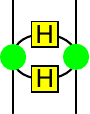} \stackrel{\eruleC}{=}
  \ninlinegraphic[0.8]{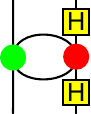} \stackrel{\eruleBp}{=}
  \ninlinegraphic[0.8]{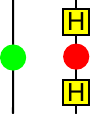} \stackrel{\eruleSii}{=}
  \ninlinegraphic[0.8]{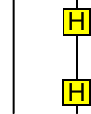} \stackrel{\eruleC}{=}
  \ninlinegraphic[0.8]{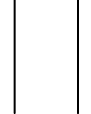}
\end{equation*}

\subsubsection{The quantum Fourier transform.}\label{Sec:QFT}

Lying at the centre of many quantum algorithms---including Shor's
famous factoring algorithm \cite{Shor:PolyTimeFact:1997}---the quantum Fourier transform is one of
the most important quantum processes.  The equations of the
diagrammatic calculus are strong enough to simulate it. 
    
To write down the required circuit, we must realise a controlled
phase gate, where the phase is an arbitrary angle $\alpha $; this is
shown below---the  control qubit is on the left. (One can prove the
correctness of this diagram using a behavioural description in a similar
fashion to the treatment of the \CX in Section~\ref{sec:cx-gate}.)

\begin{equation*}
\CZ_\alpha = \left(
  \begin{array}{cccc}
    1&0&0&0\\
    0&1&0&0\\
    0&0&1&0\\
    0&0&0&e^{i\alpha}
  \end{array}
\right) 
= 
\inlinegraphic{5.5em}{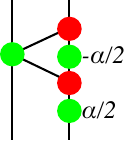}
\end{equation*}
The only  gates which are required to construct the circuit
implementing the quantum Fourier transform are the Hadamard and the
$\CZ_\alpha$---see for example \cite{NieChu}.  The circuit for the
2-qubit  QFT is shown below. 
\begin{equation*}
QFT_2 = \ninlinegraphic[0.8]{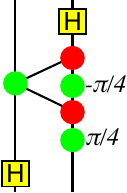}
\end{equation*}
How can we simulate this circuit?  First, we choose an input state, in this
case $\ket{10}  = \ninlinegraphic{red-pi-short-point}\
\ninlinegraphic{red-epsilondag}$; then we simply 
concatenate the input to the  circuit, and begin rewriting according
to the equations of the theory, as shown below.
\begin{eqnarray*}\fl \quad QFT_2 \circ \ket{10} = 
  \ninlinegraphic[0.8]{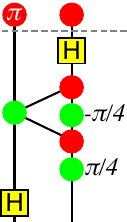} & \stackrel{\eruleK}{=} &
  \ninlinegraphic[0.8]{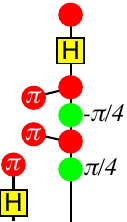} \stackrel{\eruleSi}{=}
  \ninlinegraphic[0.8]{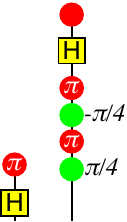} \stackrel{\eruleKii}{=}
  \ninlinegraphic[0.8]{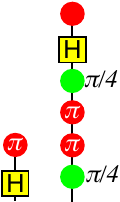} \\ &\stackrel{\eruleS}{=}&
  \ninlinegraphic[0.8]{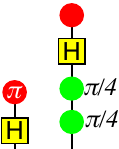} \stackrel{\eruleC}{=}
  \ninlinegraphic[0.8]{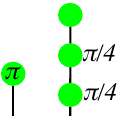} \stackrel{\eruleSi}{=}
  \ninlinegraphic[0.8]{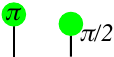} 
\end{eqnarray*}
The final diagram in the sequence is simply the tensor product 
$(\ket{0}-\ket{1})\otimes(\ket{0}+i\ket{1})$, which is indeed the
desired result.  
In passing we remark upon another feature of the
graphical language:  since the last diagram is a disconnected graph,
it represents a separable quantum state.

\subsection{Measurement-based quantum computing}
\label{sec:meas-based-quant}

Measurement-based quantum computation \cite{Jozsa:2005qa} uses the
state-changing effect of 
quantum measurements to carry out the computation, typically
propagating these changes through entangled states.  The
simplest example is the teleportation protocol \cite{BBC} which can be
viewed as the identity function computed via a Bell state.  A more
powerful model is Raussendorf and Briegel's one-way quantum computer
\cite{RB01,RBB}, which provides a computationally universal model almost
entirely based on single-qubit measurements acting on a large cluster state.

The graphical notation of the \zxcalculus is
ideal for representing these entangled states, and its equations
accurately capture the changes in these states induced by measuring
their constituent parts.

\begin{remark}
  The \zxcalculus as presented in this section,
  cannot represent the non-deterministic behaviour of
  measurements.  Rather, we replace measurements
  by projections onto some particular outcome.  One could view this as
  post-selection, but it would more accurate to understand that each
  diagram represents a one particular run of an experiment, and the
  particular outcome that was measured, rather than averaging over all
  possible runs.
%
  The restriction to pure states is not an instrinsic limitation of
  this approach.  It is a deliberate choice, made in order to simplify
  the presentation of the calculus.  The formal apparatus used here
  was introduced in \cite{CPav} to represent classical control
  structure and the branching behaviour of quantum measurements.  In
  Section~\ref{sec:non-determ-mixed} we present three extensions to
  the calculus to handle non-determinism and mixedness.
\end{remark}

\subsubsection{The teleportation protocol.}
\label{sec:telep-prot}

The teleportation protocol \cite{BBC} consists of two main components:
the preparation of the Bell state, and the Bell basis measurement.  As
described in Section~\ref{sec:interpr-zx-calc}, the (unnormalised)
Bell state is represented by a cap, and its corresponding projection
by a cup:
\[
\ket{00} + \ket{11} = \ninlinegraphic{small-cap} \,,
\qquad
\bra{00} + \bra{11} = \ninlinegraphic{small-cup} \,.
\]
Combining these two elements, we obtain an almost trivial proof of the
correctness of teleportation, in the case where Alice observes
$\ket{00} + \ket{11}$.
\[
\ninlinegraphic{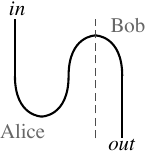}
=
\ninlinegraphic{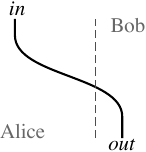}
\]
The role of classical communication is hidden in this picture, but it
is revealed by a more detailed look at the Bell basis measurement.  Let
$\alpha, \beta \in \{0,\pi\}$.  Ranging over the 4 possible
$(\alpha, \beta)$ pairs in the diagram below gives the 4 possible
outcomes of a Bell basis measurement:
\[
\{\;\bra{\mathbf{\Psi}_+}, \bra{\mathbf{\Psi}_-}, \bra{\mathbf{\Phi}_+},
\bra{\mathbf{\Phi}_-}\;\}
=
\left\{ \; \ninlinegraphic{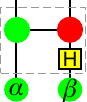} \;|\; \alpha, \beta \in \{0,\pi\}\right\}
\]
(Notice that the boxed part of the diagram is simply the circuit which
rotates the Bell basis onto the $X$-basis.)  This
description of the protocol displays the Pauli errors that are
introduced if Alice observes the other possible outcomes.
\[
\ninlinegraphic{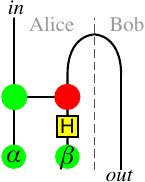}
= \;\;
\ninlinegraphic{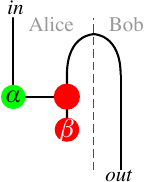}
= \;\;
\ninlinegraphic{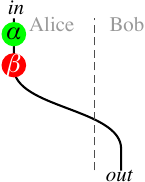}
\]
From this we can derive a complete description of the protocol, and
show, including Bob's corrections, which are classically correlated to
Alice's observations.

\begin{center}$
  \ninlinegraphic{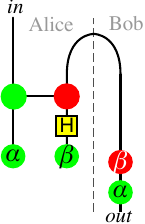} = \;\;
  \ninlinegraphic{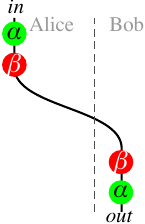} = \;\;
  \ninlinegraphic{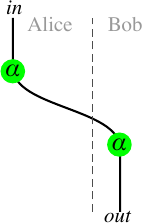} = \;\;
  \ninlinegraphic{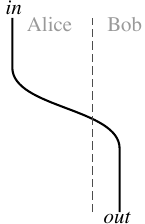}$
\end{center}
The first equation is the preceding derivation collapsed into one
step, while the last two equations use the spider rules and the fact
that $2 \alpha = 2 \beta = 0$.

\subsubsection{The state transfer protocol.}
\label{sec:state-transf-prot}

This protocol was introduced by Pedrix \cite{Perdrix} to
reduce the resources required for measurement based quantum
computing.  The core of the protocol is a measurement which
projects onto a 2-dimensional eigenspace:
\beqq
\raisebox{0.1mm}{\inlinegraphic{2.4em}{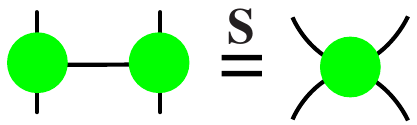}}\!
  = 
  \ \left( \begin{array}{cccc}
1&0&0&0\\0&0&0&0\\0&0&0&0\\ 0&0&0&1
  \end{array}\right)={\rm P}_{Z\otimes Z}
\eeq
It is easily seen that ${\rm P}_{Z\otimes Z}$ is self-adjoint and
idempotent:
\beqq
{\rm P}_{Z\otimes Z}\circ {\rm P}_{Z\otimes Z}\ = \ \inlinegraphic{3.2em}{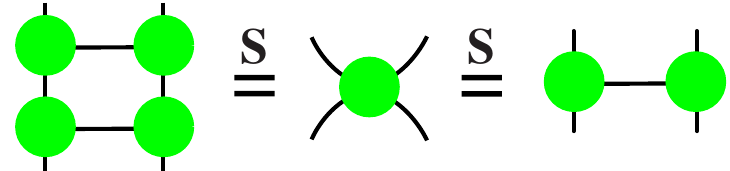}\ = \ {\rm P}_{Z\otimes Z}\ ,
\eeq
and hence a projector.   

Consider now a  protocol, which initially assume  two qubits, one
in an unknown state $\inlinegraphic{1.7em}{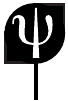} =|\psi\rangle$
and one in the state $\inlinegraphic{1.73em}{epsilondag}=|+\rangle$.
We want to transfer $|\psi\rangle$ from the first qubit to the second,
and this can be done by means of two projections:
\bc
\inlinegraphic{5.0em}{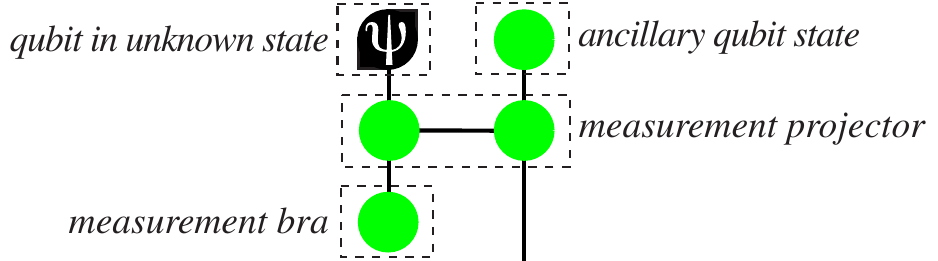} 
\ec
since by application of the  ${\bf S}$-rule we have:
\bc
$\inlinegraphic{5.0em}{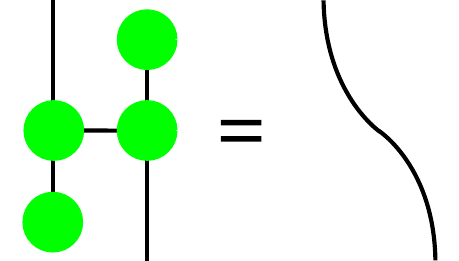}
  $.
\ec
The protocol can be extended by performing the second, single-qubit
measurement in the phase-shifted basis $\ket{0} \pm \rme^{\rmi \alpha} \ket{1}$.
\[
\hspace{-1.6cm}\raisebox{-0mm}{\inlinegraphic{5.0em}{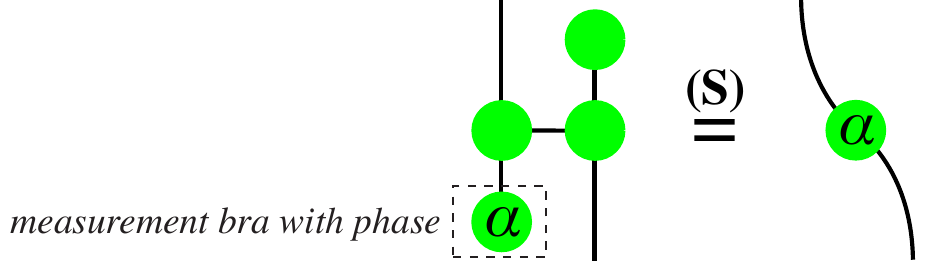}}
=\left(
  \begin{array}{cc}
    1&0\\0&\rme^{\rmi \alpha}
  \end{array}\right)
\qquad\qquad
\]
This minor change allows the protocol to apply an arbitrary
$Z$-rotation to its input; the protocol can be modified in the obvious
way to perform an $X$-rotation, and hence any single-qubit unitary.

\subsubsection{Multipartite states.}

In our graphical language, a quantum state is nothing more than a diagram
with no inputs; the outputs correspond to the individual qubits
making up the state.  The interior of the diagram---i.e.~its graph
structure---describes how these qubits are related.  This
notation is ideal for representing large entangled states.

Cluster states, which are used in measurement-based 
quantum computing \cite{RBB}, can be 
prepared in several ways and the \zxcalculus provides short
proofs of their equivalence.  For example, the original scheme
describes a \CZ interaction between qubits initially prepared in
the state $\ket{+}$;  in our notation this is $Z^0_1$, or
$\inlinegraphic{1.4em}{epsilondag}$. Hence a one-dimensional cluster state
can be presented diagrammatically as:
\bc
\inlinegraphic{5.2em}{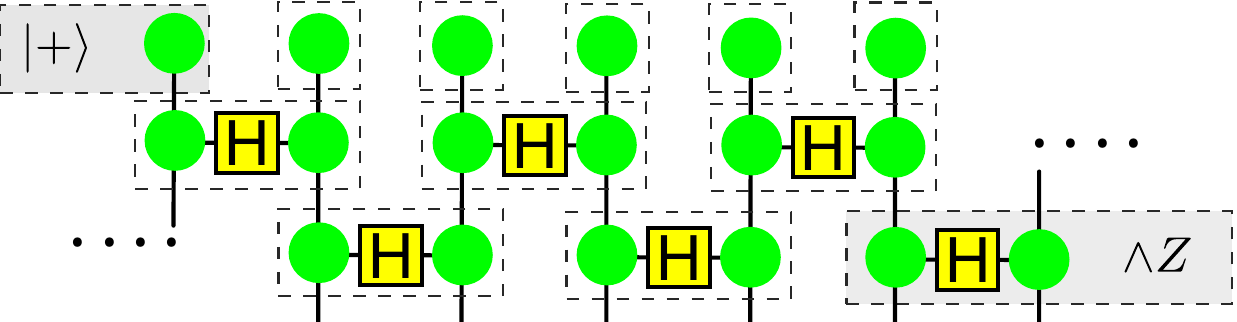}
\ec
where the boxes delineate the individual $\ket{+}$ preparations and
\CZ operations.  Alternatively, the cluster state can be prepared by
applying a Hadamard gate to one part of a Bell pair to obtain states
of the form $\ket{\Phi} = \ket{0+}+\ket{1-}$, and then ``fusing''
these entangled pairs \cite{fusion1}. The required fusion operation is
exactly 
\beqq
\inlinegraphic{2.00em}{deltadag} :{\cal Q}\otimes{\cal  Q}\to {\cal  Q}:: 
\left\{
\begin{array}{ccc}
  |00\rangle&\mapsto&|0\rangle\\
  |11\rangle&\mapsto&|1\rangle\\
  |01\rangle, |10\rangle &\mapsto& 0
\end{array}
\right.\;,
\eeq
and a 1D cluster prepared with this method looks like:
\bc
\inlinegraphic{3.6em}{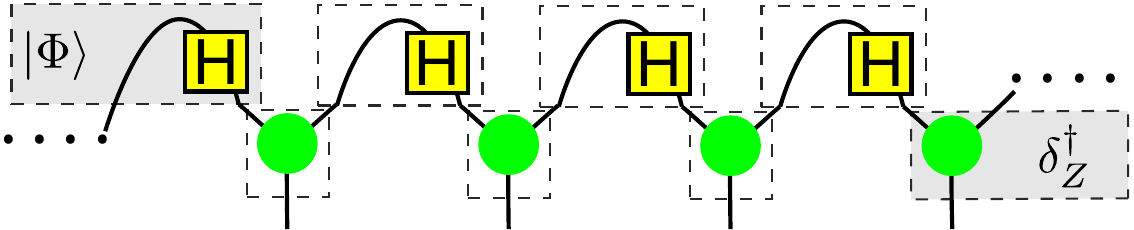}\;.
\ec
Again, dashed lines indicate the individual components.
While conventional methods require some calculation to show that
these methods of preparation produce the same state, using the
spider theorem, the two diagrammatic forms  are immediately equivalent:
\bc
\inlinegraphic{5em}{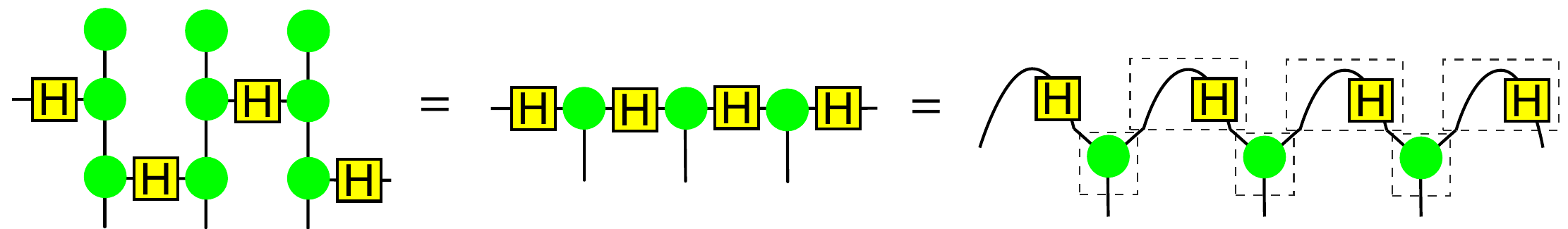}.
\ec
From the example of the 1D cluster, it's easy to see how to
construct diagrams corresponding to arbitrary graph states.  Indeed
given a graph state $\ket{G}$, with underlying graph $G$,  
we represent $\ket{G}$ by the same graph $G$, with green dots at
each vertex, and $H$ gates on each edge;  to complete the
construction we must add one output wire at each green vertex.

While graph states are important in measurement-based quantum
computation, they are not the only kind of interesting entangled
states.  As an illustration of universality of the graphical language,
we present graphical representatives of the two
non-comparable classes of genuine three qubit entangled states\footnote{
  The GHZ state cannot be converted to the W state by means of
  stochastic local operations and classical communication, nor vice
  versa.   States which can be so-interconverted are called
  \emph{SLOCC-equivalent}: up to SLOCC-equivalence the GHZ and W
  are the \emph{only} 3-qubit states with 3-party entanglement
  \cite{W_GHZ}.}.
As can be directly read from the interpretation given in
Section~\ref{sec:interpr-zx-calc}, the GHZ state is simply a
three-legged spider:
\begin{equation*}
\ket{\textrm{GHZ}} = \ket{000} + \ket{111} = \
\ninlinegraphic{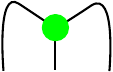}\, .
\end{equation*}
The simple form of this state hints at its importance in the
algebraic structures to be introduced later in this paper.  This
algebraic role, particularly in relation to the phase group, has been
used to explain non-locality 
\cite{CES}.  The W state, however, is less obvious:
\begin{equation*}
\ket{\textrm{W}} = \ket{001} + \ket{010} + \ket{100} = \ \ninlinegraphic{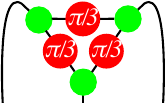}\,.
\end{equation*}
This representation supports the intuition that while
the GHZ state is a globally entangled,
the W is rather to be conceived as a pairwise entanglement
between each pair of qubits that make up the three-partite system
\cite{W_GHZ}.   
The algebraic properties of the W state have been studied elsewhere by
Kissinger and one of the authors \cite{CoeKis2010}.

\subsubsection{The one-way model}
\label{sec:one-way-model}

The graphical language is ideal for studying different models of
quantum computation in the same setting.   In this section we will
present several computations using the one-way model \cite{RB01}, and
translate them into equivalent quantum circuits using the rules of
\zxcalculus.  We use the \emph{measurement calculus} notation
introduced by Danos, Kashefi, and Panangaden \cite{DKP}, and borrow
their examples. 

For our purposes, a measurement calculus program, called a
\emph{pattern}, consists of a sequence of commands of the following
kinds:
\begin{itemize}
\item $N_i$ -- initialise qubit $i$ to the state $\ket{+}$.
\item $E_{ij}$ -- apply a \CZ operation to qubits $i$ and $j$.
\item $M^{\alpha}_i$ -- measure the qubit $i$ in the basis
  $\ket{0}\pm\rme^{\rmi \alpha}\ket{1}$. 
\end{itemize}
The commands occur in the order given: first initialisations, then
entanglement, then measurement.  Any quibit which is not initialised
is an input;  any not measured is an output.

Since, in the \zxcalculus,  measurements are replaced by projections,
the conditional operations of the measurement calculus have been omitted;
see Section~\ref{sec:non-determ-mixed} and \cite{DunPer2010} for a
more complete treatment.  We make the convention that the observed
outcome of each measurement will be the +1 outcome---that is, the
projection onto $\ket{0} + \rme^{\rmi \alpha}\ket{1}$.  With this
convention the elements of the measurement calculus can be translated
by the following table:
\begin{center}
\begin{tabular}{|c|c|c|}
\hline 
 $N_i$ & $E_{ij}$  & $M_i^{\alpha}$ \\
\hline
\ninlinegraphic{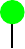} 
& \ninlinegraphic{small-CZ} 
& \ninlinegraphic{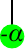}
\rule[-4.5mm]{0pt}{11mm}
\\
\hline
\end{tabular}
\end{center}

\begin{example}
  Consider a measurement-based program involving 4 qubits, which
  computes a $\CX$ gate upon its inputs.  In the syntax of the
  measurement calculus this pattern is written:
  \[
  M_2^0 M^0_4 E_{13} E_{23} E_{34} N_3 N_4.
  \]
  Reading from right to left, this specifies that qubits 3 and 4
  should be prepared in a $\ket{+}$ state, then $\CZ$ operations
  should be applied pairwise between qubits 1 and 3, 2 and 3, and 3
  and 4; finally $X$ basis measurements should be performed upon
  qubits 2 and 4.  Qubits 1 and 2 are the inputs and
  qubits 1 and 4 are the outputs.  We represent this pattern
  diagrammatically as: 
  \bc 
  \inlinegraphic{11.2em}{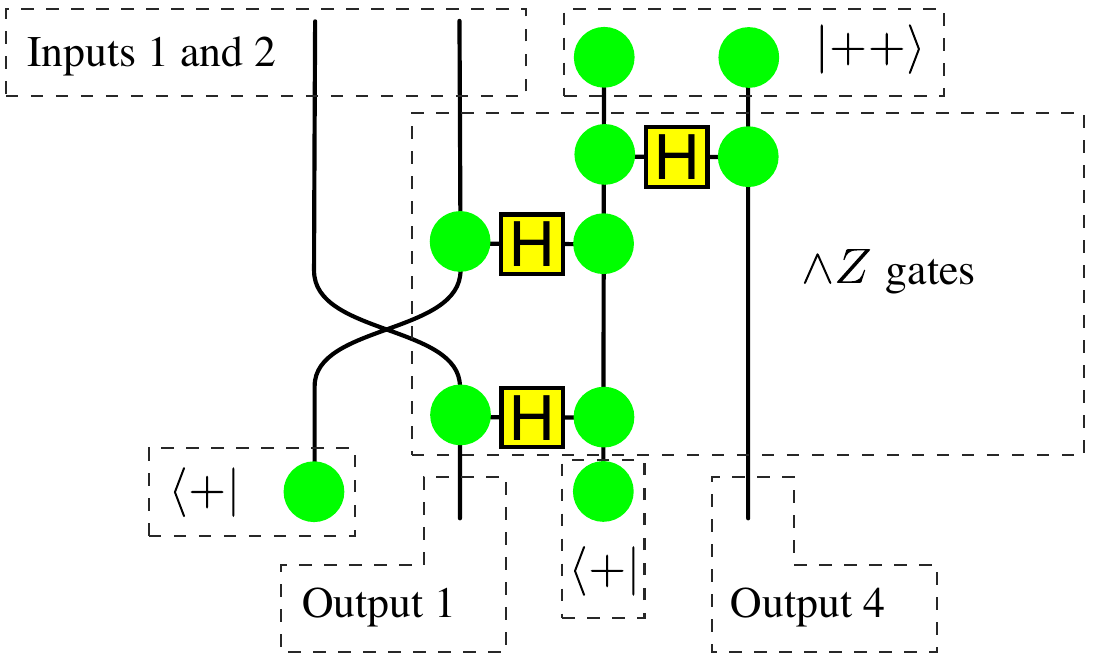} 
  \ec 
  The spider theorem allows this one-way program to be rewritten to a
  \CX gate in three steps:
  \bc \inlinegraphic{8.4em}{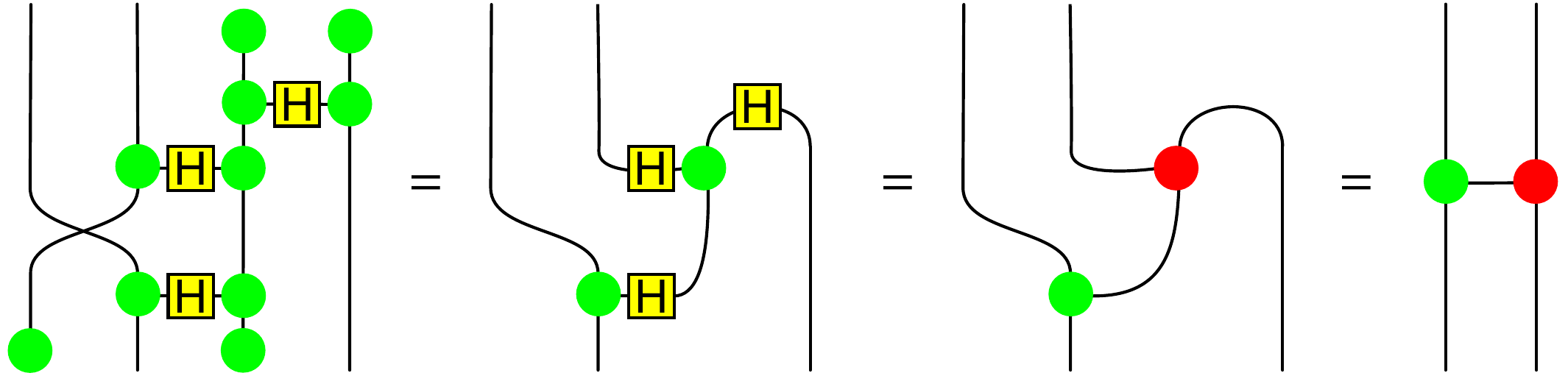}\,. 
  \ec
\end{example}

\begin{example}
  Our next example is a one-way program implementing an arbitrary
  1-qubit unitary.  Recall that any single qubit unitary map $U$ has
  an Euler decomposition as such that $U = Z_\gamma X_\beta Z_\alpha$.
  Such a unitary can be implemented by the following 5-qubit
  measurement pattern:
  \[
  M_3^{\gamma} M_2^{\beta}
  M_1^{\alpha}E_{12}E_{23}E_{34}E_{45}N_2N_3N_4N_5\,.
  \]
  The graphical form of this pattern is shown below: 
  \bc
  \inlinegraphic{7.6em}{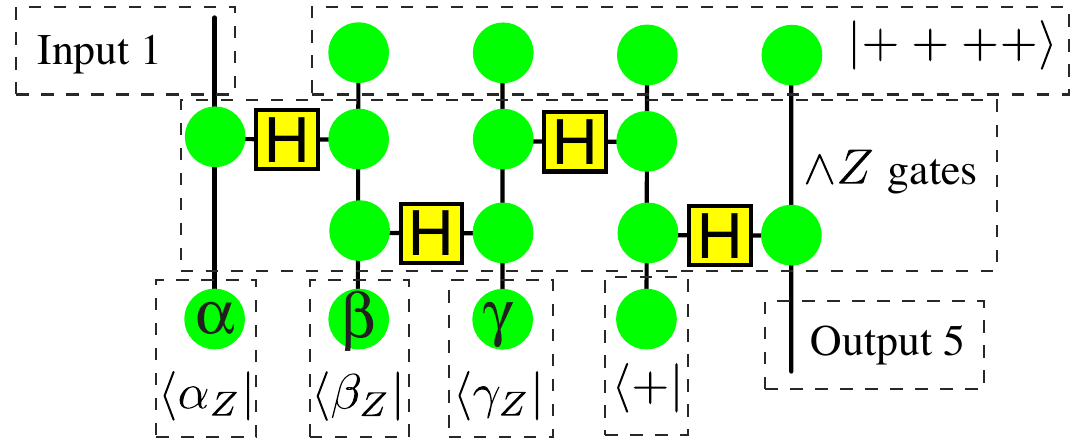}.
  \ec 
  A sequence of simple rewrites shows that the one-way program
  intended to compute such a unitary does indeed produce the desired
  map.
  \begin{equation*}\fl\quad
    \ninlinegraphic[0.7]{1wqc-U-proof-1} \stackrel{\eruleSi}{=}
    \ninlinegraphic[0.7]{1wqc-U-proof-2} \stackrel{\eruleC}{=}
    \ninlinegraphic[0.7]{1wqc-U-proof-3} 
  \end{equation*}
\end{example}

\begin{remark}
  The reader may object that the ``post-selection'' of one particular
  set of measurement outcomes reduces the number of diagrams
  significantly, and thus gives a misleading air of 
  feasibility to these techniques.  In practice the pure state version
  of the \zxcalculus needs only minor extension to handle the full
  behaviour of the one-way model, without any combinatorial explosion.
  We sketch this extension in Section~\ref{sec:non-determ-mixed}; the
  full details can be found in \cite{DunPer2010}.
\end{remark}

\section{Symmetric monoidal categories and graphical reasoning}\label{DagSMS_scalars}

The preceding sections presented the \zxcalculus as a \emph{fait
  accompli}, without any serious justification for its axioms, other
than its utility in certain calculations.  This section, and those that
follow, will put down the firm mathematical foundation upon which the
calculus is built.  This section will outline the basic concepts of
symmetric monoidal categories (\smcs) without going into too much technical detail; we
aim to provide the reader with just enough background to follow the
subsequent material, and provide many references where complete and
detailed expositions can be found.

A \emph{category} consists of \emph{objects} $A,B,C,\ldots$ and, for
each pair of objects $A,B$, a collection of \emph{morphisms}
$f,g,h,\ldots : A\to B$.  From a physical perspective, the objects can
be thought of as \emph{physical systems} and each morphism $f:A\to B$
as a physical process which transforms a system of type $A$ to a
system of type $B$.  Here, `type' should not be confused with `state'.
E.g.~type could be qubit, or field, or a certain classical system,
and each of these admits many states. For a computer scientist, the objects may be
data-types, and $f:A\to B$ would be a program accepting input of type
$A$ and producing output of type $B$.  In mathematics the objects are
typically structures of a certain kind, e.g.~sets, or groups, or
vector spaces, and $f:A\to B$ is a  structure preserving map, e.g.~a function, or
a group homomorphism, or a linear map.

Pairs of morphisms where the domain of one matches the codomain of the
other may be \emph{composed}: for each such pair, $f:A\to
\underline{B}$ and $g: \underline{B}\to C$, we write the composite $g\circ f:
A\to C$.  In the case of physical processes $g\circ f$ can be interpreted as
`process $g$ \em after \em process $f$'; in the case of structure
preserving maps composition of morphisms is just ordinary function
composition.  Composition is assumed to be associative. One also assumes the existence of units for
this composition; more precisely, for all $A$ there exist identity morphisms
$1_A: A\to A$ such 
that for all $f:A\to B$ and all $g:B\to A$ we have $f\circ 1_A=f$ and
$1_A\circ g=g$. As a physical process this would stand for the \em
void \em process, or in operational terms, ``doing nothing''.\footnote{Obviously, ``doing nothing'' in the lab is a very difficult (if not impossible) task, e.g.~preventing decoherence is the biggest stumbling block to building a quantum computer.}

In addition to the `sequential' composition operation $-\circ-$, an
\smc also comes with `parallel' composition $-\otimes-$.  For two
physical systems $A$ and $B$ there is a \em compound system \em
$A\otimes B$ and for each pair of physical processes $f:A\to C$ and
$g: B\to D$ there is a \em compound process \em $f\otimes g:A\otimes
B\to C\otimes D$.  For mathematical objects $\otimes$ then indicates a
compound mathematical object of a certain kind, built from two
`smaller' ones, e.g.~using the Cartesian product of sets, or the
direct product of groups, or the tensor product of vector spaces.  One
also assumes a \emph{unit object} $\II$  such that composing $A$
with $\II$ leaves $A$ essentially unchanged. Finally, for each pair of
objects $A$ and $B$ one assumes a \em swap morphism \em
$\sigma_{A,B}:A\otimes B\to B\otimes A$.  The remaining axioms of an
\smc then play two roles:
\begin{itemize}
\item \emph{bifunctoriality} states how the two modes of composition
  interact;
\item the existence of a number of \emph{natural isomorphisms} and
  \emph{coherence conditions} between these formalise the meaning of
  `essentially' when saying that $A\otimes\II$ is `essentially' the
  same as $A$.\footnote{ For example, while for all practical
    purposes the sets $X\times (Y\times Z)$ and $(X\times Y)\times Z$
    are equivalent, they are strictly speaking not the same: the first
    one contains elements of the form $(x, (y, z))$ while the second
    one contains elements of the form $((x,y),z)$. Making this notion
    of equivalence mathematically precise is what makes the explicit
    definition of an \smc somewhat
    heavy-handed.}
    The swap morphisms are also natural isomorphisms; these embody the 
    canonical connection between $A\otimes B$ and $B\otimes A$.\footnote{Now, $(x, y)$ and $(y,x)$ are not anymore `essentially the same', but they still are canonically connected via the operation `swapping elements'.}
\end{itemize}
All of these conditions have a straightforward physical interpretation
and are satisfied for most standard mathematical constructions of
compound objects.  

Since there are two modes of composition, \smcs
naturally lend themselves to a 2-dimensional syntax we call the
\emph{graphical} (or \emph{diagrammatic}) \emph{calculus}, where the
vertical axis corresonds the sequential composition ``$\circ$'', and
the horizontal axis to the tensor product ``$\otimes $''.  Moreover,
when expressed in the graphical language, the coherence conditions for
\smcs become trivial as a consequence of some very
powerful theorems, so play no further role in this paper.  Hence,
while below we do state the symbolic definition of a symmetric
monoidal category, it is not crucial for the remainder of this paper.
The graphical language is both clearer and closer to the physical
intuition;  the reader who prefers the graphical langauge can skip
ahead to Section~\ref{subsec:Graph_calc}.

A more detailed presentation of the physical intuition behind
\smcs can be found in
\cite{CatsI,CatsII,ContPhys}; \cite{CatsII} is an extensive tutorial
specifically written to provide the appropriate background on the kind
of category that is required for this paper.  Other tutorials that may
be of help are \cite{LNPAT,LNPBS}.  Mac~Lane's standard textbook on
category theory appeals to a mathematical audience \cite{MacLane}.

The graphical calculus for \smcs can be traced back to Penrose's work in the
early 1970's \cite{Penrose}, but was turned into a formal discipline
only after the work of and Freyd and Yetter, and Joyal and Street,
around 1990 \cite{FreydYetter,JS}.  A physicist friendly presentation
is again in \cite{CatsI,CatsII,ContPhys}, and a specifically targeted
tutorial is again \cite{CatsII}.  A recent comprehensive survey paper
on graphical languages for more general monoidal categories, which settles
a number of caveats of earlier literature, is \cite{LNPS}. The reader
interested in learning more may also find  
\cite{LNPBS,Kock,StreetBook} helpful.

\subsection{Symmetric monoidal categories}

\begin{definition}\em
  A \emph{category} \catC consists of a class of \emph{objects}
  denoted $|\catC|$, and for each pair of objects $A,B\in|\catC|$, a
  set $\catC(A,B)$ of \emph{morphisms} or \emph{arrows}.  For each
  triple $A,B, C\in|\catC|$ there is \emph{composition}
  \[
  -\circ-:\catC (A,B)\times \catC(B,C)\to \catC(A,C),
  \]
  which is associative, i.e.~$(f\circ g)\circ h= f\circ (g\circ h)$,
  and for each object $A\in|\catC|$ there is an \emph{identity}
  morphism $\id{A}:A\to A$, that is, i.e.~for all $f\in \catC(A,B)$ we
  have $f\circ \id{A} = f = \id{B}\circ f$.
\end{definition}

A morphism $f:A\to B$ has \emph{domain} $A$ and \emph{codomain} $B$.  We will
sometimes refer to objects as \emph{types}, to $A$ as the
\emph{input type}, and to $B$ as the \emph{output type}.

In order to precisely state the definition of an \smc we
need to introduce two auxilliary concepts: functors and natural
transformations.  While these definitions may seem rather abstract,
the only examples of them that will be needed are familiar ones: the
tensor product, and isomorphisms between tensor products of objects.

By explicitly stating some basic category-theoretic notions the reader
may get a sense of why even many mathematicians consider category theory 
as `very abstract'; in contrast, the diagrammatic calculus shows that 
specific parts of category theory, namely \smcs and in particular their graphical calculus,
 can make certain mathematical structures way more intuitive and easier to manipulate.

\begin{definition}\em
  Let \catC and \catD be categories.  A \emph{functor} $F: \catC \to
  \catD$ is defined by (i) for each object $A$ in  $\sizeof{\catC}$ an
  object $F(A)$ in $\sizeof{\catD}$, and (ii) for every arrow $f:A\to
  B$ in \catC an arrow $F(f) : F(A) \to F(B)$ in \catD such that:
  \[
  F(f\circ g) = F(f) \circ F(g) 
  \qquad \text{and} \qquad
  F(\id{A}) = \id{F(A)}\,.
  \]
\end{definition}

\begin{remark}
  A variation on the idea of functor is a \emph{contravariant}
  functor, which reverses the  direction of arrows;  that is, $F$
  assigns to every arrow $f:A\to B$ in \catC an arrow $F(f):F(B)\to
  F(A)$ in \catD.
\end{remark}

\begin{definition}\label{def:bifunctor}\em
  A \emph{bifunctor} is a functor 
  of two arguments $F : \catC \times \catC' \to \catD$, that is a
  functor in each argument separately, i.e., for all objects $X$ and
  arrows $f:A\to B$, $g:B\to C$ in \catC, and all objects $X'$ and
  arrows $f':A'\to B'$, $g':B'\to C'$ in $\catC'$, we have:
  \begin{eqnarray*}
    F(g,\id{X'}) \circ F(f,\id{X'}) = F(g\circ f,\id{X'})\,, \\
    F(\id{X},g') \circ F(\id{X},f') = F(\id{X},g'\circ f')\,,
  \end{eqnarray*}
  which additionally satisfies
  \begin{eqnarray*}
    F(g,\id{B'}) \circ F(\id{B},f') = F(\id{C},f') \circ F(g,\id{A'})\,,\\
    F(\id{A},\id{B'}) = \id{F(A,B')}.
  \end{eqnarray*}
\end{definition}

 In essence, a functor is a map between categories that preserves the
structure of the category, \ie composition and identities.  We will
also need maps between functors.

\begin{definition}\em 
  Let $F,G : \catC \to \catD$ be functors;  a \emph{natural
    transformation} $\tau : F \Rightarrow G$ is a family of arrows in
  \catD, $\tau_A : F(A) \to G(A)$, indexed by the objects of $\catC$, 
  such that the following square commutes:
  \begin{diagram}
    F(A) & \rTo^{\tau_A} & G(A) \\
    \dTo<{F(f)} && \dTo>{G(f)} \\
    G(A) & \rTo_{\tau_B} & G(B)\,,
  \end{diagram}
  for all arrows $f:A\to B$ in \catC.  A \emph{natural isomorphism} is
  a natural transformation where each of the $\tau_A$ is an
  \emph{isomorphism}; that is, there exists a morphism $\tau_A^{-1}$ such that $\tau_A\circ\tau_A^{-1}$ and 
  $\tau_A^{-1}\circ\tau_A$ are both identities.
\end{definition}

\paragraph{Notation and terminology:}
\label{sec:terminology}
Each directed path in the diagram above determines a composition of two maps:
$G(f) \circ \tau_A$ on the upper path, and $\tau_B \circ F(f)$ on the
lower.  The phrase ``the square commutes'' means that
both paths in this directed graph are equal, \ie $G(f) \circ \tau_A =
\tau_B \circ F(f)$.

If two objects in category are naturally isomorphic then they are
isomorphic `for structural reasons' and not because of any particular
details of the objects themselves.  The follow definition provides a
key example.

\begin{definition}\em
A  \emph{monoidal category}  $(\catC,\otimes,\II)$ is a category \catC
equipped with a  bifunctor   $-\otimes-: \catC\times \catC\to \catC$,
a distinguished unit object $\II$,   natural  unit  isomorphisms  
\[
\lambda_A : A \simeq {\rm I}\otimes A 
\quad \text{ and } \quad
\rho_A: A \simeq A\otimes{\rm I}\,,
\]
and a natural associativity isomorphism
\[
\alpha_{A,B,C}:A\otimes(B\otimes C)\simeq (A\otimes B)\otimes C\,,
\]
which are subject to certain coherence equations, which we omit. 
\end{definition}

The bifunctor $-\otimes-$ is called the \emph{tensor product} or
\emph{monoidal tensor}.  The maps $\lambda$,  $\rho$, $\alpha$ are
called the monoidal \emph{structure maps}.  A monoidal category is
called \emph{strict} when the structure maps are all identities; that
is, when the objects made isomorphic by $\lambda$,  $\rho$, $\alpha$
are in fact equal.  The following theorem by Mac~Lane justifies our omission of
the coherence equations for the  structure maps.

\begin{theorem}
  Every monoidal category is equivalent to a strict monoidal category.
\end{theorem}

For details we refer the reader to \cite{MacLane}.  Hence forward all
the monoidal categories we consider will be strict, although we will
frequently use the  symbols $\lambda_A$ and $\rho_A$ for clarity, for
example, when composing an arrow of type $B\to A$ with one of type
$A\otimes I \to C$.

\begin{definition}\em
  A \emph{symmetric monoidal category} is a monoidal category
  equipped with a natural \emph{symmetry} isomorphism 
  \[
  \sigma_{A,B}:A\otimes B\simeq B\otimes A
  \]
  such that $\sigma_{A,B}^{-1} = \sigma_{B,A}$, and again subject to
  some coherence conditions which we omit.
\end{definition}

If \catC is an \smc then $\sigma $ is counted among its
structure maps.  Unlike the other structure maps $\sigma$ cannot be
replaced by the  identity without losing essential structure.  We again
refer the reader to \cite{MacLane} for the  details of the coherence
conditions; they are summarised in the following theorem \cite{Kelly1971Coherence-in-cl}:

\begin{theorem}[Kelly-Mac~Lane]
  Let $f$ and $g$ be parallel natural isomorphisms in a symmetric
  monoidal  category, both
  constructed from identities and the structure maps by tensoring
  and composition; then $f = g$.
\end{theorem}

Essentially this result says that when one uses the structure maps to
permute the factors of a tensor product, only the permutation matters,
not how it was constructed.\footnote{The restriction to
\emph{natural} isomorphisms prevents different permutations from being
identified.  For example,  $\id{A\otimes A}$ and $\sigma_{A,A}$ cannot
be identified, despite being parallel arrows, since they are
components of \emph{different} natural transformations, namely $\id{}
\otimes \id{} : A \otimes B \Rightarrow A \otimes B$ and $\sigma : A
\otimes B \Rightarrow B \otimes A$; again, see 
\cite{MacLane} for details.}

The preceding definitions may seen rather intimidating to those
unfamiliar with category theory, but there is no need to be alarmed:
\smcs are among the most ubiquitous of
mathematical structures!

\begin{example}\label{Ex:Hilb}
  The \smc $(\fdhilb, \otimes, \mathbb{C})$,
  often written simply as $\fdhilb$, has finite dimensional Hilbert
  spaces as objects and linear maps as its morphisms, which compose by
  ordinary composition of linear maps.  The familiar Kronecker tensor
  product is the monoidal tensor, and the the field
  of complex numbers $\mathbb{C}$ ---which is a one-dimensional
  Hilbert space over  itself--- is the tensor unit.

  The requirement that the monoidal tensor be a bifunctor reduces to the
  following well-known property of linear maps:
  \[
  (f\otimes g) \circ (h \otimes k) = (f \circ h) \otimes (g \circ k).
  \]
  We indeed also have ${\cal H}\simeq \mathbb{C}\otimes {\cal H}$ via the
  natural isomorphism:
  \begin{eqnarray*}
    &&\lambda_{\cal H}:{\cal H}\to \mathbb{C}\otimes {\cal
      H}::|\psi\rangle\mapsto 1\otimes |\psi\rangle\,,
    \\
    &&\lambda_{\cal H}^{-1}\!: \mathbb{C}\otimes {\cal H}\to {\cal
      H}::c\otimes |\psi\rangle\mapsto c |\psi\rangle\,,
  \end{eqnarray*}
  where naturality means that for all $f:{\cal H}\to {\cal H}'$ the following
  diagram commutes:
  \begin{diagram}
    {\cal H} & \rTo{\lambda_{\cal H}} &  \mathbb{C}\otimes {\cal H} \\
    \dTo<f && \dTo>{\id{\mathbb{C}}\otimes f} \\
    {\cal H}' & \rTo{\lambda_{\cal H}'} &  \mathbb{C}\otimes {\cal H}' \\
  \end{diagram}
  \ie $(1_\mathbb{C}\otimes f)\circ\lambda_{\cal H}= \lambda_{{\cal
      H}'}\circ f$. In $\fdhilb$, it is easily checked that natural
  transformations are always \emph{basis-independent}.  The reader may
  consult \cite{CatsII} for a detailed description of $(\fdhilb,
  \otimes, \mathbb{C})$.
\end{example}

\begin{example}\label{ex:zx-calc-diagrams}
  Let $(\catZX, \otimes, 0)$ denote the \smc whose objects are natural numbers, and whose
  arrows $f:n\to m$ are diagrams of the \zxcalculus, as described in
  Section~\ref{Sec:ZXCalc}, with $n$ inputs and $m$ outputs.  The
  identity arrows are diagrams consisting of straight wires from inputs
  to outputs, and composition is achieved by plugging inputs to
  outputs.

  The tensor product on objects is addition $ n \otimes m := n+m$, and
  the unit object is zero: $n \otimes 0 = n + 0 = n$.  Tensor
  product of two diagrams is juxtaposition, and the identity map
  $\id{0}$ is just the empty diagram.  By its construction, \catZX is
  evidently a strict monoidal category.  We leave to the reader the task of
  constructing the symmetry maps $\sigma_{n,m}$ from crossings of
  wires.

  We remark in passing that the assignment from a diagram \diagD{} to
  its corresponding linear map $D$, described in
  \ref{sec:interpr-zx-calc}, defines a functor from \catZX to \fdhilb.
\end{example}

In the categorical setting the internal structure of the objects is hidden---abstracted away; the state spaces are effectively reduced to
labels which determine when morphisms may be composed.  However, in
\fdhilb and many other important examples, the internal structure of
the spaces may be reconstructed via the structure of the morphisms
into that space.

\begin{definition}\label{defn:points}\em
  Morphisms of type $\II\to A$ in a monoidal category ${\bf C}$
  are called \emph{points of $A$}.
\end{definition}

\begin{example}\label{ex:Hilb-points}
Any linear map $\psi:
\mathbb{C}\to {\cal H}$ is completely determined by $\psi(1)$, due to
linearity, hence there is a bijection,
\[
\fdhilb(\mathbb{C},{\cal H})\to {\cal H}::\psi\mapsto \psi(1).
\]
So the elements of $\fdhilb(\mathbb{C},{\cal H})$ are the points
 of the object ${\cal H}$.  To distinguish between the linear map
$\psi$ and the vector $\psi(1)$ we will denote the latter by
$|\psi\rangle$.    As processes, we can
think of these points $\psi: \mathbb{C}\to {\cal H}$ also as \em
preparation procedures\em.
\end{example}

A point $\Psi:\II\to A\otimes B$ is a state of the compound system
$A\otimes B$, and this state may or may not be entangled.  If it is
not entangled, then we have 
\[
\Psi=(\psi_A\otimes\psi_B)\circ\lambda_\II\,, 
\] 
that is, the state $\Psi$ factors in state $\psi_A$ of system $A$ and
state $\psi_B$ of system $B$. It is entangled if such a factorisation
does not exist.  If the category bears certain additional structure,
e.g.~compactness as described in Section~\ref{sec:corrgraphreason}, then the
existence of entangled states can be guaranteed, which in turn enables
the derivation of teleportation-like protocols \cite{AC1}.

In many categories, the points can reveal a great deal
about the arrows.  For example, in a vector space, two linear maps are
equal if they agree on a small number of points, namely a basis.  To
tell if two functions are equal, it suffices to evaluate them on every
element of their domain.  The analogous procedure  
is not possible in every category.  More precisely, a set of
points ${\cal K} \subseteq \catC(I,A)$ is called a \emph{basis}
for $A$ if for all objects $B$, and all arrows $f,g:A\to B$, we have
\[
\left[ \forall k \in {\cal K} :  f\circ k = g \circ k \right]
\text{ implies }
f = g\,.
\]
If every object of \catC has a basis, then we say that \catC \emph{has
  enough points}.  Fortunately, the examples of interest here
\emph{do} have enough points, and Section~\ref{sec:quant-mech-concr}
describes the particular forms of bases that will be of interest in
later sections.

\begin{definition}\label{defn:scalars}\em
  Let \catC be monoidal category; the arrows of type $\II\to \II$ are
  called the \emph{scalars} of \catC.  Given a scalar $c:\II\to \II$,
  we call the natural transformation with components
  \[
  c\cdot 1_A:=\lambda^{-1}_A \circ (c \otimes 1_A) \circ\lambda_A:A\to A
  \]
  the \emph{scalar multiplication} by $c$.  
\end{definition}

More explicitly, we can define
\begin{equation}\label{defeq:scalarmult}
c\cdot f := f\circ (c\cdot 1_A) 
= (c\cdot 1_B) \circ f 
= \lambda^{-1}_B \circ (c \otimes f) \circ \lambda_A
\end{equation}
to be the scalar multiplication of morphism $f:A\to B$ by the scalar $c$.

The scalars, in any monoidal category, form a commutative
monoid
with respect to composition \cite{KellyLaplaza}.  From
the definition of scalar multiplication it follows that
\begin{eqnarray}\label{defeq:scalarmult2}
(c\cdot f)\circ (c'\cdot g) & = &(c\circ c')\cdot(f\circ g)\,,
\\ \label{defeq:scalarmult3}
(c\cdot f)\otimes (c'\cdot g)& = &(c\circ c')\cdot(f\otimes g)\,.
\end{eqnarray}
Intuitively, in the language of \smcs, if a
scalar appears in the description of a morphism, it does not matter
where it appears: its effect is that of a global multiplier for the
entire morphism.

\begin{example}\label{ex:Hilb-scalars}
In $\fdhilb$ the scalars are the complex numbers.  Indeed, a linear map $c:
\mathbb{C}\to \mathbb{C}$ is completely determined by $c(1)$, due to
linearity, so there is a bijection
\[
\fdhilb(\mathbb{C},\mathbb{C})\to \mathbb{C}::c\mapsto c(1)\,.
\]
Scalar multiplication as in (\ref{defeq:scalarmult}) coincides with the usual linear algebraic notion, for which (\ref{defeq:scalarmult2}) and (\ref{defeq:scalarmult3}) indeed hold.   The commutative monoid of scalars is isomorphic to the monoid of the complex
numbers $(\mathbb{C},\cdot, 1)$.   More details are again in \cite{CatsII}.
\end{example}

\subsection{The $\dag$ functor}

Following \cite{AC1,AC2,Selinger}, we augment \smcs with additional structure that plays an essential role in
the quantum mechanical formalism. 

\begin{definition}\label{def:dagSMC}\em
  A  \emph{$\dag$-symmetric monoidal category} (\dsmc) is a symmetric
  monoidal category equipped with an identity-on-objects contravariant
  endofunctor 
  \[ 
  (-)^\dagger:{\bf C}^{op}\to{\bf C}\,, 
  \]
  which assigns to each morphism $f:A\to B$ an \emph{adjoint}
  morphism $f^\dagger:B\to A$, which coherently preserves the
  monoidal structure, that is:
  \[
  (f\circ g)^\dagger =g^\dagger\circ f^\dagger  
  \quad\ \
  (f\otimes g)^\dagger=f^\dagger\otimes g^\dagger 
  \quad\ \
  \id{A}^\dagger =\id{A}
  \quad\ \
  f^{\dagger\dagger}=f\,.
\]
Further, for the natural isomorphisms $\lambda$,
$\rho$, $\alpha$ and $\sigma$ of the symmetric monoidal structure, the
adjoint and the inverse coincide.
\end{definition}

\begin{definition}\label{def:unitarity}\em
If for an isomorphism $f:A\to B$ in a  \dsmc the adjoint and the inverse coincide, that is, $f^\dagger=f^{-1}$,  then we call it \em unitary\em.
\end{definition}

\begin{remark}
In a \dsmc, the monoid of scalars is  \em involutive\em, that is, there is an operation $\dagger:{\bf C}(\II, \II)\to {\bf C}(\II, \II)$ which satisfies 
  \[
  (c\circ d)^\dagger =d^\dagger\circ c^\dagger  
  \quad\ \ \ \ \ \ 
  \id{\II}^\dagger =\id{\II}
  \quad\ \ \ \ \ \ 
  c^{\dagger\dagger}=c\,.
\]
\end{remark}


\begin{example}\label{Ex:Hilb2}
In $\fdhilb$ the $\dag$ functor is given by the adjoints of linear algebra.  The involution for the monoid of scalars is complex conjugation. 
\end{example}

The category $\fdhilb$ is obviously not the only example of a \dsmc;
by its construction \catZX is a \dsmc.  We offer some further examples.

\begin{example}[relations]\label{ex:Spek}
  Recall that for two relations $r\subseteq X\times Y$ and $s\subseteq
  Y\times Z$ the \emph{relational composite} is again a relation:
  \[
  s\circ r:=
  \{(x,z)\mid \exists y: (x,y)\in r\,, (y,z)\in s\}\subseteq X\times Z\,.
  \]
  The category ${\bf Rel}$ which has sets as objects, relations as
  morphisms and relational composition is a \dsmc with the Cartesian
  product of sets as monoidal structure, and the relational converse
  as the $\dagger$ functor.  The unit object for the monoidal
  structure is the singleton set $ \{*\}$, since $X\times \{*\}\simeq
  X$ for any set $X$, and the monoid of scalars is now isomorphic to
  the Boolean monoid $(\mathbb{B},\wedge, 1)$, since there are only
  two relations $r: \{*\}\to \{*\}$, namely the empty relation and the
  identity relation. The involution for the monoid of scalars is now
  trivial.  We write ${\bf FRel}$ when restricting to finite sets.  In
  ${\bf Rel}$ the points of an object $X$ are not its elements but its
  \emph{subsets} ---a detailed discussion is in
  \cite{CatsI,CatsII}. While at first sight ${\bf (F)Rel}$ seems to
  have little to do with physics, it enables to encode a surprising
  amount of quantum phenomena. For example, Spekkens' toy quantum
  theory \cite{Spekkens} can be embedded within it as a sub-\dsmc
  ${\bf Spek}$ \cite{Spek,CES}.  This succinct categorical
  presentation of this toy theory is moreover the only currently
  available rigorous mathematical presentation of it.
\end{example}

\begin{example}[projective spaces]\label{ex:Proj}
  The passage from vectors to states comes with a radical change of
  the mathematical description of the spaces: from vector spaces to
  projective spaces, which drove von Neumann into the advent of
  lattice theory and quantum logic \cite{Birk,Redei}. Menwhile, 75
  years later, it is fair to say that the quantum logic research
  program failed in reaching its ambitious goals, the main problem
  being the failure to account for the tensor product description of
  compound quantum systems.  However, at the level of monoidal
  categories, which has the tensor build in as a primitive concept,
  the passage from vector spaces to projective spaces proceeds without
  any loss of the structures that play a role in this paper
  \cite{deLL}.  Let $\fdhilb_p$ be the category which has the same
  objects as $\fdhilb$ but whose morphisms are equivalence classes of
  ${\bf FdHilb}$-morphisms, given by the equivalence relation 
  \[
  f\sim g 
  \; \Leftrightarrow \; 
  \exists c\in\mathbb{C}\setminus\{0\}~s.t.~f=c\cdot g \ .  
  \] 
  In $\fdhilb_p$ the states are now indeed the rays of the Hilbert
  space, together with one point representing the zero vector. The
  points for the two-dimensional Hilbert space in $\fdhilb_p$, the set
  $\fdhilb_p(\mathbb{C},\mathbb{C}^2)$, correspond with the points of
  the Bloch
  sphere.  

  Operationally, the meaningful scalars are the probability
  amplitudes.  In ${\bf FdHilb}$ the scalars are the complex numbers,
  hence too many, and in $\fdhilb_p$ there are only two, hence too
  few.  The solution consists of enriching $\fdhilb_p$ with
  \emph{probabilistic weights}, i.e.~to consider morphisms of the form
  $r\cdot f$ where $r\in\mathbb{R}^+$ and $f$ a morphism in
  $\fdhilb_p$. Therefore, let $\fdhilb_{wp}$ be the category whose
  objects are those of ${\bf FdHilb}$ and whose morphisms are
  equivalence classes of ${\bf FdHilb}$-morphisms for 
  \[ 
  f\sim g 
  \;  \Leftrightarrow \; 
  \exists \alpha\in[0,2\pi) \text{ s.t. }   f=e^{i\alpha}\cdot g.  
  \] 
  A detailed categorical account on $\fdhilb_{wp}$ is in \cite{deLL}.

  The three categories considered above are related via inclusions:
  \begin{diagram}
    \fdhilb_p & \pile{\lOnto^{\qquad\qquad} \\ \quad \\\rInto } &
    \fdhilb_{wp} & \pile{\lOnto^{\qquad\qquad} \\ \quad \\\rInto} &
    \fdhilb
  \end{diagram}
  The theorems proven in this paper apply to all of these.
\end{example}

\begin{example}[mixed states and completely positive maps]\label{ex:CPM}
  In this paper all processes are \emph{pure} (or \emph{closed}).
  However, given any \dsmc \catC of `pure states' it is possible to
  construct a new category $\mathrm{CPM}(\catC)$ of mixed states and
  completely positive maps which is again a \dsmc.  This method,
  Selinger's CPM-construction \cite{Selinger}, will be sketched in
  Section~\ref{ex:teleportation}.
\end{example}

\subsection{Diagrammatic calculus}\label{subsec:Graph_calc} 

In diagrammatic calculus, morphisms in \smcs are represented by \emph{boxes}, input
types by  \emph{input wires}, and output types by \em output wires\em.
Identities can be represented by wires only.  Hence, $1_A:A\to A$ and
$f:A \to B$ are respectively depicted as:
\beq
\inlinegraphic{4.2em}{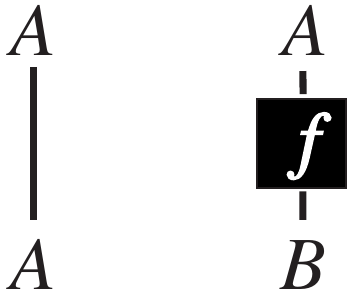}
\eeq
The input types and output type of a box can itself be compound and the unit object $\II$ is represented by no wire;  a morphism $f:A_1\otimes \ldots \otimes A_n \to B_1\otimes \ldots \otimes B_m$  and $g: \II\to B$,  $h: A\to \II$ and $k: \II\to\II$ are depicted as:
\beq\label{MorpCompound}
\inlinegraphic{4.2em}{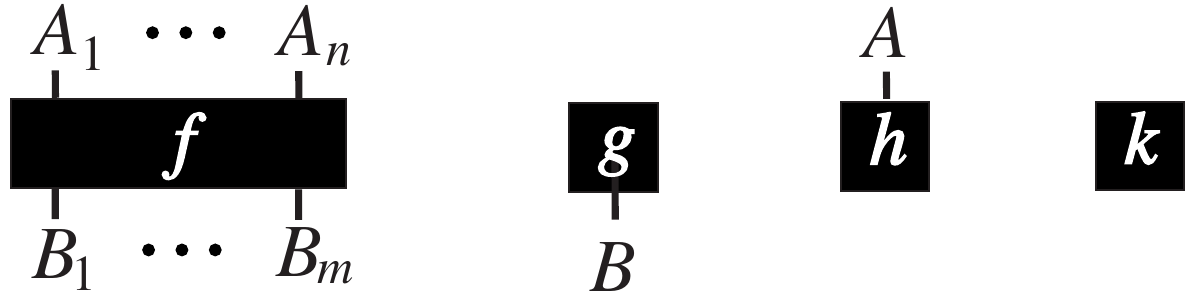}
\eeq
The identity on the monoidal unit $1_\II:\II\to\II$ is represented by, equivalently, 
an `empty picture' (be it either a wire or a box)  ---hence the graphical representation of an equation of the form $s=1_\II$ leaves the right-hand side empty.  The symmetry natural isomorphisms $\sigma_{A,B}:A\otimes B\simeq B\otimes A$ are depicted as:
\beq\label{MorpSym}
\inlinegraphic{4.2em}{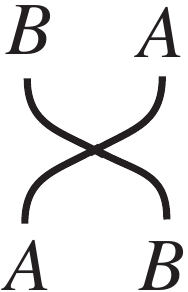}
\eeq
These boxes\footnote{Although `box' should be understood
  figuratively:  we allow ourselves other shapes as well.}, straight
wires, and crossings of wires are the only graphical elements that
make up the graphical language.  We can compose morphisms in an \smc
in two manners, and similarly, we can compose these graphical elements
in two manners,  depicted by connecting matching outputs and inputs by
wires, and tensor by juxtaposing wires or boxes side by side.  Hence
$g\circ f: A\to B\to C$ and $f\otimes g: A\otimes C\to B\otimes D$ are
respectively depicted as:
\beq
\inlinegraphic{6.0em}{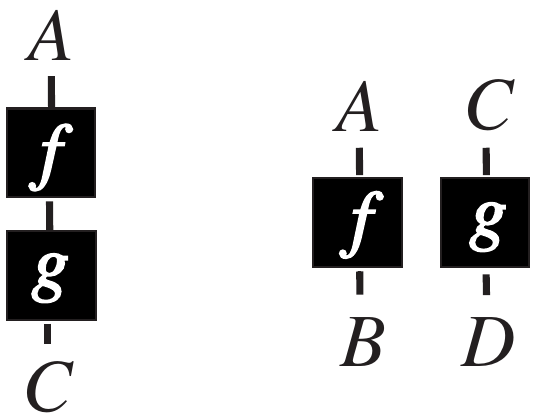}
\eeq
where $A, B, C, D$ may themselves be compound as in (\ref{MorpCompound}).
Morphisms that play special roles may of course be given special graphical representations, sometimes other than boxes.  The connection between this graphical language and the symbolic definition of \smcs is established as follows.

\begin{definition}\em\label{graph:equiv}
By \em isomorphism of diagrams \em we mean that there is a bijective correspondence between boxes and wires which preserves the manner in which boxes and wires are connected ---symmetry (cf.~\ref{MorpSym}) is interpreted as a pair of crossing wires.
\end{definition}

We will use equality  ``='' to denote isomorphic diagrams.  Examples are:
\begin{equation}
\inlinegraphic{5.6em}{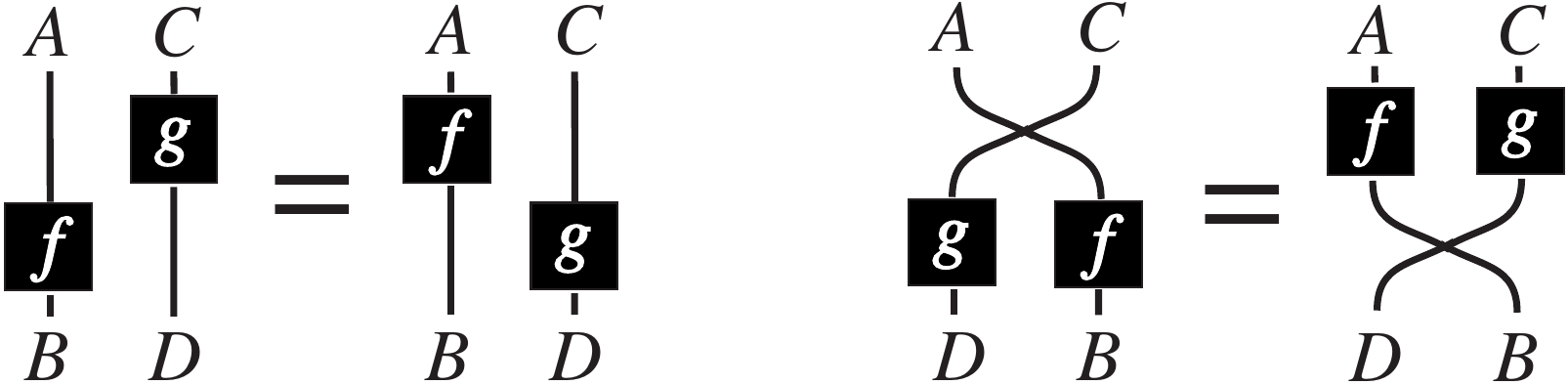}
\end{equation}
Each represents (part of) an axiom of \smcs, namely commutation of:

\begin{equation}
\xymatrix{
A\otimes C\ar[r]^{1_A\otimes g}\ar[d]_{f\otimes 1_C}& A\otimes D\ar[d]^{f\otimes 1_D}&&&
A\otimes C\ar[r]^{\sigma_{A,C}}\ar[d]_{f\otimes g}&C\otimes A\ar[d]^{g\otimes f}
\\
B\otimes C\ar[r]_{1_B\otimes g}& B\otimes D&&&
B\otimes D\ar[r]_{\sigma_{B,D}}&D\otimes B
}©
\end{equation}
This correspondence instantiates a perfect correspondence between the symbolic representation and the graphical presentation of \smcs:

\begin{theorem}[Joyal-Street \cite{JS}]\label{thm:JoyalStreet}
An equation expressed in the symbolic language of \smcs follows from the axioms of \smcs if and only if it holds up to \underline{isomorphism of diagrams} in the graphical language. 
\end{theorem}

As Selinger pointed out in \cite{Selinger}, this result straightforwardly extends to \dsmcs, when representing the adjoint by vertical reflection.  This requires breaking the symmetry of the boxes used above, and  $f: A\to B$ and $f^\dagger:B\to A$ will now be respectively depicted as:
\beq
\inlinegraphic{4.0em}{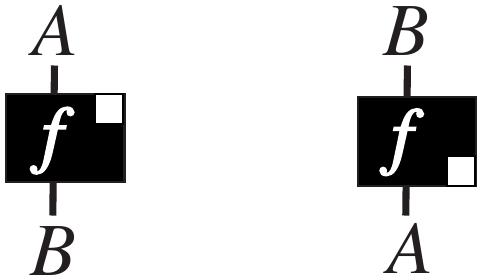}
\eeq
Note that reflecting  twice leaves the box invariant, exposing the
involutive nature of the adjoint.

\subsection{Graphical reasoning}\label{sec:corrgraphreason}

In the following sections the graphical language will be preferred to
conventional linear syntax.  Before proceding, we briefly discuss the
question of equational reasoning in the graphical language: when do
two diagrams denote the same mathematical object?  When are two
diagrams equal?  There are two basic elements: \emph{isomorphism of
  diagrams}, as discussed above; and \emph{substitution}.  Together
these principles allow us to reason by \emph{rewriting}.

The first prinicple, isomorphism, requires little elaboration beyond
the discussion of Section~\ref{subsec:Graph_calc}.  The
graphical language is a syntax for describing certain mathematical
objects, namely the arrows of symmetric monoidal categories; 
Theorem~\ref{thm:JoyalStreet} states that two diagrams denote equal
arrows when they are isomorphic (in
the sense of Definition \ref{graph:equiv}) .  

However, the axioms of \smcs are rather weak, so the isomorphism
principle will not suffice.  We must impose other equations upon our
diagrams to obtain our results.  For example, consider an arrow
$f:A\to A$.  The statement that $f$ is unitary is expressed by the
equations
\begin{equation}\label{eq:example-f-unitary}
f \circ f^\dag = \id{A} = f^\dag \circ f\,.
\end{equation}
Since this equation is a property of the particular morphism $f$, there
is no hope to absorb it into some overarching global principle, in the way that Theorem~\ref{thm:JoyalStreet}
absorbs the axioms of \smcs.  Indeed, most of the equations in this
paper are of this sort, naked identifications that must be imposed on
the langauge.  We deal with these via substitution.

From a syntactic point of view, the meaning of an equation such as
(\ref{eq:example-f-unitary}) is that whenever $f^\dag 
\circ f$ is found lurking inside some larger expression, it can be
replaced by $\id{A}$ without changing the meaning of the containing
expression.  The same method of substitution applies in the graphical
language.

The equations above can be easily translated into
diagrams:
\[
\ninlinegraphic{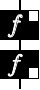} = \ninlinegraphic{small-id} = \ninlinegraphic{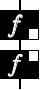}
\]
To perform the substitution, we isolate the part of the diagram
corresponding to one side of the equation, and form a new diagram by
replacing that part with the  other side of the equation.  
\[
\ninlinegraphic{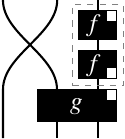}  
=  \ninlinegraphic{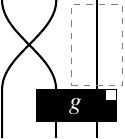}
\]
Since two equal diagrams must have the same type, such a
replacement is always possible.

In general, when presented with a diagram $d$, any vertical line
(which does not intersect any box or wire) divides the diagram into
two parts, $f$ and $g$, each of which is a valid diagram, and which
are related by the equation $d = f \otimes g$.  Similarly, any
horizontal line (which may cut wires, but avoids boxes and points
where wires cross) will divide $d$ into two sub-diagrams, $f'$ and
$g'$, which satisfy $d = f' \circ g'$.  By a sequence of such
horizontal and vertical divisions, the diagram can be cut into squares
each of which contains an atomic element of the graphical language:
a single box, a single crossing of wires, or a single straight wire.  
Each atomic square corresponds a single symbol in standard notation,
and hence the factorisation yields a symbolic expression equivalent to
the diagram.  The desired substitution can be performed on the
symbolic expression, and a new diagram constructed.  We can pass freely
from the symbolic representation to the graphical representation and
back.
\begin{diagram}
  \ninlinegraphic{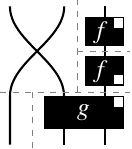} & = & \ninlinegraphic{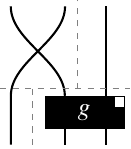}  \\
\dMapsto && \uMapsto \\
(\id{A} \otimes g)\circ (\sigma \otimes (f^\dag \circ f)) 
&  \quad=\quad & (\id{A} \otimes g)\circ (\sigma \otimes \id{A})
\end{diagram}
The above justification for substitution is admittedly rather sketchy.
Joyal and Street \cite{JS} provide rigorous presentation with all the
details.
In practice, there is no need to divide a diagram into atomic pieces:
it suffices to isolate the part to be substituted.  We will apply this
technique throughout this paper.

\subsection{Correctness of graphical reasoning in \zxcalculus}\label{sec:corrgraphreasonZX}

The cautious reader will be aware that the \ruleT-rule of the \zxcalculus in Section~\ref{Sec:ZXCalc}---`only the topology of the
diagram matters'---stretches well beyond the graphical reasoning
techniques outlined above.  As well as being a \dsmc, \catZX has a
richer structure, and enjoys a correspondingly
stronger analogue to Theorem~\ref{thm:JoyalStreet}.

\begin{definition}\em
A \emph{dagger compact (closed) category} is a
\dsmc in which for each object $A$ there exists \emph{dual} object $A^*$
and a morphism $\eta_A:\II\to A^*\otimes A$ subject to certain
coherence equations (for which see \cite{AC1,Selinger}).
\end{definition}

Throughout this paper we will take $A^*=A$; this self-duality requires
some additional coherence axioms for which we refer the reader to
\cite{SelingerSelfDual}.  

For present purposes the required equations reduce to a very simple form. Denoting $\eta_A$ graphically by 
$\inlinegraphic{1.00em}{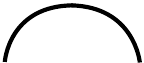}$\,, the equations that need to be satisfied are: 
\beq\label{eq:yanking1}
\hspace{-7mm}\lambda_A^\dagger\circ(\eta^\dagger\otimes 1_A)\circ(1_A\otimes \eta)\circ\rho_A=1_A\quad\mbox{\rm i.e.}\quad\inlinegraphic{3.00em}{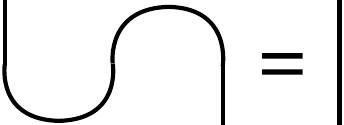}  
\eeq
\beq\label{eq:yanking2}
\hspace{-7mm}\sigma_{A,A}\circ\eta = \eta\qquad\mbox{\rm i.e.}\qquad\inlinegraphic{3.00em}{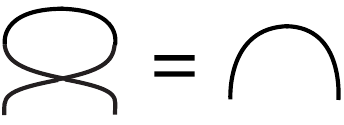}  
\eeq
  
\begin{theorem}[Kelly-Laplaza \cite{KellyLaplaza}; Selinger \cite{Selinger,SelingerSelfDual}]\label{thm:graphical-dag-compact}
  An equation expressed in the symbolic language of dagger compact
  category follows from the axioms of dagger compact categories if and
  only if it holds up to \underline{isotopy} in the graphical
  language.
\end{theorem} 

This result may be used in the \zxcalculus due the last two
equations of the \ruleSii-rule, which allows us to consider the caps and
cups as part of the overall categorical structure.  In what follows
these equations will typically not hold.  A collection of many other
theorems on diagrammatic languages can be found in \cite{LNPS}.



\section{Vector  bases and state bases of observables}
\label{sec:quant-mech-concr} 

Before diving in to the categorical treatment of observables, we 
briefly recall the relevant notions in the concrete setting of Hilbert
space quantum mechanics.  All Hilbert spaces involved will be
finite-dimensional.  We will denote the set of rays in a Hilbert space
${\cal H}$, which we 
refer to as the \emph{state space}, by ${\cal H}$ itself. To distinguish
between states and vectors, we write $\state{\psi}$ to denote the unique
state containing the (non-zero) vector $\ket{\psi}$. Similarly,
$\state{\sum_i c_i\ket{v_i}}$ is the state spanned by vector $\sum_i 
c_i \ket{v_i}$. 

All observables considered will be non-degenerate: $\hat{O} = \sum_i
\lambda_i \outp{v_i}{v_i}$.  For current purposes, the importance of
the observable is the state change $\state{\psi} \mapsto \state{v_i}$
induced by measuring it.  Since the actual values $\lambda_i$ are of
no concern here, we identify an observable with its set of
eigenstates, $\left\{\state{ v_1},\ldots, \state{ v_n}\right\}$.  We
refer to the orthonormal basis $\left\{\ket{v_1},\ldots,
\ket{v_n}\right\}$ as a \emph{vector basis}, cf.~Definition
\ref{def:statebasis} below.  To summarise:
\begin{center}
\begin{tabular}{|c|c|c|c|} 
vector & state & vector basis & observable\\
\hline
$\ket{\psi}$ & $\state{\psi}$ & $\left\{\ket{v_1},\ldots, \ket{v_n}\right\}$ & $\left\{\state{ v_1},\ldots, \state{ v_n}\right\}$
\end{tabular}
\end{center}

\begin{definition}\em
  A vector $\ket{\psi}$ (or state $\state{\psi}$) is \emph{unbiased}
  relative to a vector basis $\left\{\ket{v_1},\ldots,\ket{v_n}\right\}$ 
  (or observable $\left\{\state{ v_1},\ldots,\state{ v_n}\right\}$)
  if for all $i, j$ we have 
  \[
    \sizeof{\innp{v_i}{\psi}}=\sizeof{\innp{v_j}{\psi}}\,.
    \]
  In particular, when ${\cal H}=\mathbb{C}^D$ then 
  $\sizeof{\innp{v_i}{\psi}} =  1/\sqrt{D}$\, for all $i$.     
  Two vector bases (or two observables) are \emph{complementary} (or
  \emph{mutually unbiased}) if each vector (or state) in one of these
  vector bases (or observables) is unbiased relative to the other
  vector basis (or observable).
\end{definition}

The key physical fact here is that when a state is unbiased to some
observable, all the outcomes of that  observable are equally likely.
If classical data is encoded in the eigenbasis of some
observable, for example a classical bit encoded as a qubit in the
basis $\{\ket{0},\ket{1}\}$, then measuring an unbiased observable
will effectively \emph{erase} that data, regardless of which outcome
is observed. 

A vector basis of a Hilbert space is characterised by the following property:

\begin{proposition}\label{prop:Vect_base}
Let $\left\{\ket{v_1},\ldots,
\ket{v_n}\right\}$  be any vector basis of a Hilbert space ${\cal H}$.
Then, any linear map $f:{\cal H}\to {\cal H}'$ is completely determined by
the values $f$ takes on $\ket{v_1},\ldots,
\ket{v_n}$.  Further, no proper subset $\left\{\ket{v_1},\ldots,
\ket{v_n}\right\}$ suffices to determine $f$.
\end{proposition}

Any regular linear map induces a map from states to states, namely
\[
\hat{f}:: \state{\psi}\mapsto \state{f(\ket{\psi})}\,.
\]
However, the values that $\hat{f}$ takes on an observable
$\left\{\state{ v_1},\ldots, \state{ v_n}\right\}$ do not suffice to
fix $\hat{f}$ itself. For example, let $\theta,\theta'\in [0,2\pi)$,
and define a family of  linear maps, relative to basis
$\{\ket{0},\ket{1}\}$, by the matrix:
\[
f_\theta = 
\left(
  \begin{array}{rl}
    1&\ 0\\ 0&\  e^{i\theta}\!\!
  \end{array}
\right)\,.
\]
Every $\hat{f_\theta}$ leaves both $\ket{0}$ and $\ket{1}$ invariant, while  
\[
\hat{f}(\state{+})
= \state{ \ket{0} +  e^{i\theta} \ket{1}} 
\not = \state{ \ket{0} +  e^{i\theta'} \ket{1}}
= \hat{f_{\theta'}}(\state{+})
\]
whenever $\theta\not=\theta'$.
 Is there an analogue to Proposition~\ref{prop:Vect_base}? 
 Can $\hat{f}$ be characterised by its image on some minimal set of
 states?  The answer is yes:

\begin{definition}\label{def:statebasis}\em
A set of states of the form 
\[
\mbox{observable}\,  \cup \{\mbox{unbiased state for that observable}\}
\]
is called a \emph{state basis}. The unbiased state is called the
\emph{erasing point}.   
\end{definition}

\begin{proposition}\label{prop:State_base}
Any map on states $\hat{f}$ induced by a regular linear map $f:{\cal
  H}\to {\cal H}'$ is completely determined by the values it takes on
a state basis for some arbitrary observable.  Moreover,  no proper
subset of such a set of states suffices to determine $\hat{f}$. 
\end{proposition}

\begin{proof}
  Let $f$ be a regular linear map, and let $\hat{f}$ be the
  corresponding map of states.  Let $\{\state{ v_1},\ldots, \state{
  v_n}\}\cup\{\state{ s }\}$ form a state basis, and suppose that
  $\hat{f}$ takes known values upon these states.  We will show this
  determines $f$ on a vector basis
  $\{\ket{\eta_1},\ldots,\ket{\eta_n}\}$ of ${\cal H}$, up to a
  common, overall phase.  

  Set $\ket{\eta_i}= \innp{v_i}{s}\ket{v_i}$ and let ${\cal H}''$ be
  the subspace spanned by $f(\ket{\eta_1}),\ldots, f(\ket{\eta_n})$.
  Since $f$ is regular $\{f(\ket{\eta_1}),\ldots,f(\ket{\eta_n})\}$ is
  a basis for ${\cal H}''$, and let $\innp{-}{-}_{\diamond}$ denote
  the inner-product on ${\cal H}''$ for which
  $\{f(\ket{\eta_1}),\ldots,f(\ket{\eta_n})\}$ is orthonormal. Then
  the codomain restriction of $f$ to $({\cal H}'',
  \innp{-}{-}_{\diamond})$ is unitary.  Relying on this we have
  $f(\ket{\eta_i})=f(\innp{v_i}{s}\ket{v_i})= \innp{v_i}{s}
  f(\ket{v_i}) = \innp{f(\ket{v_i})}{f(\ket{s})}_{\diamond}\,
  f(\ket{v_i})$.
  This expression is completely determined by $\hat{f}(\state{ v_i})$
  and $\hat{f}(\state{ s})$ up to a phase factor contributed by
  $f(\ket{s})$, but this phase factor is the same for all
  $f(\ket{\eta_i})$.  

  It now follows that, given an arbitary state 
  $\state{\psi}=\state{\sum_i c_i\ket{\eta_i}}$, we have
  $\hat{f}(\state{\psi})
  =  \state{f(\ket{\psi})} 
  =  \state{f(\sum_i c_i\ket{\eta_i})}
  = \state{\sum_i c_i f(\ket{\eta_i})}$,
  where each $f(\ket{\eta_i})$ is determined upto the common
  phase.  This phase is therefore a global phase for the vector
  $\sum_i c_i f(\ket{\eta_i})$, hence   $\hat{f}(\state{\psi})$ yields a
  unique state, and so  $\hat{f}$ is well-defined on all states.

  It is easily seen that no proper subset of
  $\{\state{ v_1},\ldots, \state{ v_n}, \state{ s }\}$ is sufficient to
  completely determine $\hat{f}$.
\end{proof}\vspace{-3.5mm}


State bases and vector bases are related by the following proposition: 

\begin{proposition}\label{Base_State_iso}
Let ${\cal S}$ be the set of all state bases for ${\cal H}$, let
${\cal V}$ be the set of all vector bases for ${\cal H}$, and let
${\cal V}/\sim$ be the set of equivalence classes $[-]_\sim$ in ${\cal
  V}$ for the equivalence relation `equal up to an overall phase'
i.e. 
\bc$
\left\{\ket{v_1},\ldots, \ket{v_n}\right\}\sim\left\{\ket{w_1},\ldots, \ket{w_n}\right\}
\ \Leftrightarrow\ \exists \theta \mbox {such that } \forall j: \ket{v_j}= e^{i\theta}\cdot \ket{w_j}.
$\ec
There is a bijective correspondence 
\bc$
\xymatrix@=.80in{
{\cal S}\ar@/^0.4em/[r]^{sv}&\ar@/^0.4em/[l]^{vs}{\cal V}/\sim
}
$\ec
where 
\bit
\item $sv:\left\{\state{ v_1},\ldots, \state{ v_n}\right\}\cup\{\state{ s }\}\mapsto[\left\{\innp{v_1}{s}\ket{v_1},\ldots, \innp{v_n}{s}\ket{v_n}\right\}]_\sim$
\item $vs:[\left\{\ket{v_1},\ldots, \ket{v_n}\right\}]_\sim\mapsto\left\{\state{ v_1},\ldots, \state{ v_n}\right\}\cup\{\state{\sum_i \ket{v_i}}\}$.
\eit
\end{proposition}
\begin{proof}
Note that $sv$ is indeed well-defined in the sense that
its prescription does not depend on the choice of the respective 
vectors $\ket{v_1}, \ldots,\ket{v_n},\ket{s}$  in the states $\state{
v_1},\ldots, \state{ v_n},\state{ s}$. Also $vs$ is easily seen to be well-defined. Verifying that these maps are
each other's inverse is straightforward. 
\end{proof}\vspace{-3.5mm}

\begin{example}
On the Bloch sphere the erasing point lies on the equator for the antipodal points that represent the observable. E.g.~for the ${\sf Z}$-observable $\{\state{0},\state{1}\}$ and the erasing point 
$\state{+}:=\state{ \ket{0}+\ket{1}}$ we have:
\[
\inlinegraphic{8em}{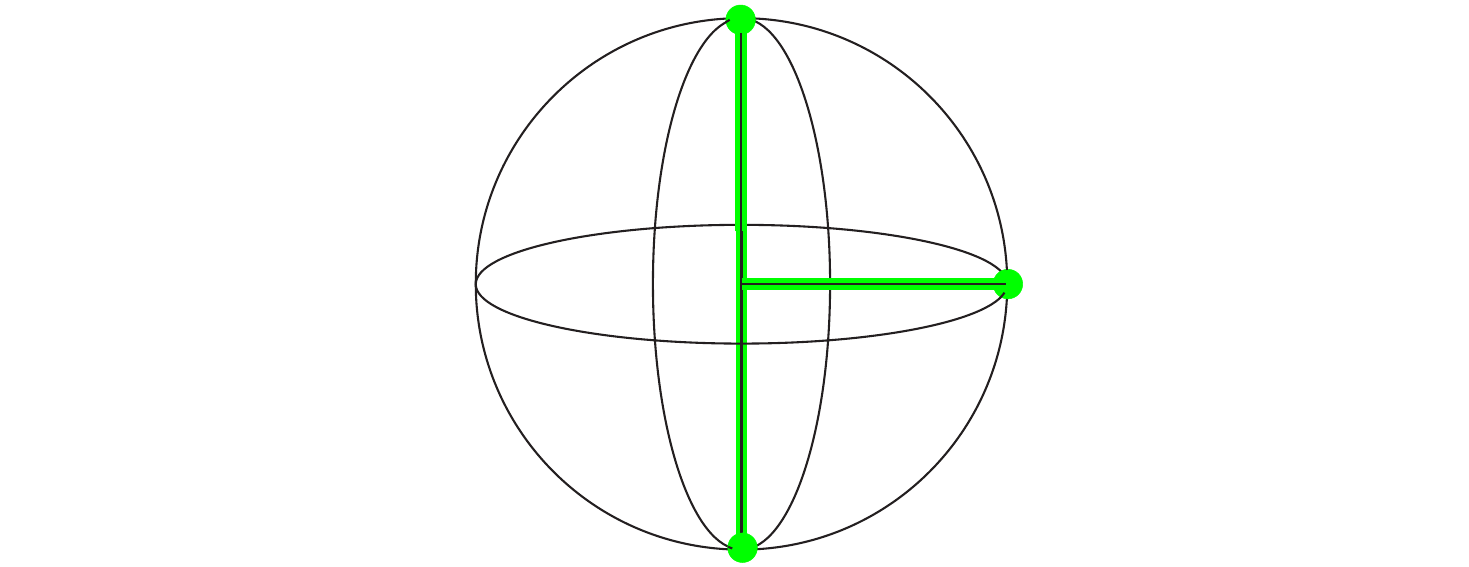}
\]
so the observable and the erasing point together make up a T-shape.
\end{example}

Given a vector basis, we can turn it into an observable by forgetting
the phases of each of the basis vectors, which we formalise by
passing to equivalence classes.  To construct a vector basis from an
observable, we have to choose a phase for each basis element. This
construction factors over the construction of state bases as follows: 
\bc$
\xymatrix@=1.30in{
\mbox{observable} \ar@/^0.2em/[r]^{\mbox{\it\scriptsize\ \ choose erasing point}}\ar@/^2.6em/[rr]^{\mbox{\it \scriptsize choose individual phases}} & 
\ar@/^0.2em/[l]^{\mbox{\it \scriptsize\ \ forget erasing point}} \mbox{state basis}\ar@{}[r]|{\mbox{\scriptsize Prop.~\ref{Base_State_iso}}} \ar@/^0.5em/[r]^{\mbox{\it \scriptsize choose overall phase\ \ \ \ }} & 
\ar@/^0.5em/[l]^{\mbox{\it \scriptsize forget overall phase\ \ \ \ }}\ar@/^2.8em/[ll]^{\mbox{\it \scriptsize forget  individual phases}} \mbox{vector basis}
}
$\ec

\begin{definition}\em\label{def:coherence}
A pair of mutually unbiased vector bases (MUVBs), ${\cal
  V}=\{\ket{v_1},\ldots,\ket{v_n}\}$ and ${\cal
  W}=\{\ket{w_1},\ldots,\ket{w_n}\}$, are called \emph{coherent} iff 
\[
{1\over\sqrt{n}}\sum_i v_i\in{\cal W}
\ \ , \ \ 
{1\over\sqrt{n}}\sum_i w_i\in{\cal V}\,. 
\]
Two mutually unbiased state bases (MUSBs) are \emph{coherent}
iff the erasing point of each  is contained in the observable of the
other.
\end{definition}

These notions of coherence 
coincide along the bijection of Proposition~\ref{Base_State_iso}:

\begin{proposition}\label{prop:coherence}
 If ${\cal V}$ and ${\cal W}$ are coherent MUVBs
then $vs([{\cal V}]_\sim)$ and $vs([{\cal W}]_\sim)$ are coherent
MUSBs, and if ${\cal S}$ and ${\cal T}$ are coherent MUSBs then there
exist ${\cal V}\in sv({\cal S})$ and ${\cal W}\in sv({\cal T})$ such
that ${\cal V}$ and ${\cal W}$ are coherent MUVBs. 
\end{proposition}
\begin{proof}
The first statement follows directly from the definition of $vs$.  Let
${\cal S}=\{\state{ v_1},\ldots, \state{ v_n},\state{ w_1}\}$ and ${\cal 
  T}=\{\state{ w_1},\ldots, \state{ w_n},\state{ v_1}\}$  be coherent MUSBs;
for ${\cal V}\in sv({\cal S})$ and ${\cal W}\in sv({\cal T})$ we have
$\sum_i {\cal V}=\sum_i \innp{v_i}{w_1}\ket{v_i}=\ket{w_1}$ and
$\sum_i {\cal W}=\ket{v_1}$.  Hence we obtain coherence if
$\innp{w_1}{v_1}\ket{w_1}=\ket{w_1}$ and
$\innp{v_1}{w_1}\ket{v_1}=\ket{v_1}$, that is, $\innp{v_1}{w_1}=1$. 
This can be realised by choosing an appropriate overall phase for
${\cal V}$ relative to ${\cal W}$. 
\end{proof}\vspace{-3.5mm}

Pairs of observables arise from  coherent bases:

\begin{theorem}\label{thm:CoherentBases}
For each pair of MUVBs $\{\ket{v_1},\ldots,\ket{v_n}\}$ and
$\{\ket{w_1},\ldots,\ket{w_n}\}$ there exists a pair of coherent MUVBs
$\{\ket{v_1'},\ldots,\ket{v_n'}\}$ and $\{\ket{w_1'},\ldots,\ket{w_n'}\}$
that induces the same observables i.e.~$\state{ v_i}=\state{
v_i'}$ and $\state{ w_i}=\state{ w_i'}$ for all $i$. 
\end{theorem}
\begin{proof}
Given a pair of MUVBs forget all phases to obtain the corresponding
pair of induced observables. Then adjoin as an erasing point to each
of these a state of the other, in order to obtain coherent MUSBs.
Now we can rely on Proposition~\ref{prop:coherence} to obtain 
coherent MUVBs that induce the same observable as the initial one. 
\end{proof}\vspace{-3.5mm}

\section{Algebras and observables}
\label{sec:algebras-observables}

In the introduction of this paper we already indicated how conceptual
analysis leads to an algebraic characterization of bases and
observables.  Here we review this theory in technical detail.  Its
main purpose is a characterization of bases in the language of
\dsmcs---i.e.~with no reference to vectors, sums, linear combinations
etc.~---allowing bases to be defined in purely diagrammatic terms.
As we shall also see, the diagrammatic presentation of bases admits
very simple calculational rules in terms of so-called `spiders'---
indeed, those that play a central role in the \zxcalculus.

\subsection{Monoids, comonoids, and observable structures}
\label{sec:mono-comon-observ}

Recall that a \emph{monoid} is a triple $(M,\bullet, 1_\bullet)$ with $M$ a set,  $\bullet$ an associative multiplication, and  $1_\bullet\in M$ is its unit.  The multiplication is a map:
\beq
m_\bullet:M\times M\to M::(x,y)\mapsto x\bullet y
\eeq
and we can also represent the unit as a map 
\beq
e_\bullet:\mathbb{I}\to M::\star\mapsto 1_\bullet
\eeq
where $\mathbb{I}:=\{\star\}$ is a singleton set.  The associativity and unit laws of the monoid can now be re-written in terms of composition of maps and cartesian product:
\beqn
m_\bullet\circ (m_\bullet \times 1_M)=m_\bullet\circ (1_M\times m_\bullet)\,,\\
m_\bullet\circ (e_\bullet \times 1_M)\simeq m_\bullet\circ (1_M \times e_\bullet)\simeq 1_M
\eeqn
where $1_M:M\to M$ is the identity map on  $M$.  Now,  $(M,m_\bullet,e_\bullet)$ is \emph{commutative} if 
  \beq 
  m_\bullet\circ\sigma_{M,M}=m_\bullet
  \eeq
where $\sigma_{M,M}: (x, y)\mapsto (y, x)$.

More generally, commutative monoids
can be defined \emph{internally} in any \smc.

\begin{definition}\label{def:internal-monoid}\em
  An \emph{internal commutative monoid}  in an \smc  is a
  triple $(M, m,e)$, consisting of an object $M$, equipped with a 
  \emph{multiplication} $m :M\otimes M\to M$, and a \emph{unit}
  $e:\II\to M$,  satisfying 
  \beqn
    m\circ (m\otimes 1_M)
    = m\circ (1_M\otimes m) \,,
    \\
    m\circ (e\otimes 1_M)\circ\lambda_M
    = m\circ (1_M \otimes e)\circ\rho_M
    = 1_M\,,\\
      m\circ\sigma_{M,M}=m \,.   
  \eeqn 
  \end{definition}

By reversing the types and the order of composition we obtain a new concept.

\begin{definition}\label{def:internal-comonoid}\em
  An \emph{internal cocommutative comonoid} in an \smc is a triple
  $(X,\delta,\epsilon)$ consisting of an object $X$ equipped with  a
  \emph{comultiplication} $\delta:X\to  X\otimes X$ and a
  \emph{counit} $\epsilon:X \to \II$, satisfying 
  \beqn\label{eq:comon1}
    (\delta\otimes 1_X)\circ\delta
    = (1_X\otimes\delta)\circ\delta\,,
    \\ 
    \label{eq:comon2}
    \lambda^{-1}_X\circ(\epsilon\otimes 1_X)\circ\delta
    = \rho^{-1}_X\circ(1_X\otimes\epsilon)\circ\delta=1_X\,,\\
\label{eq:comon3}
  \sigma_{X,X}\circ\delta=\delta\ .  
  \eeqn
\end{definition}

Note that in a $\dsmc$ each internal commutative monoid is also an internal cocommutative comonoid, for $\delta:=m^\dagger$ and $\epsilon:=e^\dagger$, and vice versa. 

Now consider a set $X$ and let $\delta:X\to X\times X$ be the function which \em copies \em entries, i.e.~$\delta::x\mapsto(x,x)$.  
Since $\delta$ is a function it is also a relation, namely
\beq
\delta:=\{(x,(x,x))\mid x\in X\}\subseteq (X\times X)\times X\,,
\eeq
and as a relation it admits a \em relational converse\em, obtained by exchanging the two entries in the pairs which make up that relation. The relational converse to $\delta$, 
\beq\label{eq:relbasis1}
m:=\{((x,x),x)\mid x\in X\}\subseteq (X\times X)\times X\,,
\eeq
relates pairs $(x,x)\in X\times X$ to $x\in X$, while it does not relate pairs $(x,y)\in X\times X$ for $x\not=y$ to anything.  
Let $\epsilon:X\to\mathbb{I}$  be the function which \em erases \em entries i.e.~$\epsilon:x\mapsto \star$.  When conceived as a relation $\epsilon$ admits a relational converse 
\beq\label{eq:relbasis2}
e:=\{(\star,x)\mid x\in X\}\subseteq\mathbb{I}\times X\,, 
\eeq
which now relates $\star\in\mathbb{I}$ to each $x\in X$.
The copying/erasing pair $(\delta,\epsilon)$ is a comonoid in ${\bf Rel}$, and the pair $(m,e)$ consisting of their respective converses is a monoid in ${\bf Rel}$.  The pair $(m,\delta)$  moreover satisfies another remarkable  property: 
\beq
\delta\circ m= (1_X\times m)\circ(\delta\times 1_X)
=\{(x,x),(x,x)  \mid x\in X\}\,.
\eeq

\begin{remark}
This interesting property first appeared in the literature as part of
Carboni and Walters' axiomatisation of the category ${\bf Rel}$ in
\cite{CarboniWalters} where they introduced the notion of a
\emph{Frobenius algebra}  in an \smc  ${\bf C}$,  as a quintuple of 
morphisms
\beq
(X, d:X\otimes X\to X,e:\II\to X,\delta:X\to X\otimes X,\epsilon:X\to\II)
\eeq
where $(X,m,e)$ is an internal commutative monoid and
$(X,\delta,\epsilon)$ is an internal cocommutative comonoid, which
together satisfy the \emph{Frobenius law}--- see 
(\ref{eq:Frobenius}) below.  
\end{remark}

\begin{definition}\label{def:classicalstruc} {\rm\cite{CPav}} \em 
An \em observable structure \em in a \dsmc is a triple
\bc$
\left(\,A\ ,\ \delta= \raisebox{1pt}{\inlinegraphic{1.8em}{delta}} :A\to A\otimes A\ ,\ \epsilon=\ \inlinegraphic{1.55em}{epsilon}:A\to \II\,\right)
$\ec
which:
\ben
\item is a cocommutative comonoid; the defining equations
  (\ref{eq:comon1},\ref{eq:comon2},\ref{eq:comon3}) are depicted as:
\beq\label{eq:monoidlaws}
\hspace{-17mm}\inlinegraphic{3.90em}{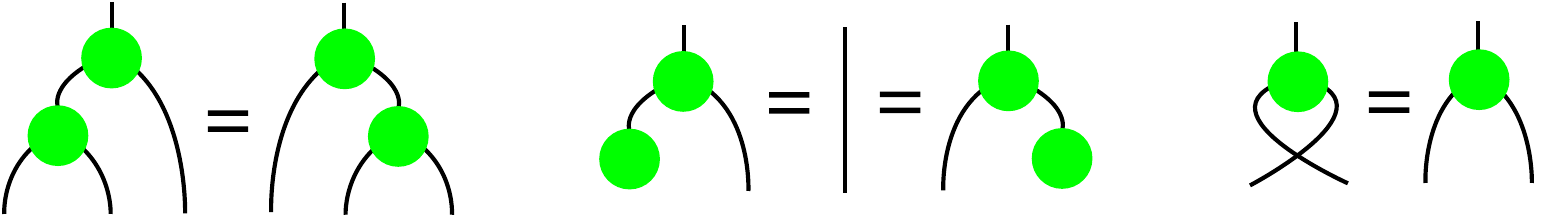}
\eeq
\item satisfies the \em Frobenius law\em, that is,
\beq\label{eq:Frobenius}
\hspace{-7mm} (\delta^\dagger\otimes 1_A)\circ(1_A\otimes\delta)=\delta\circ\delta^\dagger
\quad\mbox{\rm i.e.}\quad\inlinegraphic{3.6em}{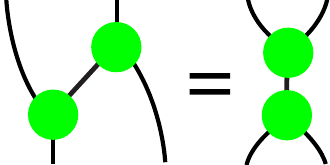}  
\eeq
\item  is \em special\em, that is,
\beq
\hspace{-7mm}\delta^\dagger\circ\delta=1_A \quad\mbox{\rm i.e.}\quad\inlinegraphic{3.6em}{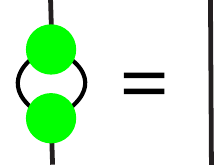} .
\eeq
\een
\end{definition}

\begin{example}
The unit object $\II$ canonically comes with an observable structure: 
\beq
\delta_\II:=\lambda_\II:\II\simeq\II\otimes\II\ , \
\epsilon_\II:=1_\II\,.
\eeq  
\end{example}

\begin{example}\label{example:FromInHilb}
As indicated in the introduction to this paper,  any orthonormal basis $\{\ket{\psi_i}\}_i$ for a Hilbert space   
${\cal H}$ induces an observable structure by considering the linear maps that respectively `copy' and `uniformly erase' the basis vectors:
\beq\label{eq:HilbClasOb}
\delta:{\cal H}\to{\cal H}\otimes{\cal H}::\ket{\psi_i}\mapsto 
\ket{\psi_i}\otimes\ket{\psi_i}\ , \ \epsilon:{\cal
  H}\to\mathbb{C}::\ket{\psi_i}\mapsto 1\,.
\eeq
Moreover, this observable structure is `basis capturing': we can recover the basis vectors from which we constructed $\delta$ as the solutions to the equation
\beq
\delta(\ket{\psi})=\ket{\psi}\otimes\ket{\psi}\,.
\eeq
This also shows that the basis $\{\ket{\psi_i}\}_i$ is faithfully encoded in the linear map $\delta$ alone, and that $\epsilon$ does not carry any additional data.
\end{example}

Conversely, all observable structures in $\fdhilb$ arise from bases:

\begin{theorem}{\rm\cite{CPV}}
All observable structures in {\rm$\fdhilb$} are of the form {\rm(\ref{eq:HilbClasOb})}.  
\end{theorem}

So observable structures provide an \emph{axiomatic
characterisation} of
bases in \dsmc-language, with no reference to the
linear structure of the underlying vector spaces.

\begin{example}\label{WPobservable-structures}
  Each observable structure in $\fdhilb$ induces an observable
  structure in $\fdhilb_{wp}$ in the obvious manner.  But, in the
  light of Proposition \ref{Base_State_iso}, the correspondence in
  $\fdhilb$ between observable structures and vector bases, becomes
  one between observable structures and state bases in $\fdhilb_{wp}$.
  Note that \beq \delta:{\cal H}\to{\cal H}\otimes{\cal
    H}::\state{\psi_i}\mapsto \state{\ket{\psi_i}\otimes\ket{\psi_i}}
  \eeq does \emph{not} define a unique $\delta$ anymore.  In addition
  we need to specify that $\state{\sum_i\ket{\psi_i}}$, the erasing
  point of the state basis, provides the unit for the
  comultiplication: \beq
  \delta^\dagger\left(-\otimes\state{\sum_i\ket{\psi_i}}\right)=1_{\cal
    H}\,.  \eeq
\end{example} 

\begin{corollary}\label{ProjectiveBase}
Observable structures in {\rm$\fdhilb_{wp}$} are in bijective
correspondence with state bases via the correspondence outlined in
Example \ref{WPobservable-structures}. 
\end{corollary}

\begin{example}
The somewhat surprising fact that there are observable structures in
${\bf FRel}$ other than those of the form
(\ref{eq:relbasis1},\ref{eq:relbasis2}) was noted by Edwards and one
of the authors in \cite{Spek}. The observable structures in ${\bf
  FRel}$ have been classified by Pavlovic in \cite{Dusko}, in terms of
groups. 
The notion of observable structure also applies to non-standard
quantum-like theories. For example, it provides a generalised  notion
of basis for Spekkens' toy theory \cite{Spek}, despite the lack of an
underlying vector space structure.
\end{example}


When the monoidal structure is strict, which is always the
case in the graphical language, observable 
structures obey the following remarkable theorem, which follows  from
results in \cite{Kock,Lack} ---a direct derivation is in \cite{CPaq}.
  
\begin{theorem}[normal form]\label{thm:spider} 
Let $\delta_n : A\to A^{\otimes n}$ be defined by the recursion 
\beq
\delta_0 = \epsilon \,, \qquad
\delta_n =  (\delta_{n-1} \otimes \id{A})  \circ \delta  \qquad i.e.~\quad\underbrace{\inlinegraphic{3.15em}{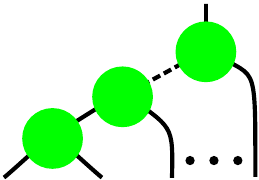}}_n\,.
\eeq
If $f : A^{\otimes n} \to A^{\otimes m}$ is a morphism
generated from the observable structure $(A,\delta,\epsilon)$, the
symmetric monoidal structure maps, and the adjoints of all of these,
and if the graphical representation of $f$ is connected, then we have
\beq
f = \delta_m \circ \delta_n^\dag  \qquad\qquad i.e.~\qquad\quad\underbrace{\overbrace{\inlinegraphic{6.00em}{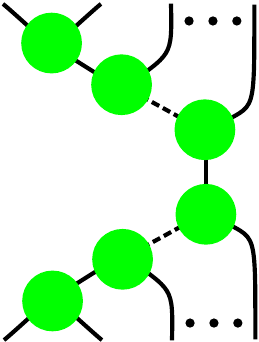}}^n}_m\,.
\eeq
Hence, $f$ only depends on the object $A$ 
and the number of inputs $n$ and outputs $m$.
\end{theorem}

\begin{theorem}[spider rules]\label{thm:spiderrule} 
When representing the unique morphism  with $n$ inputs and $m$ outputs of Theorem~\ref{thm:spider} as an  `$n+m$-legged spider': 
\bc$
\underbrace{\overbrace{\inlinegraphic{3.0em}{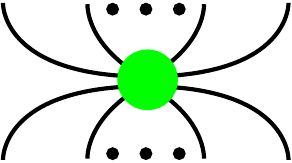}}^n}_m\ , 
$\ec
 then these spiders obey the following composition rule: 
\beq\label{fuse1}
\inlinegraphic{5.8em}{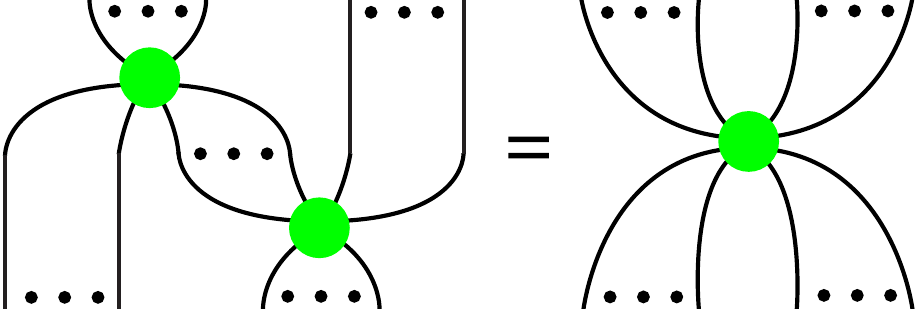}. 
\eeq
i.e.~when two spiders `share legs' then these two spiders `fuse' into a single spider. Also,   
the $1+1$-legged spider must be equal to the identity:
\beq\label{fuse2}
\inlinegraphic{2.9em}{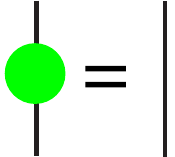}\ ,
\eeq
and thirdly, spiders are invariant under `leg swapping':
\beq\label{fuse3}
\inlinegraphic{4.8em}{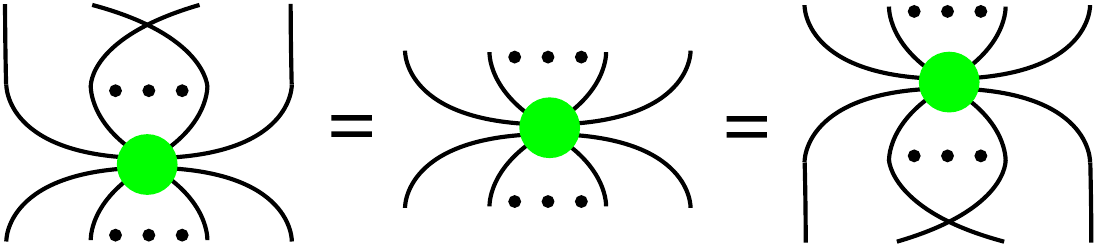}\ .
\eeq 
Conversely, given a family $\{\delta_m^\dag\circ\delta_n|n, m \in\mathbb{N}\}$ of morphisms, 
the equations defining an observable structure can be recovered from  (\ref{fuse1}),  (\ref{fuse2}) and (\ref{fuse3}),
hence providing an alternative characterization of observable structures now purely in terms of spiders.
\end{theorem}  


\begin{example}
We have:
\beqa
&&\hspace{-1.5cm}(1_A\otimes \delta^\dagger_2)\circ(\delta_2\otimes\delta_2)\circ (1_A\otimes\sigma_{A, A}\otimes 1_A)
\circ(\delta^\dagger_2\otimes\delta^\dagger_2)\circ(\delta_2\otimes 1_A)\\
&&\hspace{1.4cm}=\inlinegraphic{7.0em}{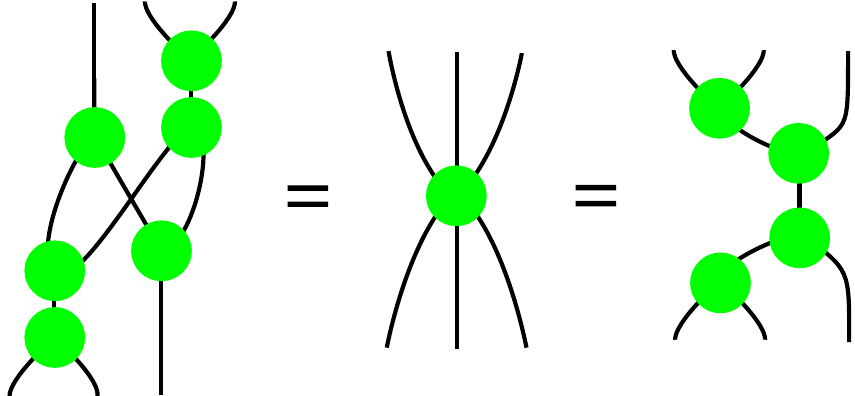}=\ \delta_3 \circ \delta_3^\dag. 
\eeqa
\end{example}

The spider rules enabled us to define the \zxcalculus of section
\ref{Sec:ZXCalc} in terms of spiders obeying the rules (\ref{fuse1})
and (\ref{fuse2}), rather than algebras.  The swapping of remaining
rule (\ref{fuse3}) was eliminated by (implicitly) allowing freedom of
how to spiders may be connected, absorbing the symmetry into the graph
structure (see Example~\ref{ex:zx-commutative-monoid}).  Here we have
been more explicit.  The angles labeling the spiders of the
\zxcalculus will be explained below in
Section~\ref{sec:gener-spid-theor}.

\subsection{Induced $\dag$-compact structure}
\label{sec:induc-comp-struct}


\begin{definition}\em
A \em self-dual $\dag$-compact structure \em is a pair 
\[
(A,\eta:\II\to A\otimes A)
\]
which satisfies equations
(\ref{eq:yanking1}) and (\ref{eq:yanking2}).
\end{definition}

\begin{proposition}
Every observable structure on an object $A$ defines a self-dual $\dag$-compact structure when setting $\eta= \delta^0_2$.
\end{proposition}
\begin{proof}
Equations (\ref{eq:yanking1}) and (\ref{eq:yanking2}) follow from equations (\ref{fuse1}) and (\ref{fuse3}) respectively.
\end{proof}\vspace{-3.5mm}


\begin{remark}\label{compactdontcoincide}
The $\dag$-compact structures induced by different observable structures may
or may not coincide \cite{Coecke-Paquette-Perdrix}.  For example,
while $\dag$-compact structures induced by the observable structures that copy the
$\{\ket{0},\ket{1}\}$ and $\{\ket{+},\ket{-}\}$ coincide, this is not
the case for the bases $\{\ket{0},\ket{1}\}$ and 
$\{\ket{0}+ i \ket{1}, \ket{0}- i \ket{1}\}$.  Returning to the
\zxcalculus, the two equations of the \ruleSii-rule are quite
different to 
each other.  The equations 
\begin{center}
  \inlinegraphic{2.7em}{zx-rule-straight}
\end{center}
are true because of the counit law for the observable structure, and the
fact there is only one identity map.  On the other hand the equations
\begin{center}
  \inlinegraphic{1.5em}{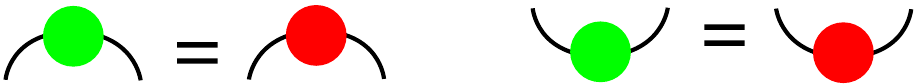}
\end{center}
are not a consequence of the axioms of an observable structure, but a fact specific to the $Z$ and $X$ bases. 
Only when no confusion is possible we simplify the notation for the
$0+2$-legged spider,  as in the \zxcalculus:
\begin{center}
  \inlinegraphic{1.45em}{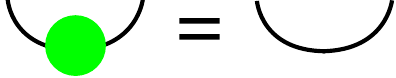}.
\end{center}
The dots will be retained when disambiguation is required.
\end{remark}

\begin{definition}\label{def:transpose-conjugate}\em
  Let $f:A\to B$ be a morphism.
  Its \emph{transpose}, 
  $f^*:B\to A$, and its \emph{conjugate}, $f_*:A\to B$,  relative to observables structures on $A$ and  $B$,  are defined as: 
\[
f^*:=(1_A\otimes \eta_B^\dagger)\circ (1_A\otimes f \otimes 1_B)\circ(\eta_A\otimes
1_B), 
\]
\[
f_*:=(1_B\otimes \eta_A^\dagger)\circ (1_B\otimes f^\dagger\otimes 1_A)\circ(\eta_B\otimes
1_A)\,.
\]
\end{definition}

Diagrammatically, we denote $f^*$ and $f_*$ respectively as: 
\par\vspace{1mm}\noindent
\begin{minipage}[b]{1\linewidth}
\centering{\includegraphics[width=230pt]{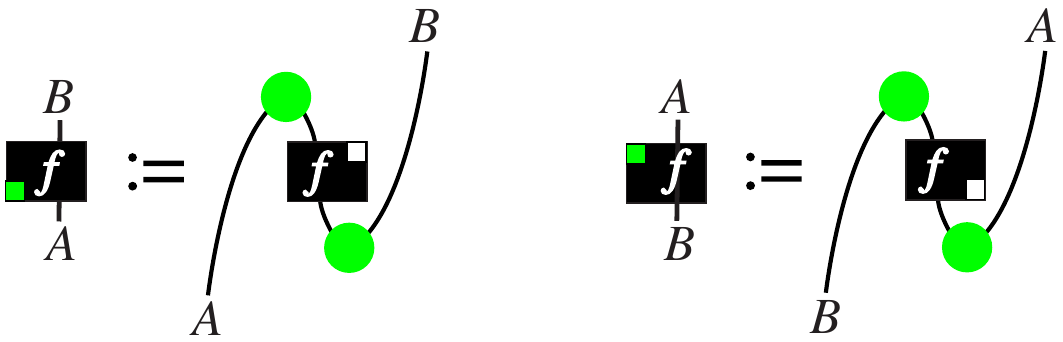}}      
\end{minipage}
\par\vspace{1mm}\noindent
The green squares indicate the dependency of $f^*$ and $f_*$ on the
observable structures on $A$ and $B$; when there is no risk of confusion we omit
the colouration and rely simply upon the position of the
indicated corner to distinguish between $f$, $f^\dag$, $f^*$, and $f_*$. 
We follow \cite{Selinger} in representing the conjugate by horizontal reflection and the transpose by a 180$^\circ$ rotation: 
\par\vspace{1mm}\noindent
\begin{minipage}[b]{1\linewidth}
  \centering
\ninlinegraphic[0.3]{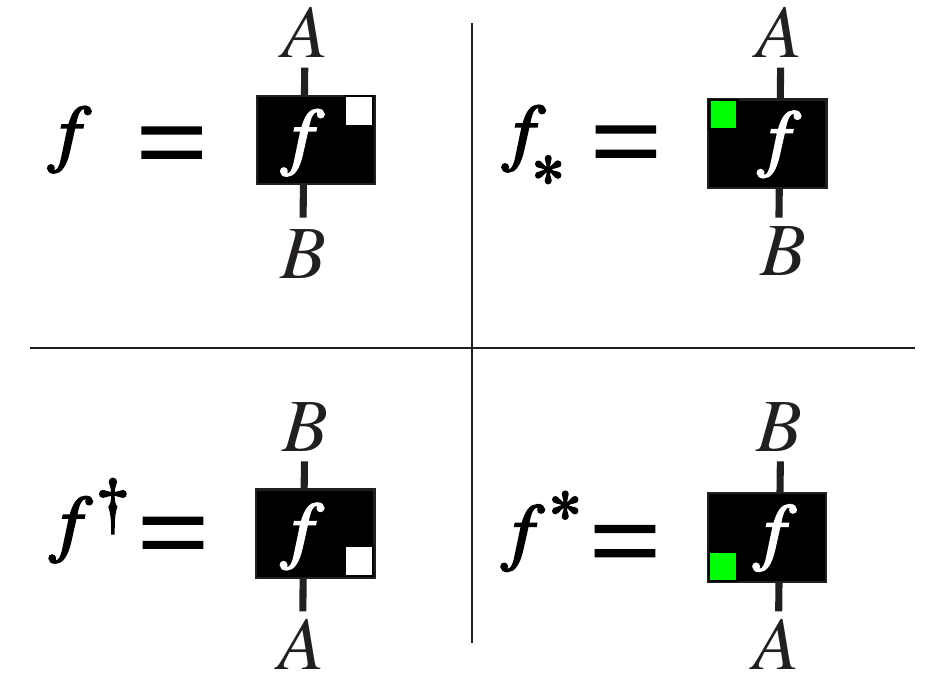}
\end{minipage}
\par\vspace{1mm}\noindent
which is consistent with the fact that 
\[
f_*=(f^\dagger)^*=(f^*)^\dagger\,,\qquad \mbox{or equivalently\,,}\qquad  f^\dagger=(f_*)^*=(f^*)_*\,.
\] 

\begin{corollary}\label{col:self-conjugare}
For  an observable structure all spiders are self-conjugate.
\end{corollary}

\begin{example}
In \fdHilb every observable structure corresponds to a basis, as per
Example \ref{example:FromInHilb}.  The linear function $f_*$ is
obtained by conjugating the entries of the matrix of $f$ when
expressed in the bases  corresponding to the observable structures on $A$ and $B$;  $f^*$ is obtained by
transposing the matrix of $f$.
\end{example}

\begin{definition}\em
Let $A$ be an object in a \dsmc which comes with an observable
structure, and hence an induced $\dag$-compact structure $(A,\eta)$. The \em
dimension \em of $A$ is ${\rm dim}(A):=\eta^\dagger\circ\eta$,
represented graphically by a circle:
\par\vspace{1.5mm}\noindent
\begin{minipage}[b]{1\linewidth}
\centering{\includegraphics[width=60pt]{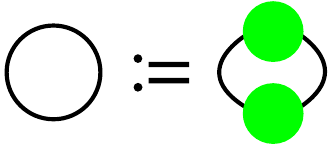}}      
\end{minipage}
\end{definition}

\begin{lemma}\label{ref:dimension}
Dimension is independent of the choice of observable structure.
\end{lemma}
\begin{proof}
We will depict the two distinct observable structures in green and red respectively.
Then, repeatedly relying on $\dag$-compactness we have:
\bc
\inlinegraphic{8.5em}{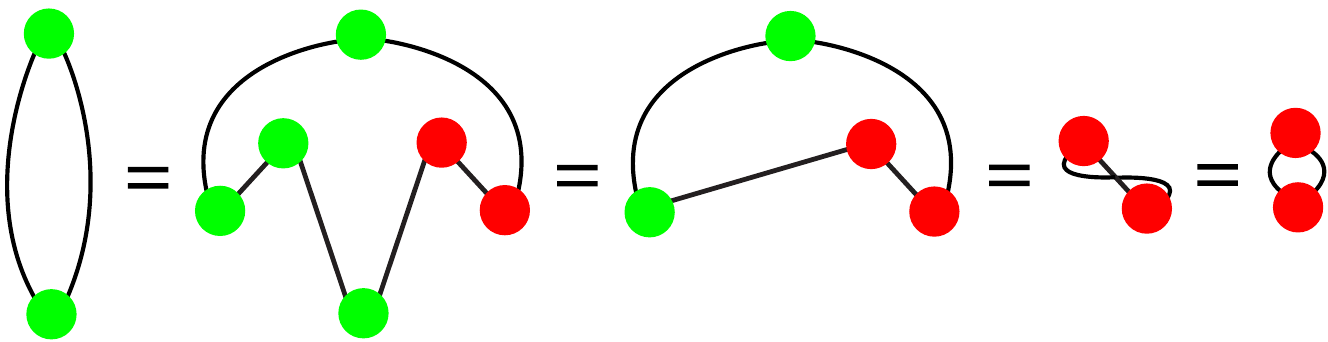}
\ec
so the circles induced by the two observable structures 
coincide. 
\end{proof}\vspace{-3.5mm}

\begin{remark}
  In the language of $\dag$-compact categories, the circle is formed
  by taking the trace of the identity morphism; in a finite
  dimensional Hilbert space this will always give the dimension, hence
  the terminology $\dim A$.
\end{remark}

\subsection{Classical points and generalised bases}
\label{sec:class-points-gener}

We now provide a category-theoretic counterpart to the role played by basis
vectors/states in $\fdhilb$ and ${\bf FdHilb}_{wp}$.

\begin{definition}\label{def:classpoint}\em
Given an observable structure $(A,\delta,\epsilon)$, a morphism
$k:\II\to A$ is a  \em classical point \em iff it is a \em self-conjugate comonoid
homomorphism\em, that is, graphically: \smallskip
\beq\label{eq:classpoint1}
\hspace{-0.5mm}\inlinegraphic{2.0em}{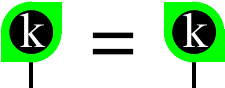}
\eeq
\beq\label{eq:classpoint2}
\hspace{-2.0mm}\inlinegraphic{3.em}{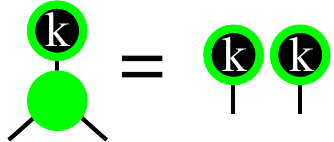}
\eeq
\beq
 \label{eq:classpoint3}
\inlinegraphic{3.0em}{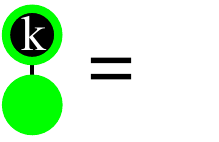} 
\eeq
\end{definition}

\begin{remark}\label{rem:corners}
The notation \inlinegraphic{1.3em}{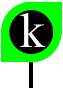}$\!$  reflects `sensitivity to conjugation' while the notation 
\inlinegraphic{1.3em}{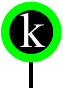}$\!$ reflects `invariance under conjugation'.  We used \inlinegraphic{1.3em}{ClassPointcor}$\!\!$ in (\ref{eq:classpoint1}) to express invariance under conjugation, and given this fact, we used 
\inlinegraphic{1.3em}{ClassPoint}$\!\!$ in (\ref{eq:classpoint2}) and (\ref{eq:classpoint3}).
\end{remark}

\begin{proposition}
Classical points are normalised.
\end{proposition}
\begin{proof}
Since each classical point $k:\II\to A$ is self-conjugate its adjoint
$k^\dagger: A\to \II$ and its transpose $k^*: A\to \II$ coincide.
Hence we have:
\bc
\inlinegraphic{4.0em}{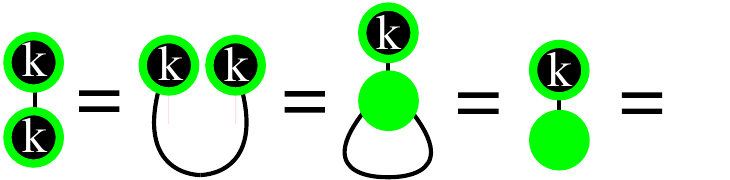}
\ec
that is, $k^\dagger\circ k=1_\II$.
\end{proof}\vspace{-3.5mm}

\begin{example}
In ${\bf FdHilb}$ the classical points for an observable structure are
exactly the basis vectors $\{\ket{v_1}, \ldots, \ket{v_n}\}$ and in ${\bf FdHilb}_{wp}$
they constitute the corresponding observable 
$\{\ket{v_1},  \ldots, \ket{v_n}\}$.  
Hence, while in ${\bf FdHilb}$ the classical points completely determine an
observable structure, this is not the case in  ${\bf FdHilb}_{wp}$,
where it is the classical points together with an erasing point that
determine an observable structure. 
\end{example}

The following is a category-theoretic generalisation of a notion of
basis, either in the sense of Proposition~\ref{prop:Vect_base} or in
the sense of Proposition~\ref{prop:State_base}, which respectively
applies to the categories ${\bf FdHilb}$ and ${\bf FdHilb}_{wp}$.   

\begin{definition}\label{def:pointbasis}\em
An observable  structure $(A,\delta,\epsilon)$ with classical points ${\cal K}$ is called a \em vector basis \em iff for all objects $B$ and all morphisms $f,g:A\to B$ we have 
\beqq
\big[
\forall k\in{\cal K} : f\circ k= g\circ k\ 
\big]
\Rightarrow\ f=g\,.
\eeq
It is called a \em state basis \em iff for all $B$ and all $f,g:A\to B$ we have
\beqq
\big[ 
\forall k\in{\cal K} \cup\{\epsilon^\dagger \} : f\circ k= g\circ k\
\big]
\Rightarrow\ f=g\,.
\eeq
\end{definition}

One can easily construct new observable structures by combining old
ones, as the in the following proposition.

\begin{proposition}\label{prop:TensorObsStruc}
Two  observable structures $(A,\delta_A,\epsilon_A)$ and $(B,\delta_B,\epsilon_B)$ canonically induce an observable structure on $A\otimes B$ with
\beqq
\delta_{A\otimes B}= (1_A\otimes\sigma_{A,B}\otimes 1_B)\circ (\delta_A\otimes \delta_B) \quad
\mbox{i.e.} \quad \inlinegraphic{2.5em}{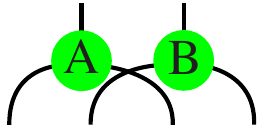} ,
\eeq
and
\beqq
\epsilon_{A\otimes B}=\lambda_\II^\dagger\circ(\epsilon_A\otimes\epsilon_B)\qquad
\mbox{i.e.} \qquad \inlinegraphic{1.8em}{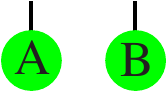}\,.
\eeq
Moreover,  if $k:\II\to A$ is a classical point for $(A,\delta_A,\epsilon_A)$ and $k':\II\to B$ is a classical point for $(B,\delta_B,\epsilon_B)$ then $(k\otimes k')\circ\lambda_\II$ is a classical point for 
$(A\otimes B,\delta_{A\otimes B} , \epsilon_{A\otimes B})$.
\end{proposition}

\begin{definition}\label{def:monoidalliftofbasis}\em
We say that  the monoidal tensor  \em lifts vector bases  \em iff for all vector bases $(A,\delta_A,\epsilon_A)$ with classical points ${\cal K}$ and 
 $(B,\delta_B,\epsilon_B)$ with classical points ${\cal K}'$, all objects $C$, and all morphisms $f,g:A\otimes B\to C$, we have that
\beqq
\big[
\forall (k,k')\in{\cal K}\times{\cal K}':f\circ (k\otimes k')= g\circ
(k\otimes k')
\big]
\ \Rightarrow\ f=g\,.
\eeq
-- hence it follows that the observable structure $(A\otimes
B,\delta_{A\otimes B} , \epsilon_{A\otimes B})$ is also vector
basis. Similarly,  the monoidal tensor   \em lifts state bases \em iff 
\beqq
\big[
\forall (k,k')\in({\cal K}\times{\cal
  K}')\cup\{(\epsilon_A^\dagger,\epsilon_B^\dagger)\}:
f\circ (k\otimes k')= g\circ (k\otimes k')\ 
\big]
\Rightarrow\ f=g\,.
\eeq
-- hence it follows that $(A\otimes B,\delta_{A\otimes B} , \epsilon_{A\otimes B})$ is also state basis. 
\end{definition}

Since observable structures induce $\dag$-compact structures we have the following.

\begin{proposition}\label{prop:bases_lift}
Monoidal tensors always lift vector bases and state bases.
\end{proposition}
\begin{proof}
We have:\vspace{-3mm}
\beqa
&&\hspace{-2cm}\inlinegraphic{5.6em}{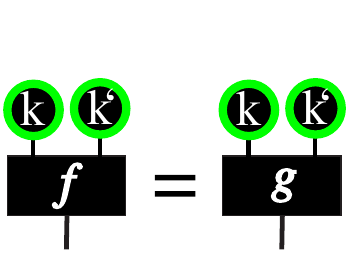}\raisebox{-4.5mm}{\ $\Leftrightarrow$\ }
\inlinegraphic{5.6em}{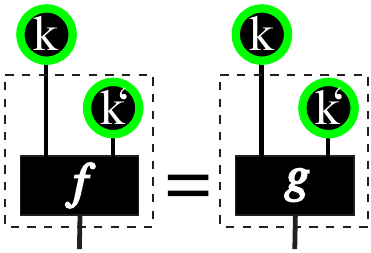}\raisebox{-4.5mm}{\ $\stackrel{*}{\Leftrightarrow}$\ }
\inlinegraphic{5.6em}{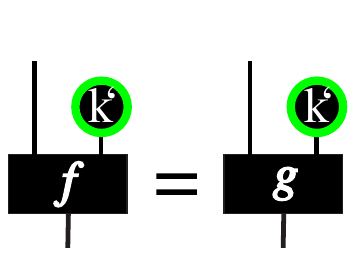}\\
&\hspace{-2cm}\raisebox{-3.2mm}{\ $\Leftrightarrow$\ }\hspace{2cm}&\hspace{-2cm}
\inlinegraphic{4.7em}{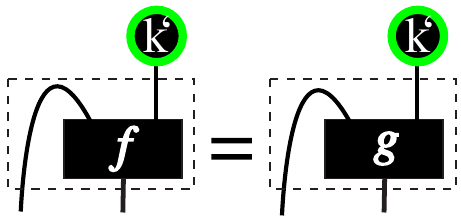}\raisebox{-3.2mm}{\ $\stackrel{*}{\Leftrightarrow}$\ }
\inlinegraphic{4.7em}{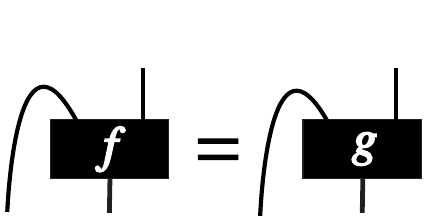}\raisebox{-3.2mm}{\ $\Leftrightarrow$\ }
\inlinegraphic{4.7em}{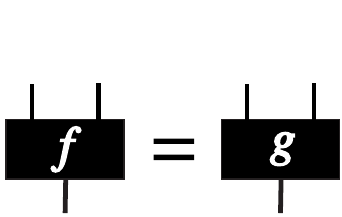}
\eeqa
 where the marked  equivalences rely on the vector/state basis assumption.
\end{proof}\vspace{-3.5mm}

\section{Phase shifts  and a generalised spider theorem}
\label{sec:gener-spid-theor}

In the preceding section we saw that observable structures correspond
to bases of our state space;  now we introduce an abstract notion of
\emph{phase} relative to a given basis, and generalise
Theorems~\ref{thm:spider} and \ref{thm:spiderrule} to incorporate  such phase shifts.

\subsection{A monoid structure on  points}

\begin{definition}\em 
Let $(A,\delta,\epsilon)$ be an observable structure in a \dsmc ${\bf C}$ and let 
${\bf C}(\II,A)$ be the underlying set of points.  We define a \em multiplication on points\em
\beqq
-\odot-:{\bf C}(\II,A)\times {\bf C}(\II,A)\to {\bf C}(\II,A)
\eeq
by setting, for all points $\psi,\phi \in {\bf C}(\II,A)$,
\beqq
\psi \odot \phi := \delta^\dag \circ (\psi\otimes\phi)\circ\lambda_\II
\qquad\mbox{\rm i.e.}\qquad 
\inlinegraphic{2.9em}{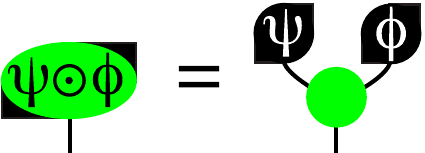}. 
\eeq
\end{definition}

\begin{remark}
As already explained in Remark \ref{rem:corners}, the shape of the points reflects that
they may not be invariant under conjugation.
\end{remark}

\begin{proposition}
$({\bf C}(\II,A),\odot,\epsilon^\dagger, (-)_*)$ is an involutive commutative monoid.
\end{proposition}
\begin{proof}
Associativity, commutativity, and that $\epsilon^\dag$ is the monoid's unit, i.e.:
\bc
\inlinegraphic{4.4em}{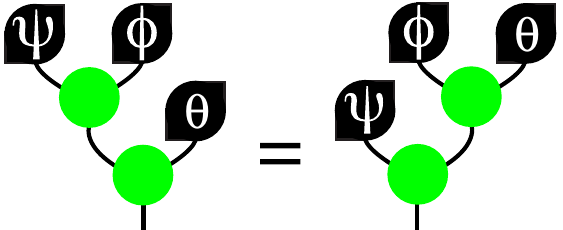} \quad\inlinegraphic{2.9em}{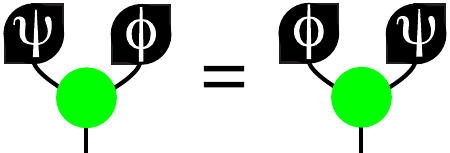}\quad
\inlinegraphic{3.2em}{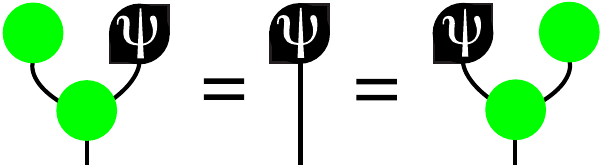} 
\ec
follow immediately from internal monoid laws for $(A,\delta^\dag,\epsilon^\dag)$--- see (\ref{eq:monoidlaws}) above --- and conjugutation is an involution since $\delta^\dag$ is self-conjugate--- see Corollary \ref{col:self-conjugare}.
\end{proof}\vspace{-3.5mm}

\begin{example}
  Let $\delta_Z:\mathbb{C}^2 \to \mathbb{C}^2 \otimes \mathbb{C}^2$ be
  defined by $\delta : \ket{i} \mapsto \ket{ii}$.  The induced
  multiplication $\odot_Z$ is just point-wise multiplication in the
  standard basis:
\[
\left( \begin{array}{c}    a\\b  \end{array} \right) 
\odot_Z
\left(  \begin{array}{c}  a'\\b'  \end{array} \right) 
= 
\delta_Z^\dag\left(  
\left(\begin{array}{c}    a\\b  \end{array} \right) 
\otimes
\left(\begin{array}{c}    a'\\b'  \end{array} \right)
\right) 
= \left(  \begin{array}{c}  aa'\\bb'
  \end{array} \right)   \,.
\]
Indeed, the same will be true for any observable structure, provided
we write the vectors in the corresponding basis.
\end{example}

As well as giving an involutive commutative monoid on the  points of $A$, we can
use $\delta^\dag$ to lift this monoid up to the endomorphisms of $A$.

\begin{definition}\em\label{defn:Lambda}
For $(A,\delta,\epsilon)$ an observable structure in a \dsmc ${\bf C}$ let
\beqq
\Lambda:\catC(I,A) \to \catC(A,A) 
\eeq
be defined by setting, for each point $\psi \in {\bf C}(\II,A)$,
\beqq
  \Lambda (\psi) =  \delta^\dag \circ (\psi\otimes\id{A})
\qquad\mbox{\rm i.e.}\qquad 
\inlinegraphic{3.1em}{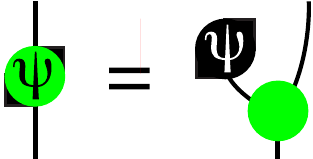}\,,
\eeq
and we denote the range of $\Lambda$ by $\Lambda(A,A)$.
\end{definition}

\begin{proposition}\label{lem:lambda-basic-properties}
The map $\Lambda$ is an isomorphism of involutive commutative monoids:
\beqq
\left({\bf C}(\II,A),\odot,\epsilon^\dagger, (-)_*\right)\simeq\left(\Lambda(A,A), \circ, 1_A, (-)^\dagger\right)\,.
\eeq
that is, explicitly:
\beqq
\Lambda(\psi\odot\phi) =     \Lambda(\psi) \circ \Lambda(\phi)
\qquad 
\Lambda(\epsilon^\dag) = \id{A}
\qquad
\Lambda(\psi_*)=\Lambda(\psi)^\dag,
\eeq
and hence commutativity is inherited:
\beqq
\Lambda(\psi) \circ \Lambda(\phi)=\Lambda(\psi\odot\phi)=\Lambda(\phi) \circ \Lambda(\psi)
\quad i.e.\quad
\inlinegraphic{3.4em}{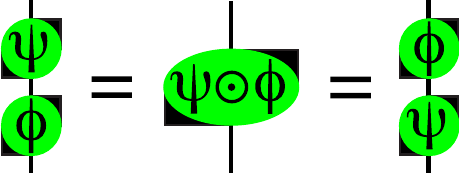}\,.
\eeq
\end{proposition}
\begin{proof}
Preservation of monoid multiplication and unit  follow directly from the unit and commutativity law of the internal monoid. By the spider rules 
we have:
\bc
\inlinegraphic{3.8em}{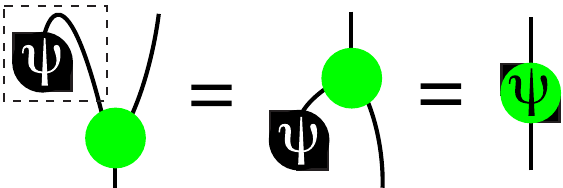}
\ec
that is, conjugation of points is mapped to the adjoint of endomorphisms.
\end{proof}\vspace{-3.5mm}

Note that the notation of endomorphisms in $\Lambda(A,A)$ is invariant 
under 180${}^\circ$ rotations.   This is justified by the following
proposition.

\begin{proposition}
Each $\Lambda(\psi)\in\Lambda(A,A)$ is equal to its transpose.
\end{proposition}
\begin{proof}
We have:
\bc
\inlinegraphic{3.8em}{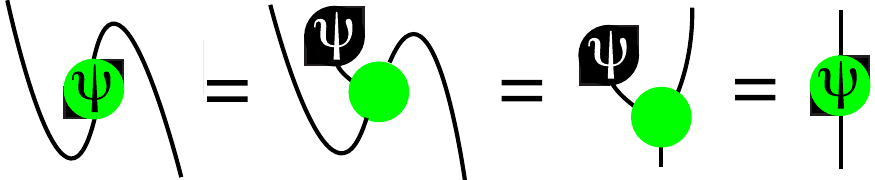}
\ec
were we relied on the spider rules.
\end{proof}\vspace{-3.5mm}

\begin{proposition}\label{phases-flow around}
The endomorphisms in $\Lambda(A,A)$ obey
\beq
\Lambda(\psi) \circ \delta^\dag = 
  \delta^\dag \circ (\id{A} \otimes \Lambda(\psi)) =
  \delta^\dag \circ (\Lambda(\psi) \otimes  \id{A})
\eeq
 i.e.\,
\beq
      \inlinegraphic{3.2em}{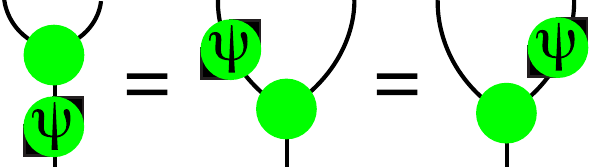}\,.
\eeq
\end{proposition}
\begin{proof}
By the spider rules  all three diagrams normalise to:
  \bc
    \inlinegraphic{2.6em}{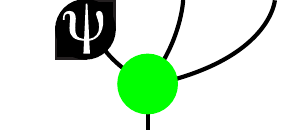}\vspace{-1.5mm}
  \ec
so they are indeed all equal.
\end{proof}\vspace{-3.5mm}

The following proposition shows that the inner-product structure on
points is subsumed by the commutative involutive monoid structure on
points.

\begin{proposition}\label{col:inner-prod}
For points $\psi,\phi:I\to A$ in $\catC$ we have: 
  \beq
  \langle \phi|\psi\rangle:= \phi^\dag \circ  \psi = \epsilon \circ (\phi_* \odot \psi)\,.
  \eeq
\end{proposition}
\begin{proof}
Using the definition of the transpose for $\phi^\dag= (\phi_*)^*$ we obtain
\bc
\inlinegraphic{4.0em}{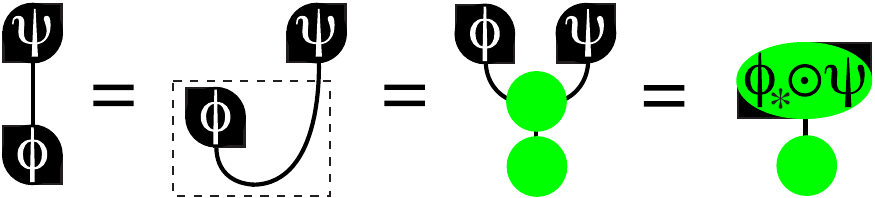}
\ec
where again we used the spider rules.
\end{proof}\vspace{-3.5mm}

\subsection{The decorated spider theorem}

\begin{theorem}[normal form with points]\label{thm:gen-spider}
If $f : A^{\otimes n} \to A^{\otimes m}$ is a morphism  generated
from the observable structure $(A,\delta,\epsilon)$, the symmetric
monoidal structure maps, and the adjoints of all of these, \underline{points $\psi_i : I\to A$} (exactly one occurrence for each $i$), 
and if the graphical representation of $f$ is connected then we have
\bc$
  f =   
  \delta_m \circ 
  \Lambda \Bigl( \bigodot_i \psi_i \Bigr)   \circ           
  \delta^\dag_n  \qquad\quad i.e.~\quad\underbrace{\overbrace{\inlinegraphic{8.2em}{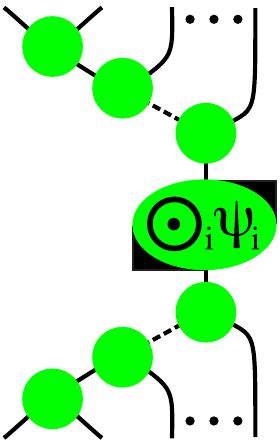}}^n}_m\, .
$\ec
\end{theorem}

\begin{proof}
  If neither $\delta$ nor $\delta^\dag$ occurs in $f$, then result is
  trivial.  Otherwise, all points $\psi_i$ occurring in $f$ may be
  lifted to $\Lambda(\psi_i)$; by virtue 
  of Proposition~\ref{phases-flow around},  these commute
  freely with the observable structure, hence can all be collected
  together.  The result now follows by applying Theorem~\ref{thm:spider}.
\end{proof}\vspace{-3.5mm}

Theorem~\ref{thm:gen-spider} is a strict generalisation of
Theorem~\ref{thm:spider}: diagrams with equal numbers of inputs $n$ and
outputs $m$ are equal whenever the product of points occurring in them
is also equal. The theorem gives a specific normal form to which this entire class of
diagrams is equal; we also have corresponding \emph{decorated spider} rules.

\begin{theorem}[decorated spider rules]\label{thm:gen-spiderrule}
When setting 
\bc
\includegraphics[height=8.2em]{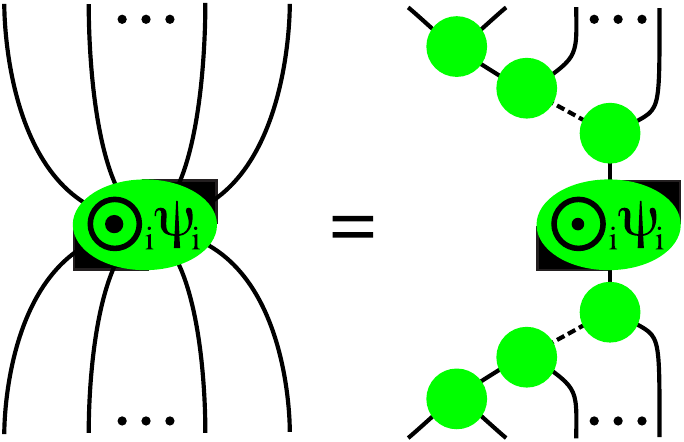}
\ec
then these decorated spiders obey the following composition rule:
\bc
\inlinegraphic{5.8em}{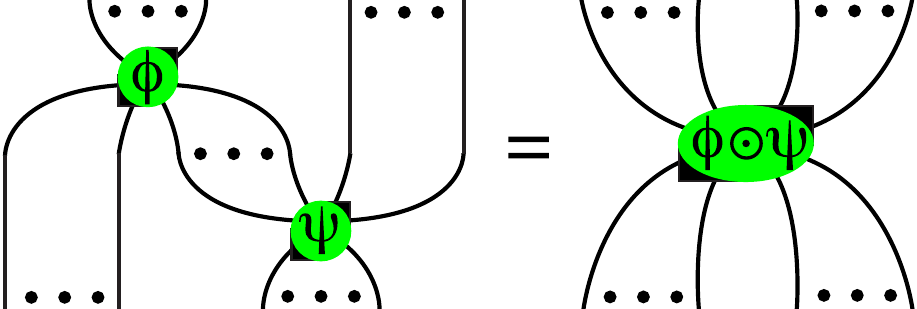}. 
\ec
i.e.~when two decorated spiders `share legs' then these two spiders `fuse' together into a single decorated spider provided we  `multiply decorations'.
\end{theorem}


This is of course the form of the spider rule  in the \zxcalculus.

\subsection{Unbiased points}

\begin{example} 
Let $(\delta, \epsilon)$ be the observable structure corresponding to
the standard basis of $\mathbb{C}^{n}$, and consider 
$|\psi\rangle=\sum_i c_i|\,i\,\rangle$. 
When written in the basis fixed by $(\delta, \epsilon )$,
$\Lambda(\psi)$ consists of
the diagonal $n\times n$ matrix with $c_1,\ldots, c_n$ on the diagonal. 
Hence, $\Lambda(\psi)$ is unitary 
if and only if  $\bar{c_1}c_1 = \ldots = \bar{c_n}c_n=1$,
that is, if and only if $\ket{\psi}$ is unbiased for $\{\ket{ 1},\ldots, \ket{
n}\}$ (upto a normalising constant).
\end{example}

This situation admits generalisation to arbitrary \dsmcs.

\begin{definition}\em\label{defn:unbiased-abstract}
We call a point $\alpha : \II \to A$ \emph{unbiased relative to an
observable structure} $(A,\delta,\epsilon)$ if there exists a scalar $s:\II\to \II$ such that:
\beqq
s \cdot \alpha \odot \alpha_* = \epsilon^\dag
\qquad \text{i.e.}\qquad
\inlinegraphic{2.8em}{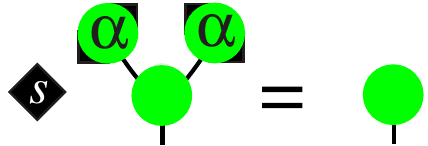}\,.
\eeq
\end{definition}

\begin{example}
Since by the spider rules the point $\epsilon^\dag$ satisfies this definition, every observable structure has at least one unbiased point, namely its unit.
\end{example}

\begin{lemma}\label{lm:unbiasedscalar}
  If an unbiased point $\alpha$ is normalised, i.e.~$\alpha^\dagger\circ\alpha=1_I$, then the scalar $s$ in
  the above definition is equal to $D=\dim{A}$.  Hence, if on the other hand $|\alpha|^2:=\alpha^\dagger\circ\alpha=D$ then this scalar is $1_\II$.
\end{lemma}
\begin{proof}
We have:
  \bc
  \inlinegraphic{3.8em}{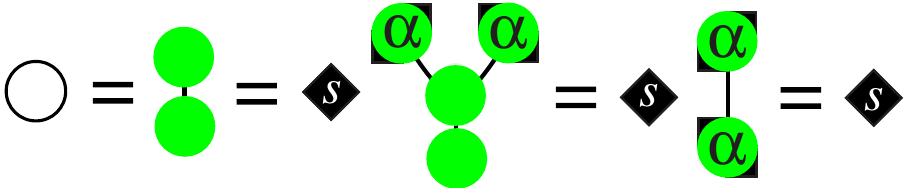}
  \ec
  where we relied on Proposition~\ref{col:inner-prod}.
\end{proof}\vspace{-3.5mm}

The expression $\alpha \odot \alpha_*$ denotes `convolution' of $\alpha$
with itself;  since the point $\epsilon^\dag$ represents the uniform
distribution over the basis defining $\delta$, this
definition  indeed captures the usual understanding of what it means for
a vector to be unbiased with respect to  a basis.  The following shows
that  this is exactly correct.

\begin{lemma}\label{lem:scalar-unbiased}
  Let $\alpha, k : \II \to A$ be points of $A$ such that $\alpha$ is
  unbiased and normalised, and $k$ is classical, for
  $(A,\delta,\epsilon)$;  then 
  \beqq\label{ref:unbwithscal}
D\cdot |\langle k|\alpha\rangle|^2:= D \cdot (k^\dag\circ \alpha)\cdot (\alpha^\dag\circ k)   =
 1_\II
    \quad\text{i.e.}\quad 
  \inlinegraphic{2.95em}{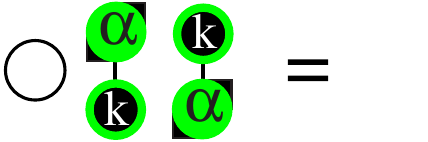}.
  \eeq
\end{lemma}
\begin{proof}
We have:
  \bc
  \inlinegraphic{4.10em}{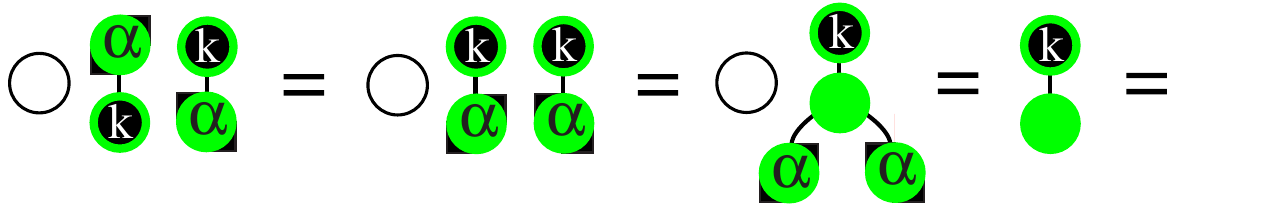}
  \ec
  where we relied on the fact that scalars are always self-transpose, used (\ref{eq:classpoint2}),  used the adjoint to the unbiasedness law, and finally used (\ref{eq:classpoint3}).
\end{proof}\vspace{-3.5mm}

\begin{remark}
To retrieve the usual
definition of unbiasedness, 
\beqq
\sizeof{\innp{z}{\alpha }} =\frac{1}{\sqrt{\dim{A}}}\,, 
\eeq
we simply divide; however,  since we operate in an arbitrary \dsmc, the scalars form
a commutative monoid rather than a group, so dividing is not always possible. Hence 
the form (\ref{ref:unbwithscal}) for the unbiasedness law.
\end{remark}

\subsection{The phase group}\label{sec:phase-group}

\begin{proposition}\label{prop:unbiased-is-unitary}
  A point $\alpha$ of length $|\alpha|^2=D$ is unbiased iff $\Lambda(\alpha)$ is
  unitary.
\end{proposition}
\begin{proof}
Due the commutativity property of $\Lambda$ in Proposition~\ref{lem:lambda-basic-properties}
we need check only one equation
to show that $\Lambda(\alpha)$ is  unitary, namely,
\beqq
  \Lambda(\alpha)\circ \Lambda(\alpha)^\dagger=1_A
  \qquad\text{i.e.}\qquad 
  \inlinegraphic{3.65em}{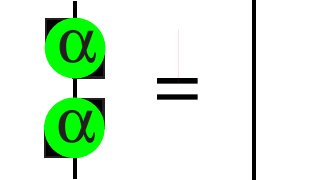}.
\eeq
Suppose that $\alpha$ is unbiased; then by
the spider rules we have
\bc
\inlinegraphic{4.2em}{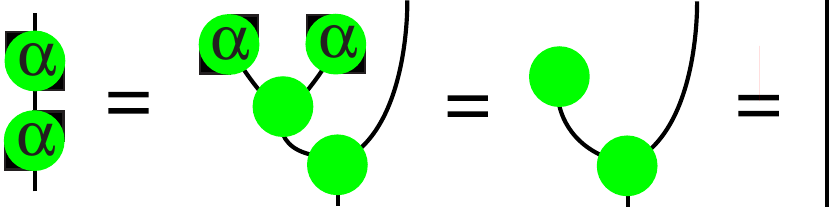},
\ec
as required. The other direction of the  proof is essentially the same.
\end{proof}\vspace{-3.5mm}

Since unitary maps are invertible, they form a group, and this group
structure transfers back onto the unbiased points.

\begin{theorem}\label{prop:inverse}
If in the isomorphism of involutive commutative monoids of
Proposition~\ref{lem:lambda-basic-properties} we restrict ourselves to
unbiased points relative to the observable structure of length
$|\alpha|^2=D$, and unitary endomorphisms, then we obtain an
isomorphism of abelian groups, with the involution as the inverse.
\end{theorem}
\begin{proof}
This immediately follows from Proposition~\ref{prop:unbiased-is-unitary} and the fact that for a unitary morphism the adjoint is the inverse.
\end{proof}\vspace{-3.5mm}

\begin{definition}\em
The abelian group of endomorphisms of Theorem~\ref{prop:inverse} is
called the \emph{phase group}, and its elements are called \emph{phase
shifts}.
\end{definition}

\begin{remark}
We chose $|\alpha|^2=D$ since this results in the inverse
taking a simple form; fixing another length would also have
given us an abelian group structure.
\end{remark}

\begin{example}\label{ex:ZXphasegroup}
Consider the observable structure $(\mathbb{C}^2,\delta_Z,\epsilon_Z)$ in ${\bf FdHilb}$,
defined by
\[
\delta_Z : \ket{i} \mapsto \ket{ii} ,
\qquad\qquad 
\epsilon_Z : \ket{0} + \ket{1} \mapsto 1\,.
\]
Its classical points are $\{\ket{0},\ket{1}\}$.  The unbiased
points for $(\mathbb{C}^2,\delta_Z,\epsilon_Z)$ are of the form 
$\ket{\alpha_Z} = \ket{0} + e^{i \alpha }\ket{1}$, and $\ket{\alpha_Z}\odot_Z\ket{\beta_Z} = \ket{(\alpha + \beta)_Z}$, hence
the group of unbiased points is isomorphic to the interval
$[0,2\pi)$ under addition modulo $2\pi$.  We have
\[
\Lambda^Z(\alpha) = \left(
  \begin{array}{cc}
    1&0\\0&e^{i \alpha}
  \end{array}\right),
\] 
and in particular, $\Lambda^Z(\pi) ={\sf Z}$.  In other words, the phase shifts of the observable structure
 are obtained by rotating along the equator of
  the Bloch sphere: 
  \[
  \inlinegraphic{8em}{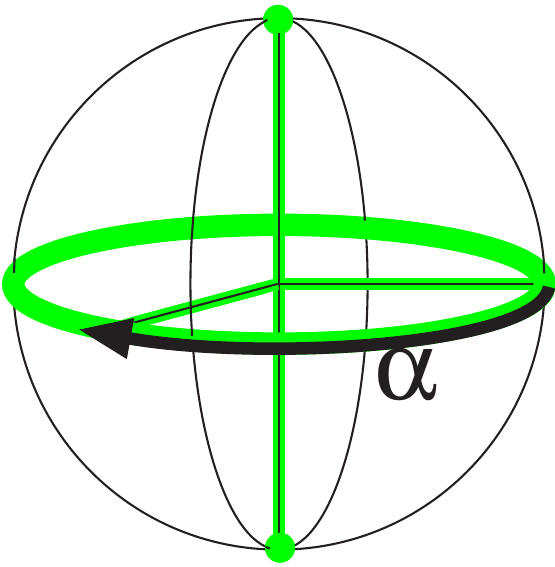}.
  \]
In particular, we have
  $\ket{\theta_1}\odot\ket{\theta_2}=\ket{\theta_1\!+\theta_2}$,
  that is, the operation $\odot$ is simply addition
  modulo $2\pi$, which is an abelian group with minus as
  inverse.  This group of phases played a key role in proving universality of the \zxcalculus in Section~\ref{sec:universality}.
\end{example}

\begin{example}\label{ex:stabspek}
Phase groups can provide an algebraic witness for physical differences
between theories.   For example, as shown in \cite{CES}, the toy model
category ${\bf Spek}$ (cf. Example \ref{ex:Spek}) and the category
${\bf Stab}$ (a restriction of ${\bf FdHilb}$ to the  qubit stabilizer
states) are essentially the same except for the phase groups of their
respective qubits.  In the case of ${\bf Spek}$ the phase group
is the Klein four group $\mathbb{Z}_2\times \mathbb{Z}_2$, while for
${\bf Stab}$ the 
phase group is the cyclic four group $\mathbb{Z}_4$.  This difference in phase
groups is closely connected to the fact that while states in  ${\bf
  Spek}$ always admit a local hidden variable representation, in ${\bf
  Stab}$ there are states which don't, namely the GHZ state
\cite{MerminGHZ}.
\end{example}

\begin{example}
Using decorated spider notation we can set
\[
\inlinegraphic{2.0em}{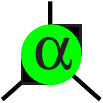}:=\bigl(\Lambda(\alpha)\otimes\Lambda(\alpha)\bigr)\circ\delta\circ\Lambda(\alpha)^\dagger
\qquad\mbox{\rm and} \qquad
\inlinegraphic{1.7em}{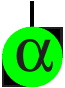}:=\epsilon\circ\Lambda(\alpha)^\dagger,
\]
and for $\alpha$ unbiased relative to $(A,\delta, \epsilon)$, again by the decorated spider rules, it follows that these
morphisms define an observable structure.  Hence, each  element of 
the phase group transforms the given observable structure into  a new observable structure.  
\end{example}

\section{Complementarity is equivalent to the Hopf law}
\label{SEC:Complementarity}

For  observable  structures $(A, \delta_Z, \epsilon_Z)$ and $(A,\delta_X,\epsilon_X)$ in a \dsmc we will denote the corresponding induced $\dag$-compact structures respectively as $(A, \eta_Z)$ and $(A,\eta_X)$.
First we study complementarity for pairs of observable structures on the same object with coinciding induced $\dag$-compact structures, and then we study the slightly more involved case of non-coinciding  induced $\dag$-compact structures--- cf.~Remark \ref{compactdontcoincide}. 

\subsection{Observable structures with coinciding $\dag$-compact structures} 

First we define complementarity for observable structures in arbitrary \dsmcs in a manner which makes explicit reference to their classical points, simply in analogy with the usual definition in Hilbert space quantum theory, and then we show that this definition can be equivalently restated without any reference to points.

\begin{definition}\em\label{defn:complementary-observables}
Two observable  structures  $(A, \delta_Z, \epsilon_Z)$ and $(A,\delta_X,\epsilon_X)$
in a \dsmc  are called  \emph{complementary} if they obey the following rules:
  \begin{itemize}
  \item[{\bf comp1}] whenever $z:I\to A$ is classical for $(\delta_Z,\epsilon_Z)$ it is
    unbiased for $(\delta_X,\epsilon_X)$\,;
  \item[{\bf comp2}] whenever $x:I\to A$ is classical for $(\delta_X,\epsilon_X)$ it is
    unbiased for $(\delta_Z,\epsilon_Z)$\,.
  \end{itemize}
\end{definition}
We abbreviate \emph{complementary observable structures} as \coss.  

\par\medskip\noindent
{\it Notation.} 
Graphically we distinguish two distinct observable structures in terms
of their colour.  To emphasise that a classical point $k:\II\to A$ is copied by an
observable structure of one colour, say green, while unbiased with
respect to an observable structure of another colour, say red, we
denote them by:
\[
  \inlinegraphic{2.0em}{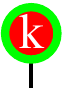}
\]
that is, the outside colour indicates which observable structure
copies this point,  while the inner colour shows to which
observable structure the point is unbiased.

The fact that we denote these points in a manner which is invariant
under conjugation is a consequence of the trivial observation that classical points of one colour are not only self-conjugate for `their own colour' (cf.~Corollary \ref{col:self-conjugare}), but also self-conjugate for `another colour' provided induced $\dag$-compact structures coincide: 

\begin{proposition}\label{prop:concideCCthenSelfCong}
If $(A, \delta_Z, \epsilon_Z)$ and $(A,\delta_X,\epsilon_X)$ are observable structures with 
\beqn
&&\ \ \, \eta_Z=\eta_X\qquad\mbox{\rm
  i.e.}\qquad\inlinegraphic{2.7em}{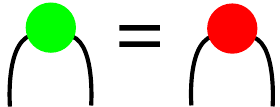}  ,
\eeqn
then for classical points $k:I\to A$ of  $(\delta_Z,\epsilon_Z)$ and $k':I\to A$ of  $(\delta_X,\epsilon_X)$ we have 
\beqn
&&k=k_{*_X}=(k^\dagger\otimes 1_A)\circ\eta_X
\qquad\,\mbox{\rm  i.e.}\qquad
\inlinegraphic{1.8em}{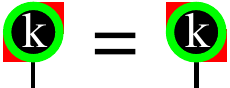},\\
\quad
&&k'=k'_{*_Z}=(k'^\dagger\otimes 1_A)\circ\eta_Z 
\qquad\mbox{\rm  i.e.}\qquad
\inlinegraphic{1.8em}{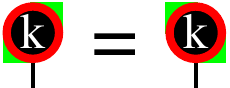}.
\eeqn
\end{proposition}


In the  updated notation for classical points of \coss, the comonoid homomorphism laws governing classical points become:
\[
\inlinegraphic{3.em}{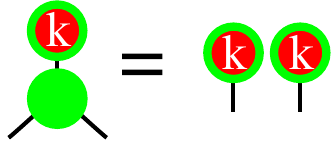} 
\quad\ \ 
\inlinegraphic{3.0em}{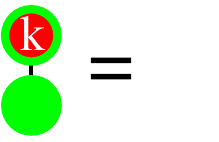} 
\qquad\ \ 
\inlinegraphic{3.em}{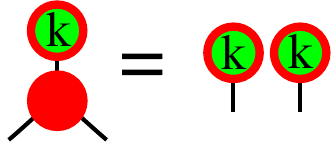} 
\quad\ \ 
\inlinegraphic{3.0em}{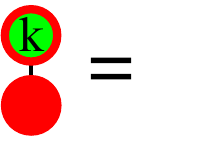} 
\]
and the mutual unbiasedness conditions become:
\[
\inlinegraphic{3.em}{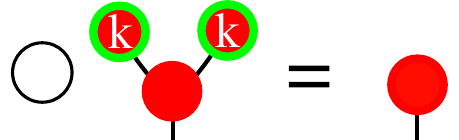} 
\qquad\qquad\qquad\, \ \ 
\inlinegraphic{3.em}{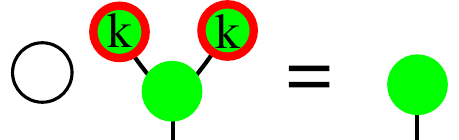}.
\]

\begin{theorem}[complementarity $\Rightarrow$]\label{propHopf_1}
Let $(A, \delta_Z, \epsilon_Z)$ and $(A,\delta_X,\epsilon_X)$ be two observable structures whose induced $\dag$-compact structures coincide.   If they obey 
\beq\label{eq:hopf}
D\cdot\delta_X\circ \delta_Z^\dagger = \epsilon_X^\dagger\circ \epsilon_Z
\qquad\mbox{\rm i.e.}\qquad   \inlinegraphic{4.7em}{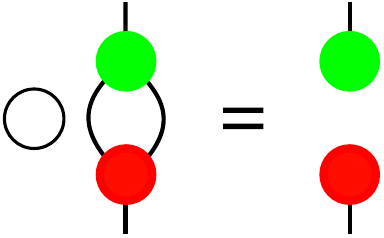}\,,
\eeq
then  they are complementary observable structures. We call (\ref{eq:hopf}) the `Hopf law'. 
\end{theorem}
\begin{proof}
We have to show that {\bf comp1} and {\bf comp2} of Definition \ref{defn:complementary-observables} hold.  
Since:
\[
  \inlinegraphic{6.4em}{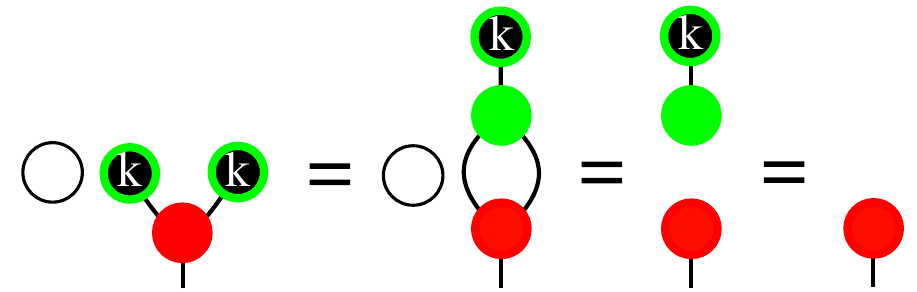}
  \]
  {\bf comp1} holds; by exchanging the colours we obtain {\bf comp2}.
\end{proof}\vspace{-3.5mm}

The converse to Theorem~\ref{propHopf_1} also holds if one of the observable structures involved  is either a vector or state basis--- cf.~Definition \ref{def:pointbasis}.

\begin{theorem}[complementarity $\Leftarrow$]\label{propHopf_2}
If  $(A, \delta_Z, \epsilon_Z)$ and $(A,\delta_X,\epsilon_X)$ are complementary observable structures, and if at least one of these is either a vector basis or a state basis, then the Hopf law of Theorem~\ref{propHopf_1} holds.
\end{theorem}
\begin{proof}
We need to show that when applying the left-hand side and the
right-hand side of the Hopf law to an element of the basis that both
sides are equal, for all the elements of the basis. 
For  the case of a vector basis we have:
\[
  \inlinegraphic{6em}{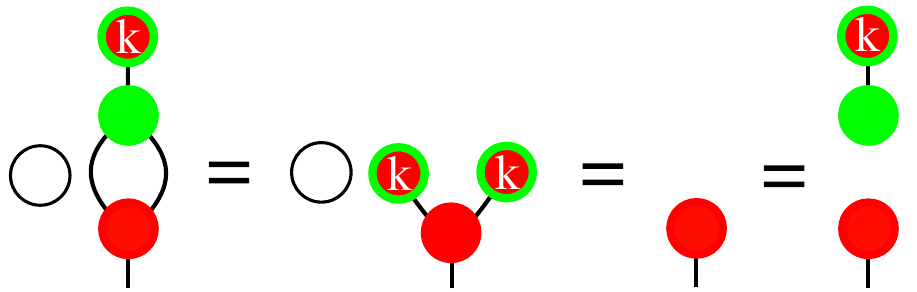}
  \]
and for the case of a state basis we moreover have:
\[
  \inlinegraphic{6em}{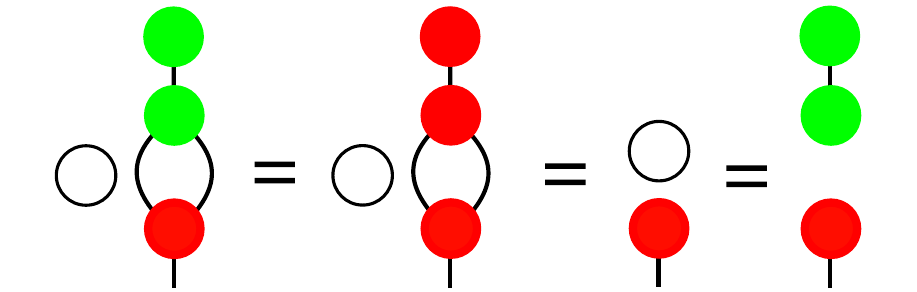}
  \]
where the first equation relies on coinciding $\dag$-compact structures.
\end{proof}\vspace{-3.5mm}

\subsection{The general case: dualisers as antipodes}\label{SEC:Hopfbis}

The results of this section still hold when the  $\dag$-compact structures
induced by the two \coss do coincide,
provided we extend the observable structures formalism with dualisers,
as described in \cite{Coecke-Paquette-Perdrix} by Paquette, Perdrix and
one of the authors.   

\begin{definition}\em
The \em dualiser \em of two distinct observable structures $(A, \delta_Z, \epsilon_Z)$ and $(A,\delta_X,\epsilon_X)$ on the same object $A$ is
\beq
d_{ZX} = (\eta_Z\otimes 1_A)\circ(1_A\otimes \eta_X^\dagger) \quad\ \ \mbox{\rm i.e.}\quad \ \   \inlinegraphic{3.1em}{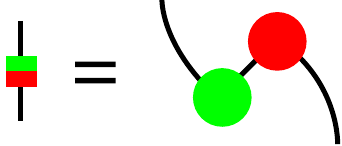}\,. 
\eeq
\end{definition}

\begin{remark}
If the induced $\dag$-compact structures of the two observable structures on $A$ happen to coincide then their dualiser is $1_A$, hence trivial.  More generally, the dualiser is easily seen to always be unitary, by $\dag$-compactness.
\end{remark}

\begin{lemma}\label{dualisercomplem}
For observable structures $(A, \delta_Z, \epsilon_Z)$ and $(A,\delta_X,\epsilon_X)$ we have
\beq
(d_{XZ}\otimes 1_A)\circ\eta_X=\eta_Z^\dagger\qquad\mbox{\rm i.e.}\qquad\inlinegraphic{2.7em}{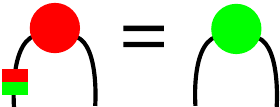}\,,
\eeq
and the equation obtained by exchanging the colours also holds.
\end{lemma}
\begin{proof}
Straightforward by $\dag$-compactness.
\end{proof}\vspace{-3.5mm}

\begin{remark}
Lemma~\ref{dualisercomplem} together with unitarity of the dualiser
provides a more concise proof of the fact that dimension does not
depend on a choice of observable structure --- cf.~Lemme \ref{ref:dimension}.
\end{remark}

\begin{lemma}\label{dualisercomplembis}
Let $k:\II \to A$ be a classical point for observable structure $(A, \delta_Z, \epsilon_Z)$ and let $(A,\delta_X,\epsilon_X)$ be another observable structure.  Then the point
\[
d_{ZX}\circ k:\II \to A\qquad\qquad\mbox{\rm i.e.}\qquad\qquad\inlinegraphic{2.4em}{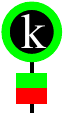}
\]
is the conjugate to $k$ for the $\dag$-compact structure induced by $(A,\delta_X,\epsilon_X)$.
\end{lemma}
\begin{proof}
The $(A,\eta_X)$-conjugate to $d_{ZX}\circ k$ is, using Lemma~\ref{dualisercomplem},
\[
\inlinegraphic{4.0em}{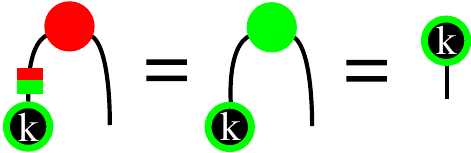}
\]
since the $(A,\eta_Z)$-transpose to $k$ is also its adjoint.
\end{proof}\vspace{-3.5mm}

\begin{theorem}\label{thm:Hopfdual}
If two observable structures obey
\beq\label{eq:hopfdual}
D\cdot\delta_X\circ(d_{ZX}\otimes 1_A)\circ \delta_Z^\dagger = \epsilon_X^\dagger\circ \epsilon_Z
 \quad\mbox{\rm i.e.}\quad    \inlinegraphic{4.7em}{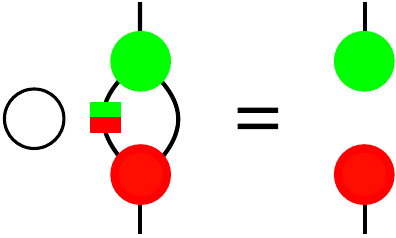},
\eeq
then  they are complementary observable structures.  Conversely, if  $(A, \delta_Z, \epsilon_Z)$ and $(A,\delta_X,\epsilon_X)$ are complementary observable structures, and if at least one of these is either a vector basis or a state basis, then the Hopf law depicted above holds.
\end{theorem}
\begin{proof}
Using Lemma~\ref{dualisercomplem} and Lemma~\ref{dualisercomplembis}, 
the proofs of Theorem~\ref{propHopf_1} and Theorem~\ref{propHopf_2}
can be easily modified to this more general situation.
\end{proof}\vspace{-3.5mm}

\begin{remark}
In the form (\ref{eq:hopfdual}) the Hopf law matches the form of the defining law of a Hopf algebra  \cite{Cartier,Kassel, StreetBook}, with the dualiser playing the role of the antipode.
\end{remark}

\section{Closed complementary observable structures}\label{sec:bialgebra}

In this section we study a special class of complementary observable
structures, which we refer as to \em closed\em.  While in the previous
section we recovered the {\bf B'}-rule of the \zxcalculus, which
captures complementarity on-the-nose, the {\bf B}-rules capture this
stronger form of complementarity for observable structures.  The main
theorem of this section provides a number of equivalent
characterisations of closedness.  The bottom line will be that certain
pairs of observable structures form a scaled variant of the usual
notion of a \em bialgebra \em \cite{Cartier,Kassel, StreetBook}.  The
defining equations of a bialgebra involve a commutation condition of
the multiplication of one algebra with the comultiplication of the
other, as well as one of the (co)multiplication of one algebra with
the (co)unit of the other.  We identify the scaled analogues to these
conditions for closed complementary observable structures in sections
\ref{sec:comcossz} and \ref{sec:coherence} respectively.  Again,
we assume that the $\dag$-compact structures induced by the complementary observable structures coincide.  

\subsection{Coherence for observable structures}\label{sec:coherence}

In this section we provide a category-theoretic generalisation of the
concrete notion of coherence for complementary vector and state
bases--- cf.~Definition \ref{def:coherence}.  The green and red
observable structures of the \zxcalculus also enjoy this property.
First we set:
\[
  \inlinegraphic{2.8em}{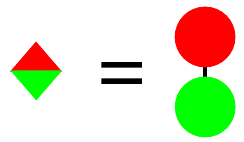}
\]
and denote this scalar by $\sqrt{D}$ for reasons we explain below.

\begin{definition}\em\label{defn:coherent-observables}
Two  observable  structures $(A,\delta_Z,\epsilon_Z)$ and $(A, \delta_{X}, \epsilon_{X})$
in a \dsmc   are called  \emph{coherent} if they obey the following two rules:
  \begin{itemize}
      \item[{\bf coher1}]  $\epsilon_{X}$ satisfies:
\[
\sqrt{D}\cdot \delta_{Z}\circ \epsilon_X^\dagger=\epsilon_X^\dagger\otimes \epsilon_X^\dagger
\qquad \, \mbox{\rm i.e.}\qquad\,
  \inlinegraphic{2.8em}{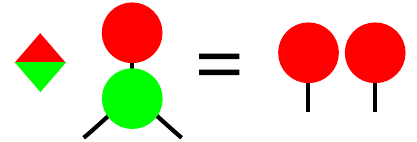}\ \quad
\]
    \item[{\bf coher2}]  $\epsilon_{Z}$  satisfies:
\[
\sqrt{D}\cdot \delta_X \circ\epsilon_{Z}^\dagger=\epsilon_{Z}^\dagger\otimes \epsilon_{Z}^\dagger
\qquad\, \mbox{\rm i.e.}\qquad\,
  \inlinegraphic{2.8em}{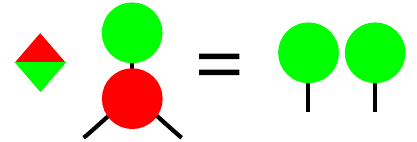}\ \quad
\]
  \end{itemize}
\end{definition}

\begin{remark}
  As already mentioned in the introduction of this section, the
  conditions {\bf coher1} and {\bf coher2} are scaled variants of two
  of the defining conditions of a bialgebra; in the language of this
  paper, these state that the erasing point of one observable
  structure is a classical point for the other. Condition {\bf coher1}
  in Definition \ref{defn:coherent-observables} states that
  $\epsilon_X^\dagger$ differs from a classical point of
  $(A,\delta_{Z},\epsilon_{Z})$ by a scalar factor of $\sqrt{D}$.  The
  choice of $\sqrt{D}=(\epsilon_Z^\dagger\circ\epsilon_{X}^\dagger)$
  for this scalar is not arbitrary but is imposed by the fact that
  $\epsilon_Z$ is the unit for the comultiplication $\delta_Z$.  To
  see this, it suffices to post-compose both sides of {\bf coher1}
  with $\epsilon_Z\otimes\epsilon_{X}$, which results in:
\[
  \inlinegraphic{4.5em}{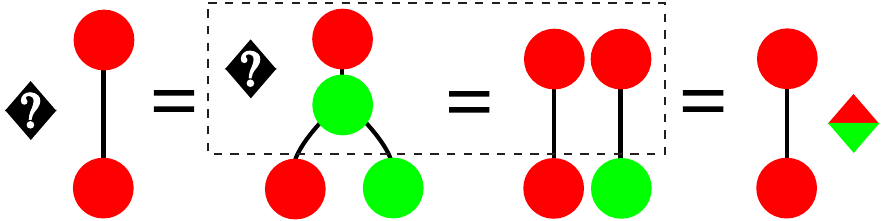}\,.
\]
Dually, condition {\bf coher2} asserts the same relationship between 
$\epsilon_{Z}^\dagger$ and $(A,\delta_X,\epsilon_X)$.
\end{remark}

\begin{example}
  For the case of ${\bf FdHilb}$ and ${\bf FdHilb}_{wp}$ this
  category-theoretic notion of coherence coincides with that of
  Definition~\ref{def:coherence}. For these cases,
  Theorem~\ref{thm:CoherentBases} indicates that the requirement of
  coherence entails no loss of generality.
\end{example}

Now we justify the notation $\sqrt{D}$ for $\inlinegraphic{1.35em}{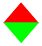}$.

\begin{lemma}\label{SQRTDis_selfconj}
If $(A, \delta_Z, \epsilon_Z)$ and $(A,\delta_X,\epsilon_X)$ are two observable structures whose induced $\dag$-compact structures coincide then we have: 
\[
\sqrt{D}^\dagger=\sqrt{D}\qquad\quad\mbox{\rm i.e.}\qquad\quad    \inlinegraphic{1.4em}{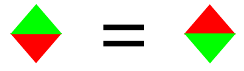}\,.
\]
\end{lemma}
\begin{proof}
By Corollary \ref{col:self-conjugare} we know that $\epsilon_Z$ are
$\epsilon_X$ are both self-conjugate, 
hence: 
\beq
\inlinegraphic{4.0em}{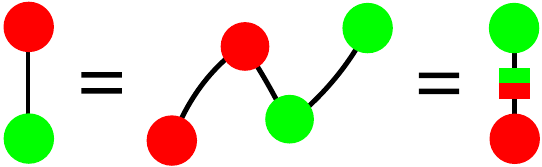}.
\eeq
The result follows by the fact that the dualiser is trivial when
the induced $\dag$-compact structures coincide.
\end{proof}\vspace{-3.5mm}

By using the notation $\sqrt{D}$ for $\inlinegraphic{1.35em}{SQRTD}$, we insinuate that 
\beq
\inlinegraphic{1.35em}{SQRTD}\
\inlinegraphic{1.35em}{SQRTD}=\sqrt{D}\cdot \sqrt{D}=D= 
\raisebox{0.02em}{\inlinegraphic{2.0em}{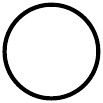}}\,,
\eeq
which by, Lemma~\ref{SQRTDis_selfconj}, becomes
\beq\label{eq:deletepointsunbiased}
\inlinegraphic{2.6em}{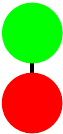}\ \inlinegraphic{2.6em}{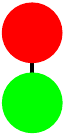}
=\inlinegraphic{1.35em}{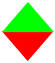}\ \inlinegraphic{1.35em}{SQRTD}=
\inlinegraphic{2.0em}{circle}\,.
\eeq
Since the `length' of both $\epsilon_Z^\dagger$ and $\epsilon_X^\dagger$ is $\sqrt{D}$--- cf.~Lemma~\ref{lem:scalar-unbiased} ---i.e.
\beq
\inlinegraphic{2.6em}{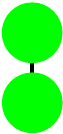}=\inlinegraphic{2.0em}{circle}
\quad,\quad
\inlinegraphic{2.6em}{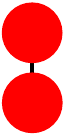}=\inlinegraphic{2.0em}{circle}\,,
\eeq
equation (\ref{eq:deletepointsunbiased}) states that $\epsilon_Z^\dagger$ and $\epsilon_X^\dagger$ are unbiased, which is a natural requirement for a pair of \coss.  Moreover, it follows from coherence of observable structures: 

\begin{proposition}\label{prop:scalarvalue}
For coherent observable structures  
on $A$ with coinciding induced $\dagger$-compact structures we have:
\beq
\sqrt{D}\cdot\sqrt{D} = D = \dim(A)
\qquad\text{i.e.}\qquad\inlinegraphic{2.0em}{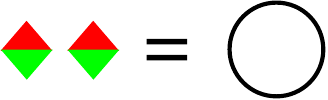}.
\eeq
\end{proposition}
\begin{proof}
We have:
\[
  \inlinegraphic{3.8em}{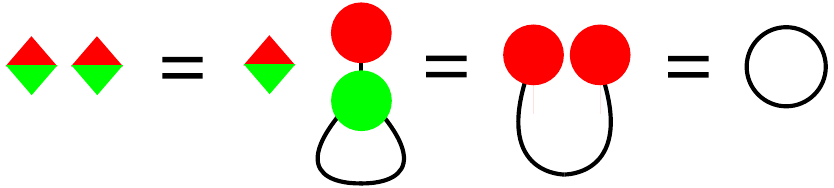}
\]
where the last step uses coincidence of induced $\dag$-compact structures.
\end{proof}\vspace{-3.5mm}

\subsection{Commutation for observable structures}\label{sec:comcossz}

Several notions of commutation may apply to observable structures.  In this section we consider three of these, of which one will complete the definition of a scaled bialgebra;
in the \zxcalculus, this is the powerful   {\bf B2}-rule.

\begin{remark}
One should clearly distinguish the notions  of commutation that we
consider in this paper from that of \em commuting observables \em as
found in most of the quantum theory literature.  The kind of
observables considered here are complementary, and are thus 
non-commuting in the usual sense.  What we wish to expose here is that
certain alternative notions of commutation, which are useful in
computations, do apply to the specific case of complementary observables.
\end{remark}

\par\medskip\noindent
{\it Notation.} 
For an observable structure $(A,\delta_Z,\epsilon_Z)$, with classical
points ${\cal K}_Z$ depicted in green, and an observable structure $(A, \delta_{X}, \epsilon_{X})$ depicted in red, we set for all $k\in{\cal K}_Z$:
\bc$
\inlinegraphic{3.0em}{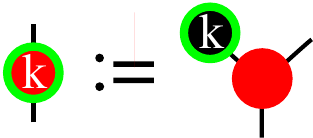}
$\ec
The use of two colours in this graphical representation reflects its
dependence on two observable structures.  We denote this morphism by
$\Lambda^X(k)$.   By Lemma~\ref{lm:unbiasedscalar} we know that, when
$k$ is unbiased for $(A, \delta_{X}, \epsilon_{X})$,  this morphism is
unitary if and only if $k^\dagger\circ k=D$.  Therefore it is more
convenient to consider classical points to have length $\sqrt{D}$
rather than being normalised.  The comonoid homomorphism laws
governing classical points then become:
\bc$
\inlinegraphic{3.em}{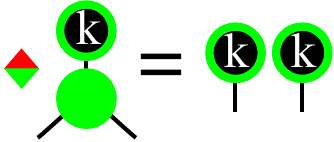} 
\qquad\ 
\inlinegraphic{3.0em}{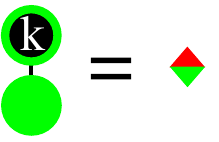} 
\qquad\ 
\inlinegraphic{3.em}{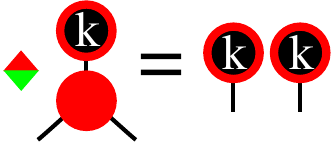} 
\qquad\ 
\inlinegraphic{3.0em}{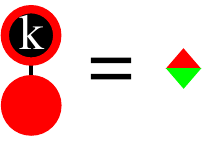}  
$\ec
where we somewhat abusively depict $\sqrt{D}$  by
$\inlinegraphic{1.0em}{SQRTD}\!$ as in the case of coherent observable
structures.  If these observables are complementary, the equations of
Definition~\ref{defn:complementary-observables} become:
\bc$
\inlinegraphic{3.em}{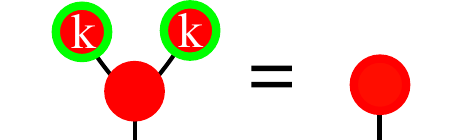} 
\qquad\qquad\qquad\, 
\inlinegraphic{3.em}{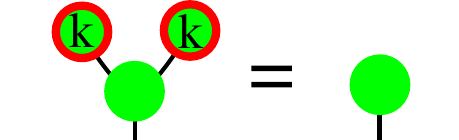} 
$\ec
\par\medskip

\begin{remark}\label{rem:notationclash1}
The similarity between the graphical notation for  $\Lambda^X(k)$ and
that of the classical points for \coss in section
\ref{SEC:Complementarity} anticipates Theorem~\ref{main:theorem} below.
\end{remark}

\begin{definition}\label{def:operator-commutation}\em 
Observable structure $(A,\delta_Z,\epsilon_Z)$ with classical points
${\cal K}_Z$, and observable structure $(A, \delta_{X}, \epsilon_{X})$
with classical points ${\cal K}_X$, obey  \em operator commutation
\em iff for all $k\in{\cal K}_Z$ and all $k'\in{\cal K}_X$:
\bc$
\Lambda^Z(k')\circ \Lambda^X(k)  =  (k'^\dagger\circ k)\cdot (\Lambda^X(k)\circ \Lambda^Z(k'))
\qquad\  \ \ \ \ \ \mbox{\rm i.e.}\qquad 
\inlinegraphic{3.6em}{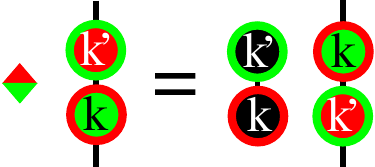}\,.
$\ec
\end{definition}

\begin{definition}\label{def:comultiplicative-commutation}\em 
Observable structure $(A,\delta_Z,\epsilon_Z)$ with classical points
${\cal K}_Z$, and  observable structure $(A, \delta_{X},
\epsilon_{X})$, obey \em comultiplicative commutation \em iff for
all $k\in{\cal K}_Z$:
\bc$
\delta_Z\circ  \Lambda^X(k)  = (\Lambda^X(k)\otimes\Lambda^X(k))\circ\delta_Z
\qquad\qquad\qquad \ \ \, \mbox{\rm i.e.}\qquad \ \,
\inlinegraphic{3.3em}{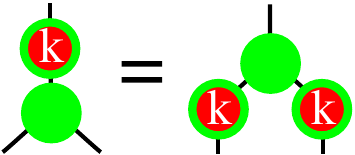}\,.
$\ec
\end{definition}

\begin{remark}
While this equation seems akin to the defining equation for classical
points it carries a lot more structure.  The reason for this is the
involvement of two observable structures, which is exposed by the
colouring.
\end{remark}

\begin{definition}\label{def:bialgebraic-commutation}\em 
Observable structures $(A,\delta_Z,\epsilon_Z)$
and $(A, \delta_{X}, \epsilon_{X})$ obey \em bialgebraic commutation \em iff:
\bc$
D\cdot (\delta^\dagger_X\otimes\delta^\dagger_X)\circ(1\otimes\sigma\otimes 1)
\circ (\delta_Z\otimes\delta_Z)=\sqrt{D}\cdot \delta_Z\circ\delta^\dagger_X
\ \  \mbox{\rm i.e.} \ \ \inlinegraphic{4.5em}{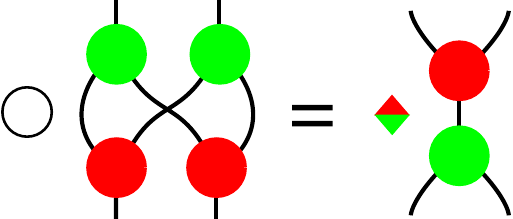}\!\!.
$\ec
\end{definition}

\begin{remark}
For coherent observable structures, by Proposition~\ref{prop:scalarvalue}, when $\sqrt{D}$ admits an inverse we can simplify the bialgebraic commutation equation to:
\bc
\inlinegraphic{4.5em}{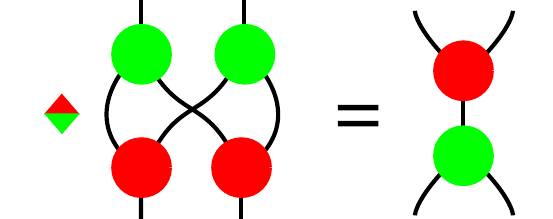}
\ec
To see that the choice of scalars is not arbitrary, we can, for example,  either assume coherence or the Hopf law for the observable structures, both resulting in:
\bc
\inlinegraphic{6.8em}{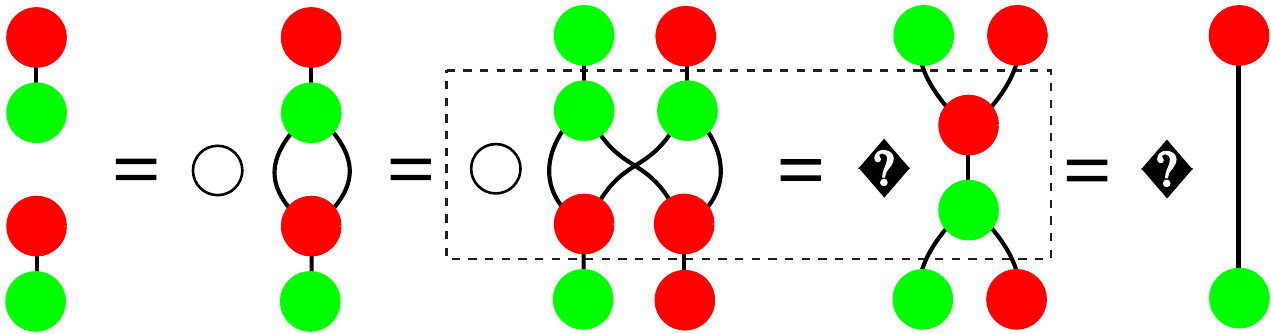}
\ec
\end{remark}


\begin{definition}\em 
A \em scaled bialgebra \em is a pair of coherent observable structures which satisfy bialgebraic commutation, that is, all together:
\[
\inlinegraphic{6.86em}{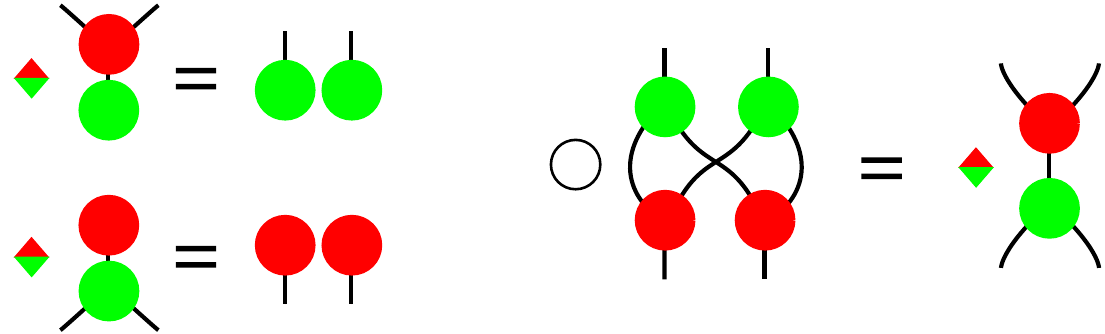}.
\]
\end{definition}

\begin{remark}
As announced above, if we remove the scalars from the  definition of a scaled bialgebra
and adjoin the equation $\epsilon_Z\circ\epsilon_X^\dagger = 1_\II$ --
which is trivial anyway when taken `up to a scalar' -- then we obtain
the usual notion of a \em bialgebra \em \cite{Cartier,Kassel, StreetBook}.
\end{remark}

\begin{theorem}\label{Bialgebrafromhopf}
Each scaled bialgebra satisfies the Hopf law.  
\end{theorem}
\begin{proof}
See the derivation of the {\bf B}'-rule in Section~\ref{sec:simplegreenredrules};
%
note that the 1st step uses $\dag$-compactness and the 4th step uses coherence i.e.~the green comultiplication copies the red unit.
\end{proof}\vspace{-3.5mm}

\begin{corollary}
If a pair of observable structures constitutes a scaled bialgebra then these are complementary observable structures.
\end{corollary}

While at first sight the three notions of commutation we have
introduced in this section look very different, in fact, they boil
down to the same thing in all of our example categories, as we shall
see in Theorem~\ref{main:theorem} below.

\subsection{Closedness for observable structures}

\begin{definition}\label{def:colsedness}\em 
The classical points ${\cal K}_Z$ of an observable structure $(A,\delta_Z,\epsilon_Z)$ are \em closed \em for another observable structure $(A, \delta_{X}, \epsilon_{X})$ iff for all $k,k'\in{\cal K}_Z$ we have 
\[
k\odot_{X} k'\,\in\,{\cal K}_Z\,.
\]
\end{definition}
From  the assumption that the induced $\dag$-compact structures coincide, since by Corollary \ref{col:self-conjugare} and Definition \ref{def:classpoint} we have that $\delta_X$, $k$ and $k'$ are all self-conjugate, it follows that the composite $k\odot_{X} k'$ is also self-conjugate. 
Hence, setting:
\[
  \inlinegraphic{2.8em}{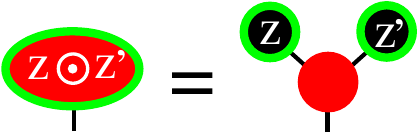}\,;
\]
the closedness requirement is depicted graphically as: 
\[
  \inlinegraphic{3.8em}{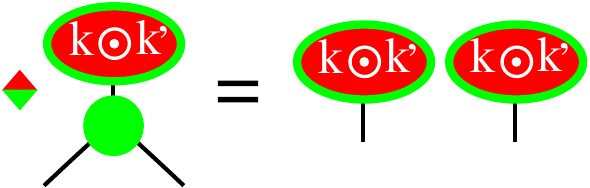}\qquad\qquad \inlinegraphic{3.8em}{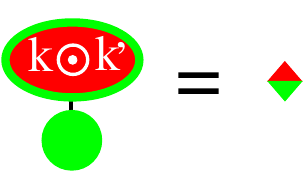}\,.
\]

\begin{remark}
If the observable structures are coherent, then the normalisation condition is also trivially satisfied. 
If classical points were taken to be normalised then we would take $\sqrt{D}\cdot k\odot_{X} k'$ rather than $k\odot_{X} k'$ in
Definition \ref{def:colsedness}.   
\end{remark}

\begin{remark}
Again, similarly to Remark \ref{rem:notationclash1}, this notation which seems to indicate that $k\odot_{X} k'$ is unbiased to $(A, \delta_{X}, \epsilon_{X})$ anticipates Theorem~\ref{main:theorem} below.
\end{remark}

We now show that on every Hilbert space we can find a pair of closed \coss, and hence by Theorem  \ref{thm:CoherentBases} we can find a pair of closed coherent \coss.

\begin{proposition}\label{Ex:closedCCCS}
In ${\bf FdHilb}$ there exist pairs of  closed
coherent \coss on Hilbert spaces
$\mathbb{C}^n$ for any dimension $n\in\mathbb{N}$.
\end{proposition}
\begin{proof}
  Without loss of generality we take the first observable structure as
  being defined by the  standard basis on $\mathbb{C}^n$,  i.e. 
  $\delta : \ket{i} \mapsto \ket{i}\otimes \ket{i}$ with the
  erasing point $\epsilon^\dag = \sum_i \ket{i}$.  Notice that the
  multiplication induced by this observable structure is
  point-wise:
  \[
  (\sum_i a_i \ket{i})\odot (\sum_i b_i\ket{i}) = \sum_i a_ib_i \ket{i}.
  \]
  We need to find a basis which contains $\epsilon^\dag$, is unbiased
  with respect to the standard basis, and is closed under $\odot$.
  It is routine to check that the family
  \[
  \ket{f_j} = \frac{1}{\sqrt{N}}\sum_k \omega^{jk}_n\ket{k}\;,
  \]
  where $j$ and $k$ range from 0 to $n-1$, and $\omega_n = e^{2\pi
    i/n}$, provides an orthonormal basis satisfying these conditions.
  We choose $\ket{0}$ as the erasing point.
\end{proof}\vspace{-3.5mm}

\begin{corollary}
  There exist pairs of closed coherent complementary observable
  structures for any dimension in ${\bf FdHilb}_{wp}$.
\end{corollary}

\begin{remark}\label{remark:hadamard}
  Thanks to Theorem~\ref{thm:CoherentBases}, to find a pair of
  coherent \coss on $\mathbb{C}^d$ it
  suffices to find any \em dephased complex Hadamard matrix\em, that is, an
  orthogonal  matrix whose entries are all complex units, and whose
  first row and column are all ones.  The columns of the  matrix will
  provide the required basis.
  The family $\ket{f_j}$ used above are a particular example: they
  form the columns of the \em $d$-dimensional Fourier
  matrix\em.  If $d=2$, $3$ or $5$ the \emph{only} dephased Hadamards
  are Fourier matrices \cite{Tadej}, hence we can conclude that every pair of coherent \coss
  in these dimensions is closed.  However this does not hold in general.
  If $d=4$, for example, 
  \[
  F_4(x) = \left(
  \begin{array}{cccc}
    1&1&1&1\\
    1&ie^{ix}&-1&-ie^{ix}\\
    1&-1&1&-1\\
    1&-ie^{ix}&-1&ie^{ix}
  \end{array} \right)
  \]
  is not closed when $x$ is irrational. Similar counterexamples can
  be constructed for dimensions $d \geq 6$.  This shows that the notion of  closed 
  \coss  is strictly stronger than that of  \coss.
\end{remark}

Since closed \coss exist for all dimensions, for most practical purposes we can assume that \coss  are both coherent and closed.   Closed \coss moreover behave well with respect to the monoidal
structure, in that the canonical induced observable structures of
Proposition~\ref{prop:TensorObsStruc}, which are defined on the tensor
space, inherit both complementarity and closedness.

\begin{proposition}\label{prop:tensor-preserved-closure}
  Let $(A,\delta_Z, \epsilon_Z)$ and $(A,\delta_X,
  \epsilon_X)$ be coherent \coss such that $(\delta_Z, \epsilon_Z)$ is closed
  with respect to  $(\delta_X, \epsilon_X)$, and let $(B,\delta_{Z'},
  \epsilon_{Z'})$ and $(B,\delta_{X'}, 
  \epsilon_{X'})$ be coherent \coss such that $(\delta_{Z'}, \epsilon_{Z'})$ is closed
  with respect to  $(\delta_{X'}, \epsilon_{X'})$; then the canonical  observable
  structure on the joint space $(A \otimes B, \delta_Z \otimes
  \delta_{Z'}, \epsilon_Z \otimes \epsilon_{Z'})$ is both complementary and closed with
  respect to $(A \otimes B, \delta_X \otimes \delta_{X'}, \epsilon_X
  \otimes \epsilon_{X'})$.
\end{proposition}

\subsection{Our main theorem on pairs of closed observable structures}

\begin{theorem}\label{main:theorem}
The following are equivalent for two observable structures:
\begin{itemize}
\item[{\bf closed}] They are closed.
\item[{\bf oper}]  They obey operator commutation.
\item[{\bf comul}] They obey comultiplicative commutation.
\item[{\bf bialge}]  They obey bialgebraic commutation.
\end{itemize}
subject to the following requirements:
\[
\xymatrix@=0.80in{
& \mbox{\bf bialge} \ar@{=>}
[d] | {none} & \\
\mbox{\bf closed} 
\ar@{=>}
[ru] | {1B (+ Coh)} 
& 
\ar@{=>}
[l] | {none} \mbox{\bf comul}
 \ar@{=>}
 [r] | {none} & 
\mbox{\bf oper} 
\ar@{=>}
[lu] | {2B (+ Coh)} 
}
\]  
where `none' stands for no additional requirements, except for the ones explicitly stated in the proof; where `1B' means that  at least one of the observable structures has either a vector basis or a state basis,  where `2B' means that this is the case for both observable structures, and `(+ Coh)' means that in the case of state bases we also require coherence.  We indicate in the proof where we assume that $\sqrt{D}$ has an inverse and where we use the fact that $\dag$-compact structures coincide.
\end{theorem}
\begin{proof}
We show all required implications graphically:
\bit
\item ${\bf bialge}\Rightarrow{\bf comul}$:
\beq\label{eq:proofalreadydone}\hspace{-2.2cm}  
\inlinegraphic{5.8em}{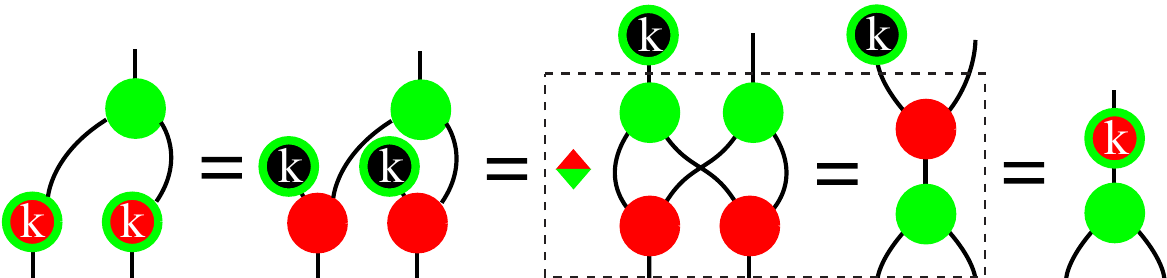}
\eeq
Here we assumed that $\sqrt{D}$ has an inverse.
\item ${\bf comul}\Rightarrow{\bf closed}$:
\beq\hspace{-2.2cm}  
  \inlinegraphic{4.7em}{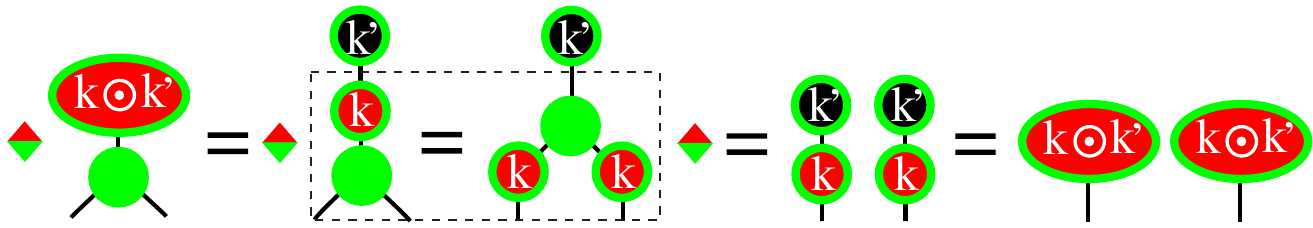}
\eeq
\item ${\bf comul}\Rightarrow{\bf oper}$:
\beq\hspace{-2.2cm}  
  \inlinegraphic{4.7em}{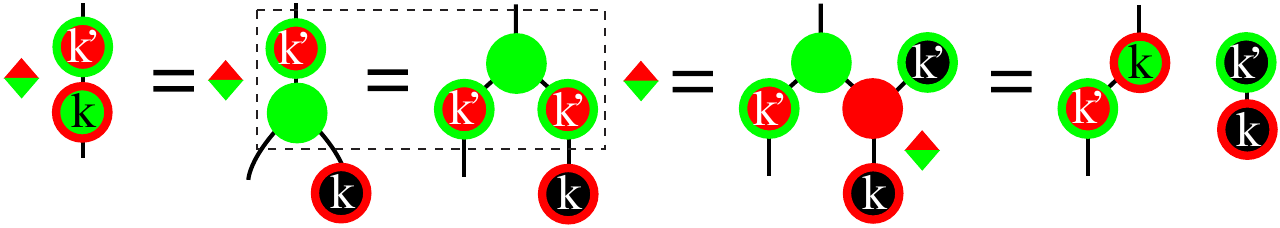}
\eeq
\item ${\bf closed}\Rightarrow{\bf bialge}$:
\beq\hspace{-2.2cm}  
  \inlinegraphic{5.8em}{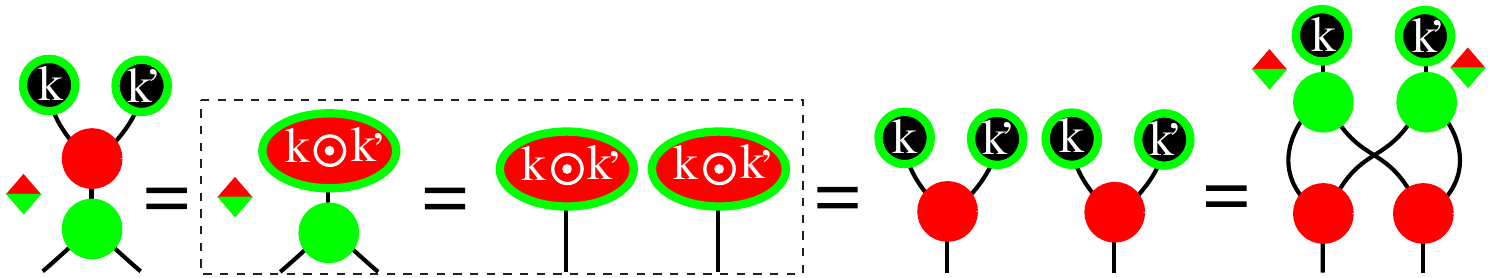}
\eeq
The assumption that  the classical points for the green
observable structure constitute a vector basis, together with the fact
that the monoidal tensor lifts to vector basis, imply bialgebraic
commutation.  By coherence we have:
\beq\hspace{-2.2cm}  
  \inlinegraphic{5.8em}{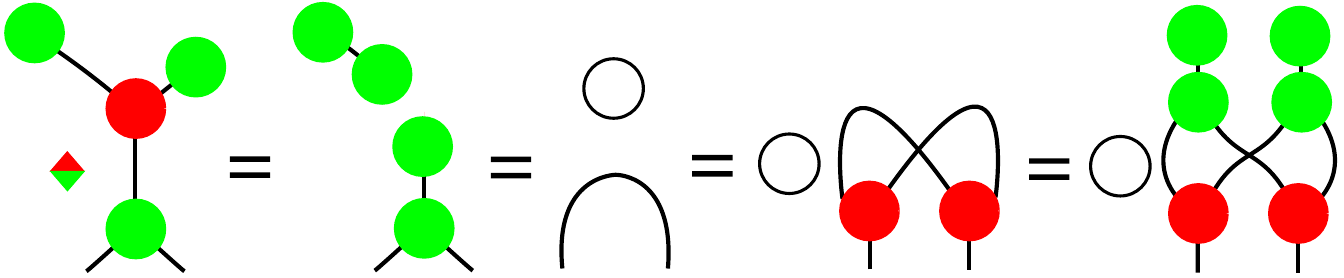}
\eeq
so the result holds when there is a state basis for the green observable structure. Steps 2--4 assume that the induced $\dag$-compact structures coincide.
\item ${\bf oper}\Rightarrow{\bf bialge}$:
\beq\label{eq:canon_bialge}
\hspace{-2.2cm} 
  \inlinegraphic{6.72em}{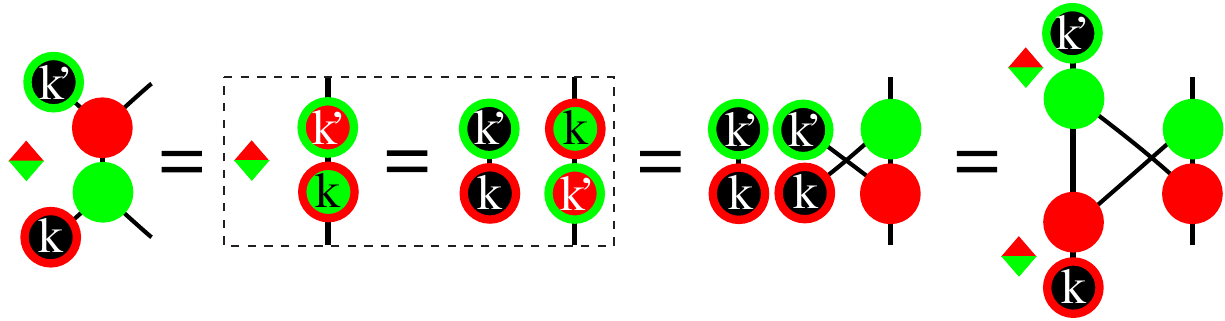}
\eeq
so by $\dag$-compactness we have:
\beq\hspace{-2.2cm}  
  \inlinegraphic{5.2em}{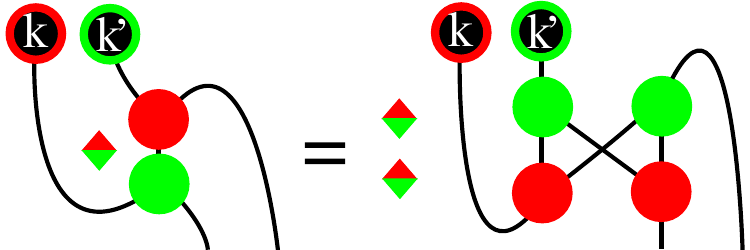}
\eeq
Under the assumption that  the classical points both for the green and
the red observable structure constitute a vector basis, together with
the fact that the monoidal tensor lifts vector bases, we have:
\beq\hspace{-2.2cm}  
  \inlinegraphic{4.2em}{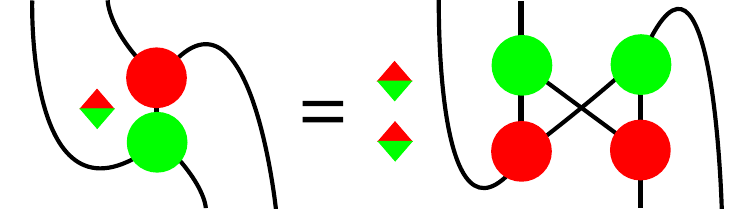}
\eeq
from which the bialgebra follows by $\dag$-compactness.   The two diamonds
are equal to a circle given that the $\dag$-compact structures coincide.
For the case that both observable structures  have a state basis it
remains to be shown that: 
\beq\hspace{-2.2cm}  
  \inlinegraphic{6.2em}{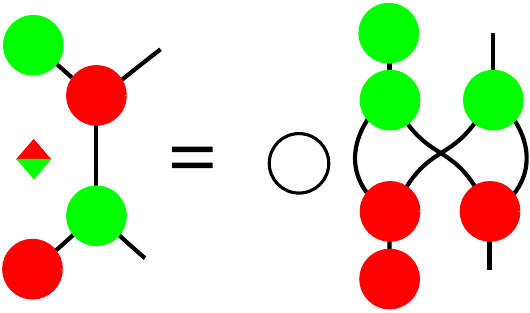}
\eeq
To do so, we now show that the equation
\beq\hspace{-2.2cm}  
  \inlinegraphic{6.2em}{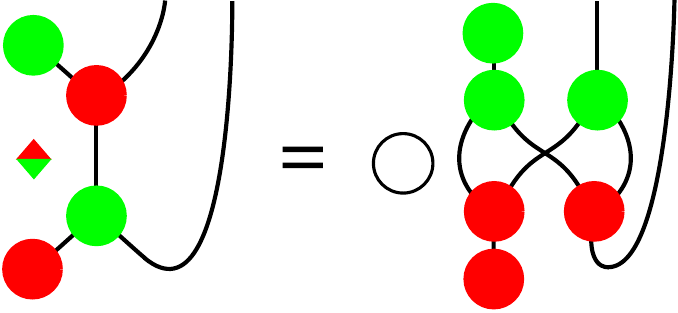}
\eeq
holds, by relying on the fact that we have a state basis for both observable structures, and as above, by again relying on $\dag$-compactness.  We have:
\beq\hspace{-2.2cm}  
  \inlinegraphic{7.23em}{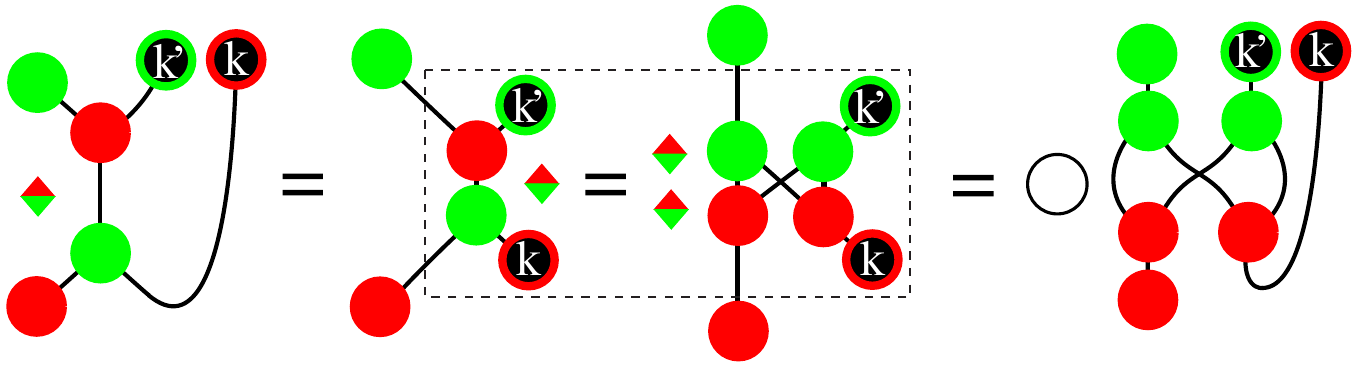}
\eeq
where in the dotted area we used derivation (\ref{eq:canon_bialge}) above. Finally:
\beq\hspace{-2.2cm}  
  \inlinegraphic{6.6em}{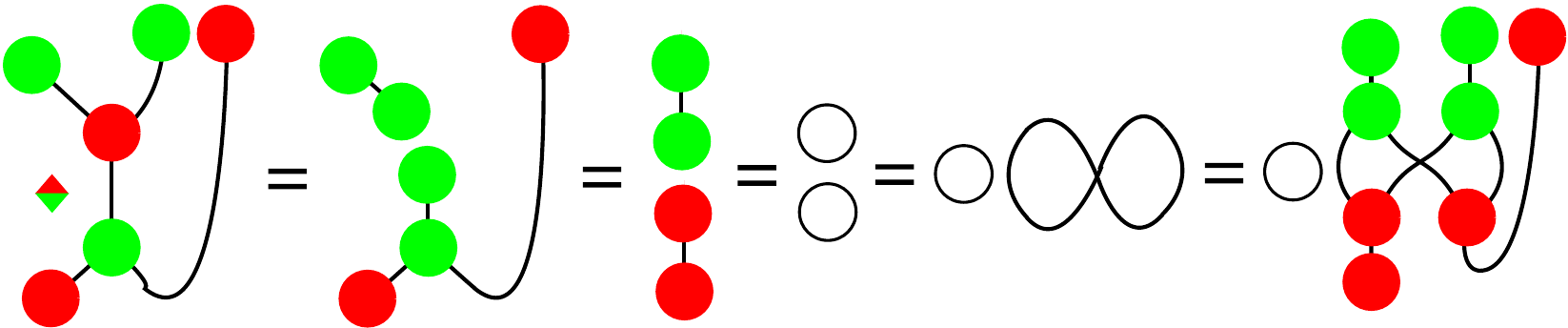}
\eeq
where the last step assumes that induced $\dag$-compact structures coincide.
\eit
This concludes this proof.
\end{proof}\vspace{-3.5mm}

\begin{remark}
We leave it to the reader to see how `2B (+ Coh)' factors into requirements for $\mbox{\bf oper} \Rightarrow \mbox{\bf comul}$ and $\mbox{\bf comul}\Rightarrow\mbox{\bf bialge}$.
\end{remark}

The examples of \coss discussed in 
Proposition~\ref{Ex:closedCCCS} satisfy all the equations stated in
Theorem~\ref{main:theorem}.  In particular, they constitute
scaled bialgebras.  These equations are strictly stronger than the
Hopf law by Theorem~\ref{Bialgebrafromhopf}, and hence all pairs of
observable structures that satisfy them are  \coss.

%
%
%
%
%

\section{Further group structure and the classical automorphisms}
\label{SEC:GroupActionEtc}

In Section~\ref{sec:phase-group} we saw how the abelian group of
phase shifts arose naturally from the presence of unbiased points for
a given observable structure.  When we have a
pair of \coss,  the two phase
groups can interfere with each other, an interaction
which arises from the special role of the classical points
within each phase group.

In the following, suppose $(A,\delta_Z,\epsilon_Z)$ and
$(A,\delta_X,\epsilon_X)$ are coherent \coss which jointly form a scaled
bialgebra.  Let ${\cal U}_Z$ denote all the unbiased points for
$(\delta_Z,\epsilon_Z)$, and let ${\cal K}_Z$ denote its classical points;
define ${\cal U}_X$ and ${\cal K}_X$ similarly.  By virtue of complementarity we
have ${\cal K}_X \subseteq {\cal U}_Z$ and ${\cal K}_Z \subseteq {\cal U}_X$.  Recall that by
Proposition~\ref{prop:inverse}, $({\cal U}_Z, \odot_Z)$ is an abelian group,
isomorphic to the phase group of $(\delta_Z, \epsilon_Z)$.

\begin{theorem}\label{thm:subgroup}
  ${\cal K}_X$ is a subgroup of $({\cal U}_Z,\odot_Z)$ if either: ${\cal K}_X$ is finite; or, if the
  two observable structures give rise to the same $\dag$-compact structure.
\end{theorem}
\begin{proof}
  ${\cal K}_X$ is always a submonoid of ${\cal U}_Z$ because of the closure and
  coherence 
  of the  two   observable structures; any finite submonoid
  is a subgroup.  Alternatively, given a point $x\in {\cal K}_X$, its
  inverse in ${\cal U}_Z$ is given by its conjugate with respect to
  the $\dag$-compact structure of $(\delta_Z, \epsilon_Z)$;  by the
  definition  of classicality, $x$ is self-conjugate with respect
  to the $\dag$-compact structure of $(\delta_X, \epsilon_X)$.  Hence, if
  these $\dag$-compact structures agree
  (cf. Proposition~\ref{prop:concideCCthenSelfCong}),  $x^{-1} =
  x$ in ${\cal U}_Z$, so ${\cal K}_X$ is a subgroup.
\end{proof}\vspace{-3.5mm}

\begin{remark}
  Shared $\dag$-compact structure is a powerful assumption. 
  The proof above indicates that in the case of coinciding $\dag$-compact
  structures, not only do the classical points within ${\cal U}_Z$ form a
  subgroup, but the resulting group is a product of copies of
  $S_2$.  In the case of qubits described by $X$ and $Z$
  spins, the $\dag$-compact structure is shared, and the resulting classical
  subgroup is just $S_2$.
\end{remark}

\begin{proposition}
For all $x\in {\cal K}_X$, $\Lambda^Z(x)$ is a left action on ${\cal U}_Z$
and, in particular, is a permutation  on ${\cal K}_X$. 
\end{proposition}
\begin{proof}
For any $\psi : I \to A$, we have
  $\Lambda^Z(x) \circ \psi = x \odot_Z \psi$ by definition;  that this
  is a permutation on ${\cal K}_X$ follows from the closure of ${\cal K}_X$.
\end{proof}\vspace{-3.5mm}

\begin{theorem}\label{thm:group-action-of-basis} 
  Suppose $k \in {\cal K}_Z$, and define $K = \Lambda^X(k)$; then $K$ is a
  group automorphism of ${\cal U}_Z$.
\end{theorem}
\begin{proof}
  Graphically we depict $K$ as:
  \beq\hspace{-1.6cm}
  K = \inlinegraphic{3.3em}{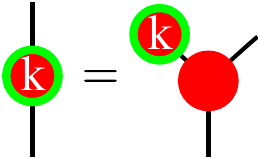}\ . 
  \eeq
  Since $k \in {\cal U}_X$, $K$ is unitary, and so is invertible.  We must
  show that if $\alpha \in {\cal U}_Z$ then also $K\circ\alpha \in {\cal U}_Z$.  This
  holds if and only if $\Lambda^Z(K\circ\alpha)$ is unitary;  we show
  this directly:
  \beq\hspace{-1.6cm}
    \inlinegraphic{10.8em}{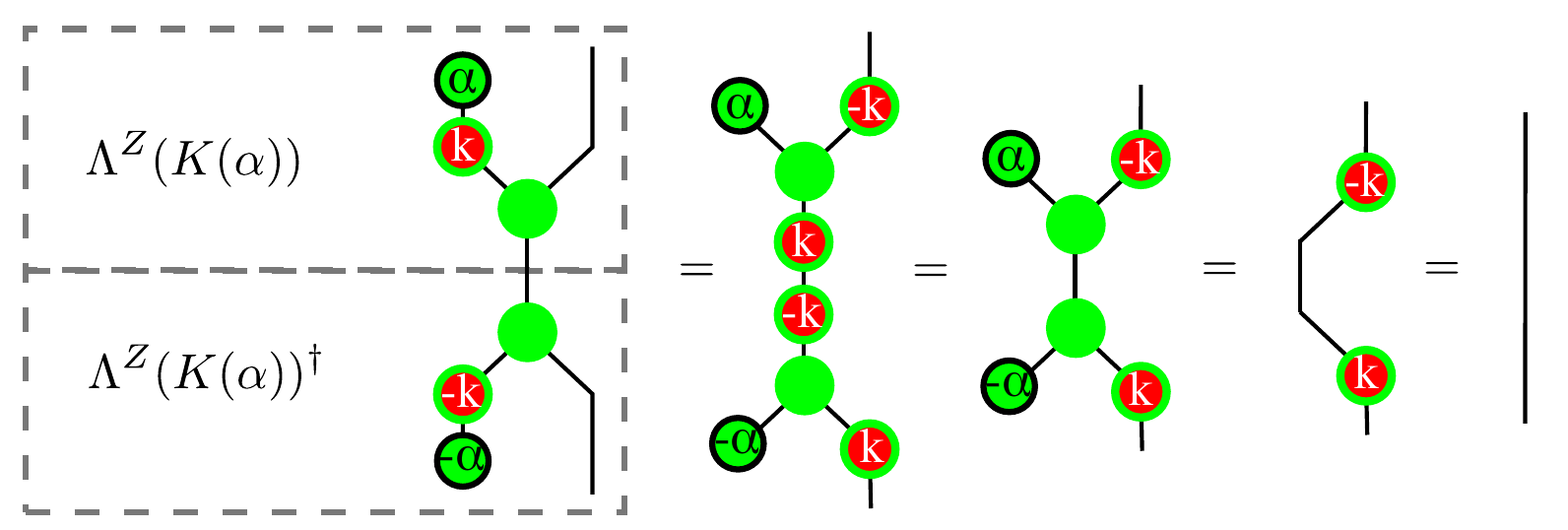}
   \eeq
 where the equations are by the comultiplication property, the
 unitarity of $K$, the unbiasedness of $\alpha $, and the unitarity of
 $K$ again.  It remains   to show that $K$ is a homomorphism of the
 group structure. 
  \begin{enumerate}
  \item $K\circ(\alpha \odot_Z \beta) 
    = (K\circ\alpha)\odot_Z(K\circ \beta)$ : 
 \beq\hspace{-2.2cm}
     \inlinegraphic{6.4em}{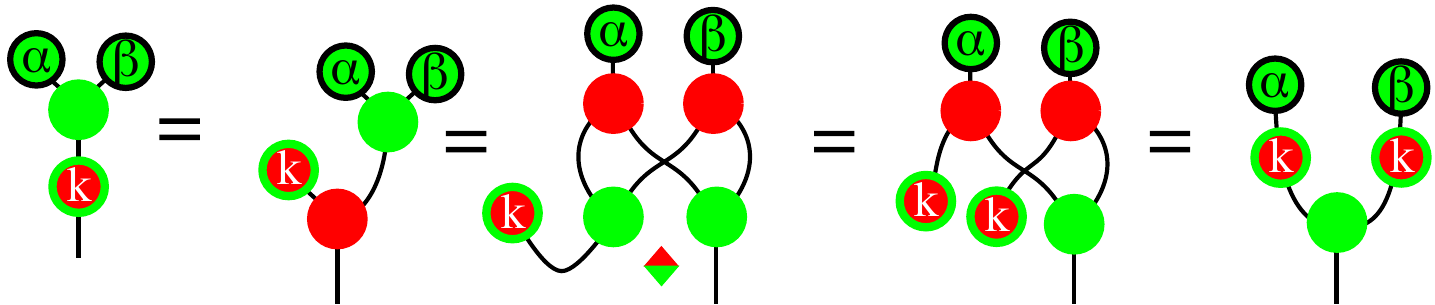} .
   \eeq
    The equations are: the definition of $K$; the bialgebra law; the
    classical property of $k$; and the definition of $K$.
  \item $K\circ\epsilon^\dag_Z = \epsilon^\dag_Z$ (upto global phase):
 \beq\hspace{-2.2cm}
     \inlinegraphic{3.6em}{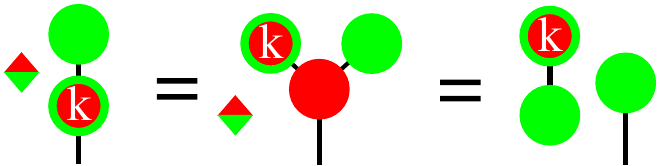}.
   \eeq
    The equation simply uses coherence of $\delta_X$ and $\epsilon_Z$;
    the result follows by dividing by the scalar factor as per
    Lemma~\ref{lem:scalar-unbiased}.
  \item $(K\circ\alpha)^{-1} = K \circ \alpha^{-1}$:
 \beq\hspace{-2.2cm}
     \inlinegraphic{5.5em}{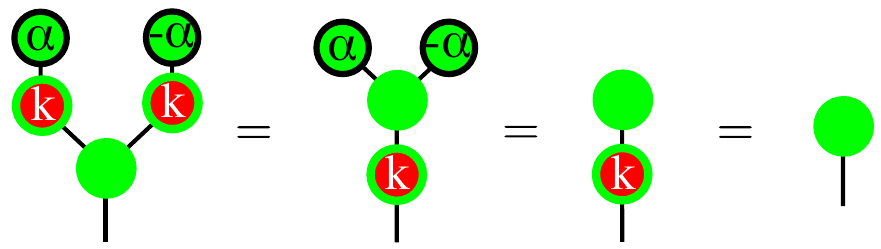}    
   \eeq
    where we relied upon the comultiplication property of $K$ and the
    unbiasedness of $\alpha$, showing that the inverse $K\circ \alpha$
    in ${\cal U}_Z$ is $K\circ \alpha^{-1}$ as required.
  \end{enumerate}
Hence $K$ is an automorphism of ${\cal U}_Z$.
\end{proof}\vspace{-3.5mm}

\begin{corollary}\label{cor:group-action}
  $({\cal K}_Z, \odot_X)$ is an abelian group of automorphisms on
  ${\cal U}_Z$ whose action is defined by $(x,z) \mapsto \Lambda^X(x)\circ z$.
\end{corollary}

The possibility that the classical points will act as
automorphisms on the corresponding phase group gives rise to
``interference'' phenomena; this is illustrated by the
example of the quantum Fourier transform of 
Section~\ref{Sec:QFT}.

In the following section we provide an example of the structure 
exposed in this section, for the spin Z and X observables, and 
show its role in the \zxcalculus.

\section{Deriving the \zxcalculus}\label{SEC:Customising}

The preceding sections derived the basic properties of complementary
observable structures in an arbitrary \dsmc.  The resulting theory,
although very rich, may feel rather abstract.  This abstraction is a
necessary consequence of working with arbitrary observable structures,
without specifying exactly what they are.  In any concrete setting,
given a fixed pair of observable structures, the remainder of the
structure can be constructed via simple direct computations.

This section will focus on a single concrete example and demonstrate
how to construct all the structures found in the earlier part of the
paper for a given pair of complementary observables.  Working in the
category of finite dimensional Hilbert spaces, we choose a pair of
complementary observable structures over $\mathbb{C}^2$---those
corresponding to the $Z$ and $X$ spin observables---and show how this
choice produces a concrete graphical theory for reasoning about
qubits.  The resulting theory is the \zxcalculus of
sections~\ref{Sec:ZXCalc} and \ref{sec:zx-calculus-use}: this section
will justify the simplified syntax of the \zxcalculus, and show that
the rules of the calculus (shown in Figure~\ref{fig:ZX-rules}) can be
derived from the theorems of the preceding sections.

As already noted in examples~\ref{Ex:Hilb} and \ref{Ex:Hilb2}, 
the category of finite dimensional Hilbert spaces is a \dsmc, and
hence the use of graphical notation is justified by Theorem
\ref{thm:JoyalStreet}.  This is already enough to justify the first
two generators of the \zxcalculus, namely the straight and crossing wires.
Since qubits are the only system of interest,
the type labels will be dropped from the edges of diagrams: all edges
are implicitly labelled by $\mathbb{C}^2$.  Having introduced the
edges, we now turn to the vertices.

The first vertices to be considered are those defining the
`green' observable structure $(\delta_Z,\epsilon_Z)$, corresponding to
the $Z$-spin observable, which is defined on $\mathbb{C}^2$ via the
linear maps,
\[
\begin{array}{ccc}
\inlinegraphic{2.15em}{delta} &\qquad&
\inlinegraphic{1.85em}{epsilon}
\\
\\
\delta_Z : \ket{i} \mapsto \ket{ii} &&
\epsilon_Z : \sqrt{2}\ket{+} \mapsto 1
\end{array}
\]
As discussed in Example \ref{ex:ZXphasegroup}, the points which are unbiased for $(\delta_Z,\epsilon_Z)$
have the form $\ket{\alpha_Z} = \ket{0} + e^{i\alpha}\ket{1}$ where
$0\leq \alpha <2\pi$, and hence the phase group consists of matrices
of the form:
\[
\Lambda^Z(\ket{\alpha_Z}) = \left(
\begin{array}{cc}
  1 &0 \\ 0 & e^{i\alpha}
\end{array}\right) = \inlinegraphic{3.8em}{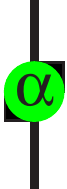}. 
\]
The group $({\cal U}_Z, \odot_Z)$ is therefore isomorphic to the circle, and
the operation $\odot_Z$ is  addition modulo $2\pi$.
The Pauli-$Z$ matrix is given by $\Lambda^Z(\pi)$.
Notice also that 
\[
\ninlinegraphic{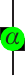}
= \left(\inlinegraphic{3.8em}{green-alpha}\right)^\dag 
= \left(
\begin{array}{cc}
  1 &0 \\ 0 & e^{i\alpha}
\end{array}\right)^\dag
=
\left(
\begin{array}{cc}
  1 &0 \\ 0 & e^{-i\alpha}
\end{array}\right)
= \ninlinegraphic{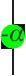}
\]
hence we can drop the ``corners'' from the diagrammatic notation and
simply write the negative angle in its place.
\[
\inlinegraphic{3.8em}{green-alpha} \mapsto
\ninlinegraphic{zx-green-alpha}
\qquad\qquad
\ninlinegraphic{small-green-alpha-dag} \mapsto 
\ninlinegraphic{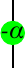}
\]
The spider rules (Theorem~\ref{thm:gen-spiderrule})
justify the first part of the \zxcalculus syntax, the family of `green'
vertices $Z^m_n(\alpha )$, along with rule \eruleSi and the first part
of rule \eruleSii.

The X-observable structure is essentially the same: defining  $(\delta_X, \epsilon_X)$ as follows:
\[
\begin{array}{ccc}
\inlinegraphic{2.15em}{red-delta} &\qquad& 
\inlinegraphic{1.85em}{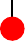}  \\
\\
\delta_X : \ket{\pm} \mapsto \ket{\pm\pm} &&
\epsilon_X : \sqrt{2}\ket{0} \mapsto 1 
\end{array}
\]
and it is easy
to verify that the corresponding phase group consists of rotations
around $X$, that is, matrices of the form
\[
\Lambda^X(\ket{\alpha_X}) = \left(
\begin{array}{cc}
 \cos \frac{\alpha}{2}  & \rmi  \sin \frac{\alpha}{2} \\ 
 \rmi  \sin \frac{\alpha}{2}  &  \cos \frac{\alpha}{2}
\end{array}\right) 
= \inlinegraphic{3.8em}{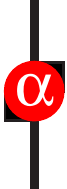}, 
\]
generated by the (unnormalised) unbiased points $\ket{\alpha_X} = \sqrt{2}(\cos
\frac{\alpha}{2}\ket{0} + \rmi\sin\frac{\alpha}{2}\ket{1})$.  We can simplify
the notation as we did for the $Z$-observable:
\[
\inlinegraphic{3.8em}{red-alpha} \mapsto \ninlinegraphic{zx-red-alpha}
\qquad\qquad
\ninlinegraphic{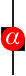} \mapsto 
\ninlinegraphic{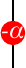}
\]
Of course, the decorated spider rules also apply to $(\delta_X,
\epsilon_X)$, and this produces the `red' family of the \zxcalculus,
the vertices $X^m_n(\alpha)$.

To complete rule \eruleSii, we appeal to one property not derived from
the formalism of complementary observables, namely the fact that
$(\delta_Z,\epsilon_Z)$ and $(\delta_X,\epsilon_X)$ induce the same
compact structure: 
\[
\left.\begin{array}{l}
\raisebox{2.5mm}{\inlinegraphic{3.5em}{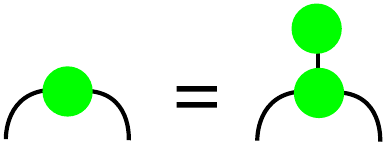}}=\delta_Z \circ \epsilon_Z^\dag\\
\raisebox{2.5mm}{\inlinegraphic{3.5em}{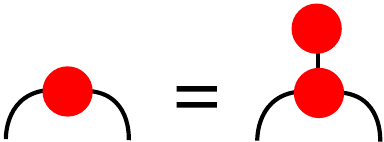}}= \delta_X \circ\epsilon_X^\dag
\end{array}\right\}=\ket{00} + \ket{11}=\raisebox{0.5mm}{\inlinegraphic{2.0em}{small-cap}} 
\]
Hence we obtain the remaining part of rule \eruleSii, as well as the bending wires.  

Further, since each object now comes with a unique $\dag$-compact
structure, we can treat the category as $\dag$-compact, and appeal to
Theorem~\ref{thm:graphical-dag-compact}.  This justifies rule \eruleT,
``only the topology matters''.

Having noted that $(\delta_Z,\epsilon_Z)$ and $(\delta_X,\epsilon_X)$
generate the same $\dag$-compact structure, we may also appeal to
Lemma~\ref{SQRTDis_selfconj}, that is:
\[
\inlinegraphic{2.6em}{deletingpointsup}\ = \ 
\inlinegraphic{1.35em}{sqrtDup}\ =  \ 
\inlinegraphic{1.35em}{SQRTD}\ = \ 
\inlinegraphic{2.6em}{deletingpoints}\,.
\]
Now, setting $\inlinegraphic{1.35em}{black-diamond} =
\inlinegraphic{1.35em}{SQRTD}$ \eruleDi, Proposition~\ref{prop:scalarvalue}
produces rule \eruleDii of the \zxcalculus.

The definition of $\delta_Z$ immediately shows that the classical
points of the $Z$ observable are $\ket{0}$ and $\ket{1}$; these points
are unbiased for $(\delta_X,\epsilon_X)$, corresponding to the angles
$0$ and $\pi$ in $({\cal U}_X,\odot_X)$.  Similarly, the classical points of
$\delta_X$ are $\ket{+}$ and $\ket{-}$, which are unbiased for
$(\delta_Z,\epsilon_Z)$, and again correspond to the angles $0$ and
$\pi$.  This being the case, we can again simplify the graphical
notation, and dispense with the two-coloured dots used in
sections~\ref{SEC:Complementarity} and \ref{sec:bialgebra} in favour
of a simpler convention:  
\begin{center}
\em a dot is unbiased for the observable structure of the same colour; \em
\end{center}
if it is labelled by $\pi$ or zero then it is classical for the other structure. 
In the \zxcalculus there is no need for dots of any other
kind, so we disallow them.  The translation between the more general
graphical language and the simplfied version used in the \zxcalculus
is summarised in table~\ref{tab:simple-transate}.
\begin{table}[h]
  \centering
  \begin{tabular}[h]{l|ccccccc}
    \textbf{General:} &
    \raisebox{1.5pt}{\inlinegraphic{2.88em}{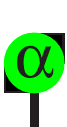}} &
    \raisebox{1.5pt}{\inlinegraphic{2.88em}{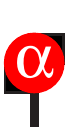}} &
    \inlinegraphic{2.4em}{CCSPoint} &
    \inlinegraphic{2.4em}{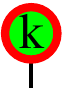} &
    \inlinegraphic{2.4em}{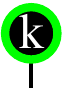} &
    \inlinegraphic{2.4em}{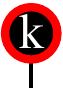} &
    \inlinegraphic{2.4em}{black-point} 
\\
\hline
\\
\textbf{Simplified:} &
    \inlinegraphic{2.88em}{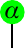} &
    \inlinegraphic{2.88em}{small-red-alpha-point} &
    \inlinegraphic{2.4em}{red-pi-short-point} &
    \inlinegraphic{2.4em}{green-pi-short-point} &
    \multicolumn{3}{c}{
      Not Allowed
    }
  \end{tabular}
  \caption{Translation between general and simplified graphical notation}
  \label{tab:simple-transate}
\end{table} 

Since each structure's classical points are unbiased for
the other structure, $(\delta_Z,\epsilon_Z)$ and $(\delta_X,\epsilon_X)$ are
\emph{complementary} as per
definition~\ref{defn:complementary-observables}.  Furthermore, we have
$\epsilon_X^\dag = \sqrt{2}\ket{0}$ and $\epsilon_Z^\dag =
\sqrt{2}\ket{+}$, so the two observable structures are also
\emph{coherent} as in definition~\ref{defn:coherent-observables},
implying rule \eruleBi of the \zxcalculus.

The classical points of $Z$ correspond to the angles $0$ and $\pi$
within the phase group of $X$, and vice versa.  Since these angles
form a two-element subgroup within the circle group,
$(\delta_X,\epsilon_X)$ and $(\delta_Z,\epsilon_Z)$ form a
\emph{closed} pair of complementary observables structures, as per
definition~\ref{def:colsedness}.  Noting that the triple
$\{\ket{0},\ket{1},\ket{+}\}$ forms a state basis for $\mathbb{C}^2$,
the conditions for Theorem~\ref{main:theorem} are satisfied: from
whence, bialgebraic commutation
(definition~\ref{def:bialgebraic-commutation}) implies rule \eruleBii,
and comultiplicative commutation
(definition~\ref{def:comultiplicative-commutation}) implies rule
\eruleKi.

\begin{table}[h]
  \centering
  \begin{tabular}{|c|c|c|c|}
    Observable & Classical Points & Unbiased Points & Phase Group
   \\
    \hline &&&\\
    $Z = (\delta_Z, \epsilon_Z)$ & 
    $\ket{0},\ket{1}$ &
    $\ket{0} + e^{i\alpha}\ket{1}$ &
    $Z_\alpha = \left(
      \begin{array}{cc}
        1&0\\0&e^{i\alpha} 
      \end{array}
    \right)$ 
    \\
    \inlinegraphic{2.2em}{delta},     \inlinegraphic{1.9em}{epsilon} &
    \inlinegraphic{2em}{red-epsilondag}, \inlinegraphic{2.2em}{red-pi-short-point}& 
    \inlinegraphic{2.8em}{green-alpha-short-point}&
    \inlinegraphic{2.8em}{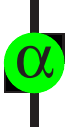}\rule{0em}{2em}
\\
    \hline &&&\\
    $X = (\delta_X, \epsilon_X)$ & 
    $\ket{+},\ket{-}$ &
    $\cos{\frac{\alpha}{2}}\ket{0} + i\sin{\frac{\alpha}{2}}\ket{1}$ & 
      $X_\alpha = \left(
        \begin{array}{cc}
          \cos \frac{\alpha}{2}  & i  \sin \frac{\alpha}{2} \\ 
          i  \sin \frac{\alpha}{2}  &  \cos \frac{\alpha}{2}
        \end{array}    \right)$ 
    \\
    \inlinegraphic{2.2em}{red-delta},     \inlinegraphic{1.9em}{red-epsilon} &
    \inlinegraphic{2em}{epsilondag},
    \inlinegraphic{2.2em}{green-pi-short-point} & 
    \inlinegraphic{2.8em}{red-alpha-short-point}&
    \inlinegraphic{2.8em}{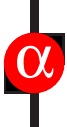}\rule{0em}{2em}
\\
\hline
\end{tabular}
\ \\ \ \\ \ \\  
  \begin{tabular}{|c|c|c|}
    Observable &  Classical Subgroup & Automorphism Action \\
    \hline &&\\
    $Z = (\delta_Z, \epsilon_Z)$ &
    $\id{}, \sf{X}$ &
    $ {\sf X} : Z_\alpha \mapsto Z_{-\alpha}$\\ 
    &&\\
    \inlinegraphic{2.2em}{delta},     \inlinegraphic{1.9em}{epsilon} &
    \inlinegraphic{2.9em}{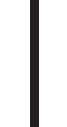}, \inlinegraphic{2.9em}{red-pi-short}& 
    \inlinegraphic{4.0em}{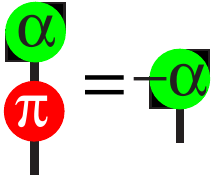}

\\
    \hline &&\\

&&\\
    $X = (\delta_X, \epsilon_X)$ &
    $\id{}, \sf{Z}$ &
    $ {\sf Z} : X_\alpha \mapsto X_{-\alpha}$\\
&&\\
    \inlinegraphic{2.2em}{red-delta},     \inlinegraphic{1.9em}{red-epsilon} &
    \inlinegraphic{2.9em}{id-short}, \inlinegraphic{2.9em}{green-pi-short}& 
    \inlinegraphic{4.0em}{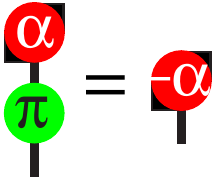}
\\
\hline
  \end{tabular}

  \caption{Summary of the group structure for qubits}
  \label{tab:qbitgrpstruct} 
\end{table}

The remaining piece of the structure to be described is the action of
${\cal K}_Z$ on ${\cal U}_Z$ discussed in
Section~\ref{SEC:GroupActionEtc}.  Since $X^1_1(\pi) = {\sf X}$ we see
that the non-trivial element of ${\cal K}_Z$ sends $\ket{\alpha_Z}
\mapsto \ket{-\alpha_Z}$; i.e. $\sf X$ assigns elements of ${\cal
  U}_Z$ to their inverses.  The action of ${\cal K}_X$ on ${\cal U}_X$
is exactly dual.  This provides the rule \eruleKii of the \zxcalculus.
The group structure is summarised in Table~\ref{tab:qbitgrpstruct}.

The preceding has shown that the  \zxcalculus, with one exception, is
readily deduced from the algebraic properties of the $Z$ and $X$
observables:  indeed everything else followed from that choice.  The
single exception is the  $H$ vertex and the associated rule \eruleC.
The addition of a special symbol for the Hadamard gate is simply to
make certain calculations easier, and it is not essential to the calculus.

\section{Non-determinism, mixed states, and classical data flow}
\label{sec:non-determ-mixed}

The graphical treatment so far has been limited to pure states, and
therefore does not capture the full behavour of quantum measurements,
decoherence, classical control, and a host of other phenomena of great
practical importance in quantum computation.

This section will present three different extensions of the
graphical lanaguage to account for mixed states, and dispense with
the simplified treatment of measurements as projections.  We briefly
describe each of these, and provide the  same example in each one: the
quantum teleportation protocol.  For full details, see the references.

The most general is Selinger's CPM construction \cite{Selinger} which,
given any category of ``pure states'' and ``pure maps'', constructs a
category of ``mixed states'' and ``completely positive maps''.  Within
the resulting category, Paquette, Pavlovic and one of the authors
defined a plethora of classicality-related concepts such as
decoherence, measurement, probability distribution, stochastic map,
function on classical data, etc., by relying on observable structure
\cite{CPaqPav} ---some of these results were already present in
Carboni and Walters' seminal paper \cite{CarboniWalters}.  An
axiomatic account on Selinger's CPM construction and classical
concepts therein was provided by Perdrix and an author of this paper
\cite{SelingerAxiom,CPer}; this introduces a new element to the
graphical calculus corresponding to the environment, and a
corresponding axiom (and some `coherence conditions').  Finally,
Perdrix and the other author of this paper have introduced a version
of the \zxcalculus parameterised by a set of variables which encode
the outcome of measurements, and the dependence of other elements upon
them.  These \emph{conditional} diagrams have been used in the context
of measurement-based quantum computation to study determinism and
information flow \cite{DunPer2010}.

These three approaches are alternatives, and depending on the
situation one may be preferable to another.


\subsection{The CPM construction and classical concepts therein {\rm\cite{Selinger,CPaqPav}}}

We extend the graphical language by a construction which is formally similar to the way in which 
classical probabilities are described by density operators.
Graphically, classical data and classical operations are represented by
a single wire, while quantum data and quantum operations are
represented by double wires.  
The passage from a double wire to a single via a dot encodes decoherence.
This encoding:
\[
\hspace{2.3cm}{\mbox{1 wire/box}
\over
\mbox{classical}}
=
{\mbox{2 wires/boxes}
\over
\mbox{quantum}}
\]
is also present in Dirac notation; 
for a mixed state
$\sum_i \omega_i |\psi_i\rangle\langle\psi_i|$ the clasical
probabilistic state $(\omega_1,\ldots,\omega_n)$ occurs only once
while the quantum states occur both as a ket $|\psi_i\rangle$  and as
a bra $\langle\psi_i|$.   Pure quantum operations, controlled pure
quantum operations, and destructive measurements are of the forms: 
\bc
\inlinegraphic{3.6em}{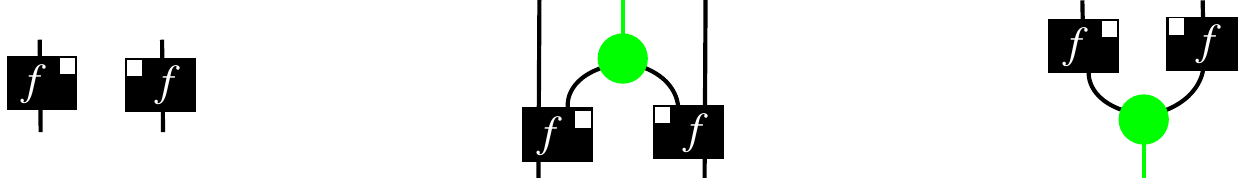},
\ec
respectively, where for clarity we choose to colour the classical
wires in the colour of the observable structure whose classical
points will encode the classical data.

In fact,  the spiders of the graphical language were initially introduced under
the name \emph{classical structures} in \cite{CPav} to model classical
data in quantum informatic protocols.   Hence, in this setting, a
single concept can account both for quantum observables,
complementarity, phases, as well as classical  information flow.   We
will be very brief on the  use of spiders in order to describe
classical data flow; we refer the reader to \cite{CPaqPav} for more
details.


\begin{example}
  \label{ex:teleportation}
  We claim that the following constitutes the quantum teleportation
  protocol, including the classical correction: \bc
  \inlinegraphic{18.0em}{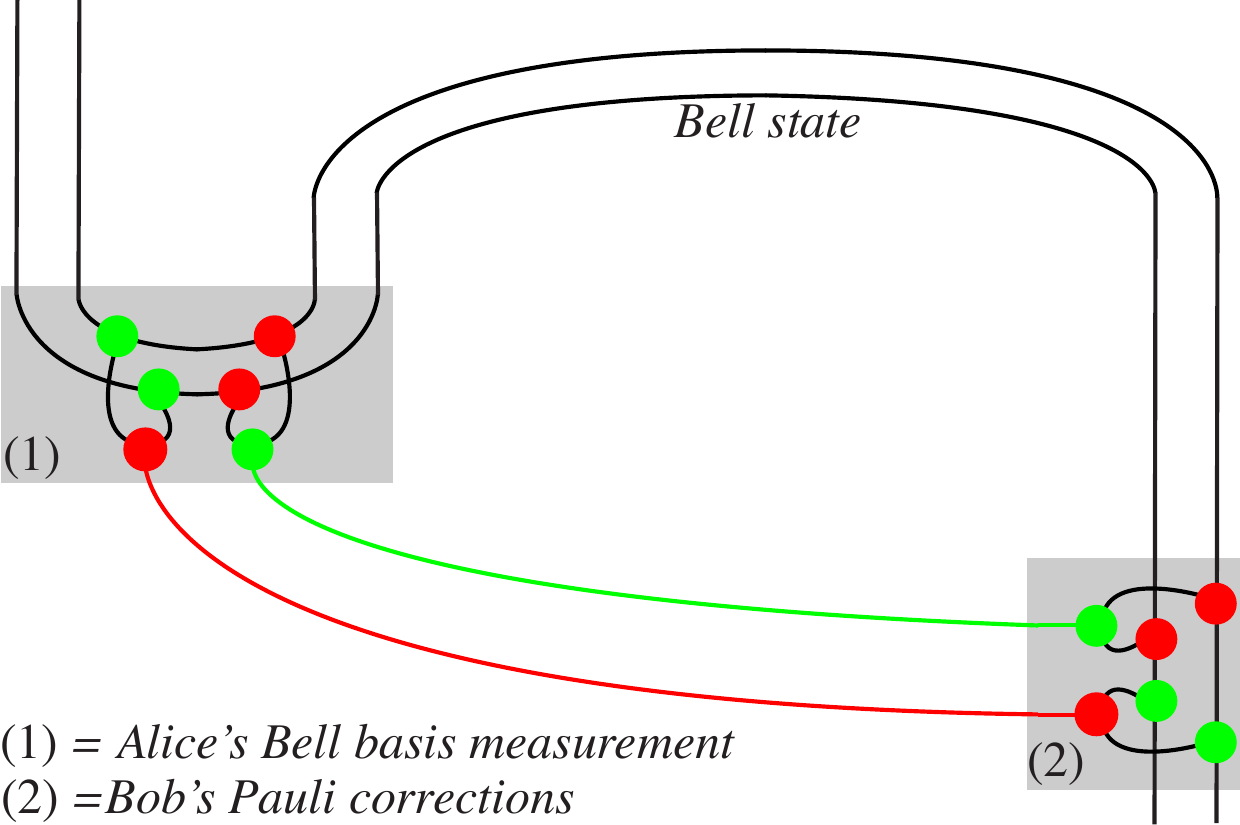}.  \ec The green and the red
  wire represent the two qubits Alice has to send to Bob to inform him
  of the measurement outcome.  The Bell state, Bob's Pauli corrections
  and Alice's Bell basis measurement can be rewritten respectively as:
  \bc \inlinegraphic{4.4em}{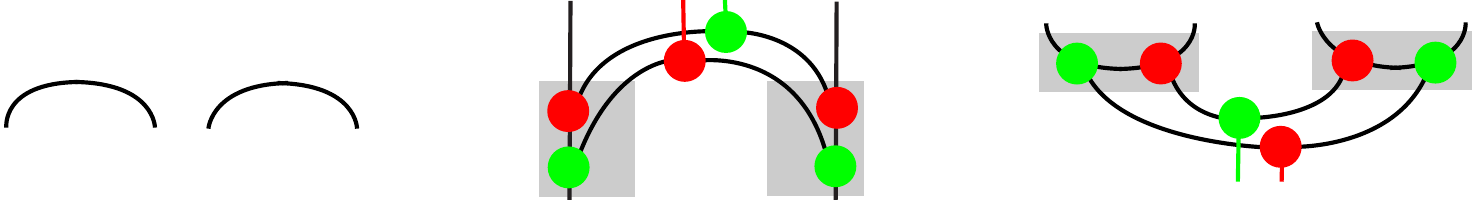}.  \ec hence, they are of
  the forms shown above.

  The picture might seem somewhat complicated; the reason for this is
  that in order to display the quantum-classical distinction
  graphically, all the quantum operations are doubled.  If we hide one
  of the copies it becomes much clearer what is going on: \bc
  \inlinegraphic{18.0em}{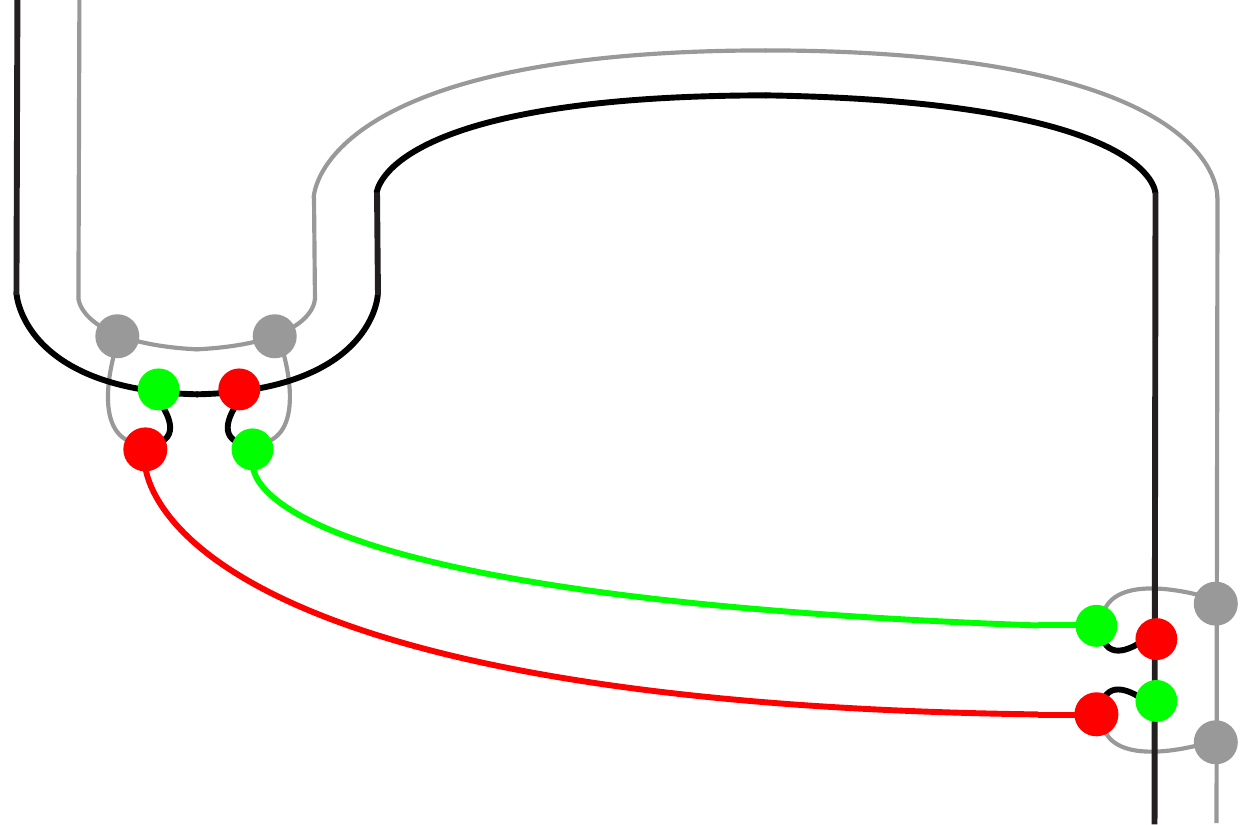}.  \ec

  We demonstrate that the measurement and corrections in this picture
  are indeed the ones we claim them to be.  By `selecting' the control
  operation, that is, by inserting a classical point at the input, we
  obtain the Pauli corrections: 
  \[
\raisebox{0.5cm}{ \inlinegraphic{6.0em}{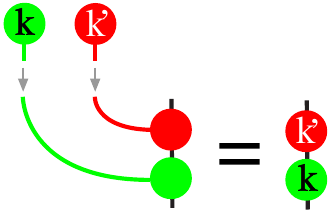}} =\left\{\begin{array}{l}
    \inlinegraphic{2.9em}{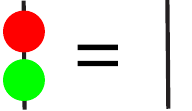}={\sf I}\vspace{2mm}\\
    \inlinegraphic{2.9em}{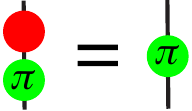}={\sf Z}\vspace{2mm}\\
    \inlinegraphic{2.9em}{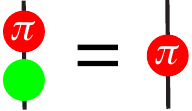}={\sf X}\vspace{2mm}\\
    \inlinegraphic{2.9em}{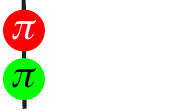}={\sf X}\circ{\sf Z}
  \end{array}\right.
\] 
Similarly, by post-`selecting' the measurement outcome, that is,
by inserting the adjoint of a classical point at the output, we obtain
the Bell-basis measurement: 
\[\hspace{-2cm}
\raisebox{-0.5cm}{\inlinegraphic{4.5em}{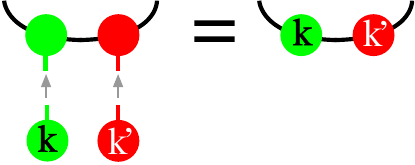}} =\left\{\begin{array}{l}
  \inlinegraphic{1.6em}{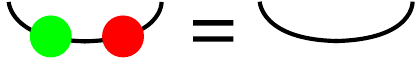}=|00\rangle+|11\rangle\vspace{2mm}\\
  \inlinegraphic{1.6em}{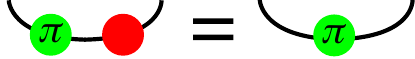}=|00\rangle-|11\rangle\vspace{2mm}\\
  \inlinegraphic{1.6em}{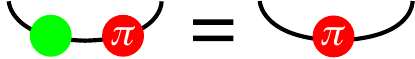}=|01\rangle+|10\rangle\vspace{2mm}\\
  \inlinegraphic{1.6em}{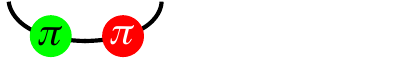}=|01\rangle-|10\rangle
\end{array}\right.\vspace{-1cm}
\]

We now compute the overall result of the teleportation protocol
diagrammatically: \bc \inlinegraphic{11.1em}{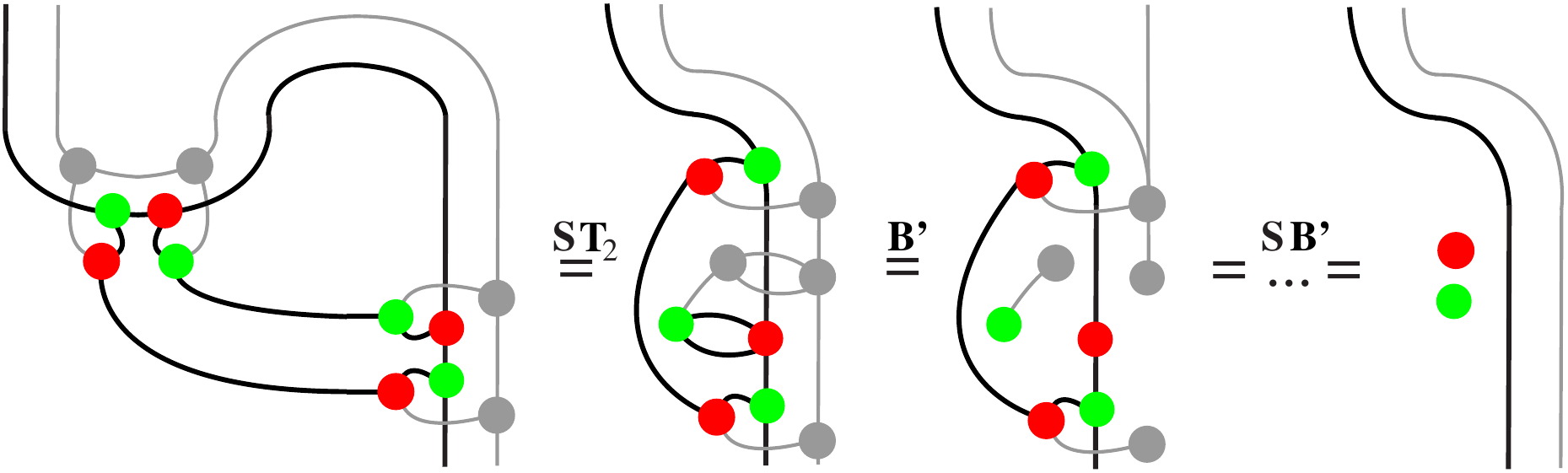}
\ec
Note how the ${\bf B}'$-rule causes the flow of classical data to
disconnect from the flow of quantum data, which of course, should not
depend on classical data anymore.  The scalars that remain at the end
are a consequence of the fact that we didn't normalise the Bell state,
nor the Bell basis measurement, nor the ${\bf B}'$-rule; if we had
done so all scalars would have cancelled out.
\end{example}

\subsection{Classicality via environment   {\rm\cite{SelingerAxiom,CPer}}}\label{sec:axioms-environment}

Roughly equivalent to the above `construction' is the following `axiomatic account'--- see \cite{SelingerAxiom} for the precise correspondence.   We consider two kinds of morphisms, `pure ones' (or `sharp') and `mixed ones'  (or `unsharp').  We represent the pure morphisms as we did throughout this paper: 
\bc
\inlinegraphic{4.8em}{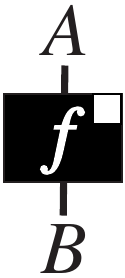}
\ec
and the mixed ones, for example, as rounded variants of the pure ones: 
\bc
\inlinegraphic{4.8em}{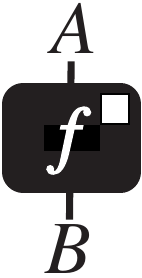}
\ec
We then introduces a new graphical element $\raisebox{-0.4mm}{\inlinegraphic{1.8em}{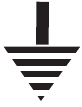}}$,
one for each object, which is coherent with the $\dag$-compact structure, and is subject to an axiom:
\beq\label{environment_axiom}
\inlinegraphic{6.2em}{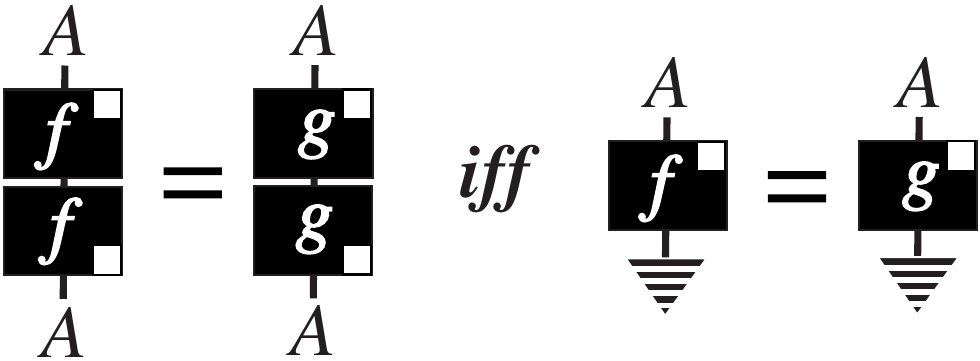}
\eeq
which is valid for all pure morphisms $f:A\to B$ and $g:A\to C$.  To obtain a precise match with the CPM construction, we also need a 
\em purification axiom \em which states that any mixed morphism can be obtained from a pure one by composing the latter with the environment.  From (\ref{environment_axiom}) it follows that \em classical channels\em, that is:
\bc
\inlinegraphic{3.6em}{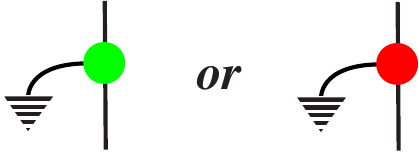}
\ec
are idempotent \cite{CPer}.  Here the color reflects how the classical data was obtained, i.e., in which measurement.  This idempotence in fact the only equation which plays some role in the teleportation protocol, which now goes as follows:
\bc
\inlinegraphic{5.2em}{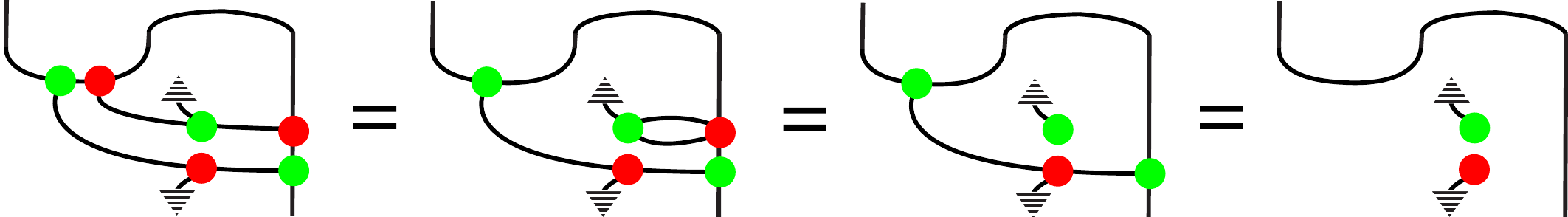}
\ec

\subsection{Conditional diagrams  {\rm\cite{DunPer2010}}}\label{sec:conditional-diagrams}

While the previous two approaches represent classical information flow
internally, as particular wires in the diagrams, the final alternative
presentation of non-determinism ignores the issue of (classical)
information flow, and focuses on classical correlations which are
mediated externally.

Let ${\cal V}$ be a set of formal variables.  A \emph{conditional
  diagram} is a diagram $\diagD$ of the \zxcalculus and, for each Z and
X vertex $v$ in \diagD, an associated subset $\mathcal{U}_v \subseteq
\mathcal{V}$, possibly empty.  A \emph{valuation} is function $f:\mathcal{V} \to
\{0,1\}$.   Each pair $(\diagD,f)$ of a conditional diagram and a
valuation determines an \emph{evaluated diagram} $\diagD_f$, which is
obtained from $\diagD$ by modifying the phase $\alpha$ at each vertex $v$
as follows:
\[
\alpha \mapsto \left\{ 
\begin{array}{cl}
  \alpha & \text{if } \prod_{u\in\mathcal{U}_v} f(u)  = 1\\
  0 & \text{otherwise}
\end{array}
\right.
\]
and forgetting the sets $\mathcal{U}_v$.  Each variable corresponds to
a two-valued measurement and each valuation $f$ corresponds to a
possible set of measurement outcomes.  The evaluated diagram
$\diagD_f$ corresponds to the process that occurs when the measurement
outcomes corresponding $f$ are observed, including both the
side-effect of the measurement itself, and any other operations which
depend classically upon its outcome.

Since each $\diagD_f$ is a diagram of the \zxcalculus, it has an
associated linear map, $D_f$, as described in
Section~\ref{sec:interpr-zx-calc}.  We construct a
superoperator for $\diagD$ by summing over all possible valuations:
\[
\mathbf{D} : \rho \mapsto \sum_{f} D_f \rho D_f^\dag.
\]
Conditional diagrams can use the same equational rules as the plain \zxcalculus, with the proviso that a rule can only be applied in $\diagD$
if it could be applied to every $\diagD_f$.  

The conditional diagrams approach is essentially a formalisation of
the informal reasoning demonstrated in Section~\ref{sec:telep-prot},
as the following example wil clearly demonstrate.

\begin{example}
  Let $\mathcal{V} = \{a,b\}$, where each variable corresponds to one
  of the single-qubit measurements comprising the Bell basis
  measurement in the teleportation protocol.  Teleportation protocol
  can then be formalised as follows:
  \begin{equation*}
    \ninlinegraphic{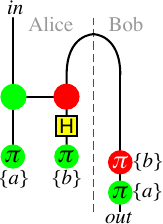} 
    = \ninlinegraphic{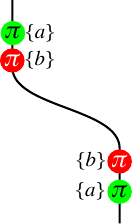}
    = \ninlinegraphic{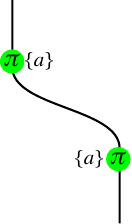} 
  \end{equation*}
  The crux of the encoding is that the subdiagram 
  \[
  \mathfrak{M} =  \ninlinegraphic{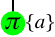}
  \]
  denotes the superoperator
  \[ 
  \rho \mapsto \bra{+}\rho\ket{+}  + \bra{-}\rho\ket{-},
  \]
  and thus correctly encodes the measurement.
\end{example}

The conditional diagram approach is very well suited to the one-way
model, since the quantum part of the system can be very complicated,
whereas the classical information flow can ususally be taken for
granted.  For many purposes, for example to translate a one-way
pattern to a circuit, we need not evaluate any valuation, and can
simply use the equational rules on the conditional diagram directly.

  The measurement calculus \cite{DKP} includes several commands which
  depend upon on classical bits, here ranged over by $s$ and $t$:
  \begin{itemize}
  \item ${}^s[M^\alpha_j]^t$ -- measure qubit $j$ in the basis
    $\ket{0} + \rme^{(-1)^s\rmi(\alpha + t\pi)}\ket{1}$.
  \item $X_j^s$ and $Z_j^s$ -- apply a Pauli ${\sf X}$ operator (or
    ${\sf Z}$) to qubit $j$, if $s = 1$; otherwise do nothing.
  \end{itemize}
  Taking $\mathcal{V}$ to be the set of measured qubits in the
  pattern, the measurement calculus can be encoded using conditional
  diagrams according to the table below.
  \begin{center}
    \begin{tabular}{|c|c|c|c|c|}
      \hline 
      $N_i$ & $E_{ij}$  & $M_i^{\alpha}$ & $X_s^i$ & $Z^i_t$ \\
      \hline
      \ninlinegraphic{small-green-epsilondag} 
      & \ninlinegraphic{small-CZ} 
      & \ninlinegraphic{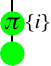} 
      & \ninlinegraphic{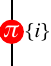} 
      & \ninlinegraphic{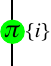} 
      \rule[-4.5mm]{0pt}{11mm}
      \\
      \hline
    \end{tabular}
  \end{center}
\noindent
  Note that conditional measurement ${}^s[M^\alpha_i]^t$ is equivalent
  to the sequence $M^\alpha_i X_i^s Z_i^t$, and therefore can be replaced
  by it.

\begin{example}
  Consider the pattern:
  \[
  X^{v_3}_4 Z^{v_2}_4 Z^{v_2}_1 M^0_3 M^0_2 E_{13} E_{23} E_{34} N_3 N_4.
  \]
  An unreliable quantum software engineer claims that it computes the
  a \CX on its inputs, regardless of the result of the 
  measurements.  We can check this claim using equational reasoning on
  the conditional diagram.  Let $\mathcal{V} = \{v_2, v_3\}$,
  corresponding to the outcomes of the measument of qubits 2 and 3.
  \begin{eqnarray*}
    \fl
    \ninlinegraphic[0.8]{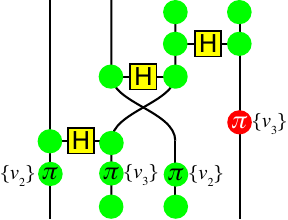} 
    &=& \!\!\!\! \ninlinegraphic[0.8]{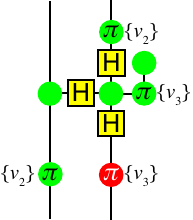} 
    = \!\!\!\!  \ninlinegraphic[0.8]{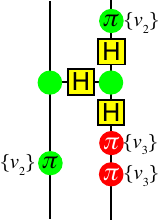} 
    = \!\!\!\! \ninlinegraphic[0.8]{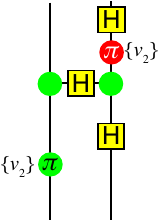}
    \\[0.5em]
    &=& \ninlinegraphic[0.8]{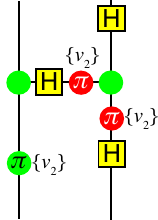} 
    =  \ninlinegraphic[0.8]{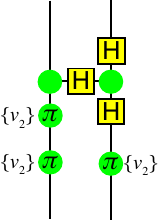} 
    \!\!\!\! =  \ninlinegraphic[0.8]{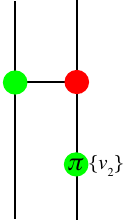} 
  \end{eqnarray*}
  Since the final diagram still has the  variable $v_2$ occuring in it,
  the original claim was false:  the result of this pattern depends
  upon the result of the measurement.  However from the diagram we can
  can easily debug the pattern by adding a final correction $Z^{v_2}_4$
  correction at the end.  The corrected pattern is:
  \[
  Z^{v_2}_4 X^{v_3}_4 Z^{v_2}_1 M^0_3 M^0_2 E_{13} E_{23} E_{34} N_3 N_4.
  \]
  These ideas are developed more fully in \cite{DunPer2010}.
\end{example}

\begin{remark}
  The notion of conditional diagrams, and their superoperator
  interpretation, can be easily generalised beyond the minimal setting
  we have presented here.  Variables could take values in an arbitrary
  set $A$---indeed not all variables need have the same set of 
  values---and the modification of the diagram based on the
  valuation function could be arbitarily more complicated without
  changing anything essential.  Finally, given an interpretation
  functor from diagrams into some $\dag$-symmetric monoidal category
  $\catC$, then a ``superoperator'' can be constructed using a
  suitable semigroup enrichment of $\catC$.
\end{remark}

\section{Conclusion}\label{sec:conclusions}

In this paper we introduced a simple and intuitive--- but at the same time universal ---graphical calculus for qubits, the \zxcalculus, and gave many example applications. We studied its mathematical underpinning in great detail, in particular:
\bit
\item We obtained a purely diagrammatic characterization of complementarity that extends to observable  structures  in arbitrary \dsmcs, in terms of the Hopf law:
\[
\inlinegraphic{4.7em}{hopflawdual}
\]
\item We identified a strong form of complementarity for observable  structures  in arbitrary \dsmcs, when the observable structures form a scaled bialgebra:
\[
\inlinegraphic{6.86em}{bialgebraALL}.
\]
We identified a number of equivalent alternative formulations:
\bc
\  \ \inlinegraphic{3.3em}{mixcoloractionII}\quad\ \ \ 
\inlinegraphic{3.6em}{mixcoloractionIV}\quad\ \ \ 
\inlinegraphic{3.8em}{Closedness1bis}
\ec
\item We identified a group structure on phases for observable  structures  in arbitrary \dsmcs, and proved a generalization of the spider rules, now involving phases:
\bc
\inlinegraphic{5.8em}{spidergencomp}. 
\ec
\eit
As already mentioned at the end of the introduction, meanwhile our results have been applied in many contexts, ranging from quantum information and quantum foundations to automated reasoning.

We end by mentioning a number of issues that require further exploration:
\bit

\item Is there an elegant extension of the \zxcalculus to a \textsc{zxy}-calculus\xspace? Note that this will necessarily involve the dualisers of Section~\ref{SEC:Hopfbis}. Would such  a \textsc{zxy}-calculus\xspace be different when either modeling ${\bf Stab}$ or ${\bf Spek}$? (cf.~Example \ref{ex:stabspek})

\item What is the precise connection between the stabiliser formalism and our graphical reasoning scheme? (Given that both are adequate tools for studying measurement-based quantum computing.) Is the stabiliser formalism and quantum error correction fully captured by the \zxcalculus?  

\item Theorem~\ref{thm:subgroup} adds extra structure with a clear physical interpretation to the phase group: while the elements of ${\cal K}$ may be seen as measurement outcomes, the corresponding cosets can be interpreted as choices of measurements. This may provide a foundation  for an axiomatic analysis of non-locality and contextuality. Some work in this direction has already been done \cite{BillNL}.

\item Can the \zxcalculus elegantly model many body states other than graph states, e.g., matrix product states or other states arising in condensed matter physics?

\item Does our characterization of complementarity extend in some way or another to observables described by infinite dimensional Hilbert spaces?  What is the connection of the above depicted laws to canonical commutation relations?

\eit

\section*{Acknowledgements}

Work supported by the authors' respective \textsc{epsrc} advanced and
postdoctoral research fellowships, by the \textsc{fnrs}, by \textsc{ec
  fp6 strep qics}, by an \textsc{fqx}i large grant and by \textsc{onr} grant, and we also
acknowledge support of Perimeter Institute for Theoretical Physics
which hosted both of the authors.  We thank Bertfried
Fauser, Prakash Panangaden, Simon Perdrix, Stefano Pironio, Marni
Dee Sheppeard and the referees for valuable feedback on an earlier version of this
paper, and to Stephen Brierly and Stefan Weigert for providing the
counter-example of Remark~\ref{remark:hadamard}.

\section{References}
\label{sec:references}




\end{document}